\documentclass[fleqn,usenatbib]{mnras}

\usepackage{newtxtext,newtxmath}

\usepackage[T1]{fontenc}

\DeclareRobustCommand{\VAN}[3]{#2}
\let\VANthebibliography\thebibliography
\def\thebibliography{\DeclareRobustCommand{\VAN}[3]{##3}\VANthebibliography}

\usepackage{graphicx}	
\usepackage{amsmath}	
\usepackage{xspace}
\usepackage{xcolor}
\usepackage{tabularx}
\makeatletter 
  \patchcmd{\NAT@citex}
    {\@citea\NAT@hyper@{%
      \NAT@nmfmt{\NAT@nm}%
      \hyper@natlinkbreak{\NAT@aysep\NAT@spacechar}{\@citeb\@extra@b@citeb}%
      \NAT@date}}
    {\@citea\NAT@nmfmt{\NAT@nm}%
    \NAT@aysep\NAT@spacechar\NAT@hyper@{\NAT@date}}{}{}

  \patchcmd{\NAT@citex}
    {\@citea\NAT@hyper@{%
      \NAT@nmfmt{\NAT@nm}%
      \hyper@natlinkbreak{\NAT@spacechar\NAT@@open\if*#1*\else#1\NAT@spacechar\fi}%
        {\@citeb\@extra@b@citeb}%
      \NAT@date}}
    {\@citea\NAT@nmfmt{\NAT@nm}%
    \NAT@spacechar\NAT@@open\if*#1*\else#1\NAT@spacechar\fi\NAT@hyper@{\NAT@date}}
    {}{}
\makeatother

\newcommand{\msun}{{\,\rm M_\odot}}

\newcommand{\kms}{\,{\rm km}\,{\rm s}^{-1}}
\newcommand{\cm}{\,{\rm cm}}

\newcommand{\Gyr}{\,{\rm Gyr}}
\newcommand{\K}{\,{\rm K}}

\newcommand{\Mpc}{\,{\rm Mpc}}

\newcommand{\LCDM}{\,\rm \Lambda CDM}

\newcommand{\thesan}{\textsc{thesan}\xspace}
\newcommand{\thesanzoom}{\textsc{thesan-zoom}\xspace}

\def\aap{A\&A}
\def\apj{ApJ}

\def\apjl{ApJ}
\def\mnras{MNRAS}
\def\araa{ARA\&A}

\def\nat{Nature}

\def\apjs{ApJS}

\title[Multi-scale SFE at high redshift]{The \thesanzoom project: Star formation efficiency from giant molecular clouds to galactic scale in high-redshift starbursts}

\author[Z. Wang et al.]{\parbox{17.5cm}{
Zihao Wang$^{1,2}$,
Xuejian Shen$^{1}$\thanks{E-mail: \href{mailto:xuejian@mit.edu}{xuejian@mit.edu}},
Mark Vogelsberger$^{1}$,
Hui Li$^{3}$,
Rahul Kannan,$^{4}$
Ewald Puchwein,$^{5}$
Aaron Smith,$^{6}$
Josh Borrow,$^{7}$
Enrico Garaldi,$^{8,9}$
Laura Keating,$^{10}$
Oliver Zier,$^{11}$
William McClymont,$^{12,13}$
Sandro Tacchella,$^{12,13}$
Yang Ni$^{14}$
and
Lars Hernquist$^{11}$
}
\\ \vspace{0.2cm} \\
$^1$ Department of Physics, Kavli Institute for Astrophysics and Space Research, Massachusetts Institute of Technology, Cambridge, MA 02139, USA \\
$^2$ School of Astronomy and Space Science, Nanjing University, Nanjing, Jiangsu 210093, People’s Republic of China \\
$^3$ Department of Astronomy, Tsinghua University, Beijing 100084, People’s Republic of China \\
$^4$ Department of Physics and Astronomy, York University, 4700 Keele Street, Toronto, ON M3J 1P3, Canada \\
$^5$ Leibniz-Institut f\"ur Astrophysik Potsdam, An der Sternwarte 16, 14482 Potsdam, Germany \\
$^6$ Department of Physics, The University of Texas at Dallas, Richardson, TX 75080, USA \\
$^7$ Department of Physics and Astronomy, University of Pennsylvania, 209 South 33rd Street, Philadelphia, PA 19104, USA \\
$^8$ Kavli Institute for the Physics and Mathematics of the Universe, The University of Tokyo, 5-1-5 Kashiwanoha, Kashiwa, 277-8583, Chiba, Japan \\
$^9$ Institute for Fundamental Physics of the Universe, via Beirut 2, 34151 Trieste, Italy \\
$^{10}$ Institute for Astronomy, University of Edinburgh, Blackford Hill, Edinburgh, EH9 3HJ, UK \\
$^{11}$ Center for Astrophysics | Harvard \& Smithsonian, 60 Garden St, Cambridge, MA 02138, USA\\
$^{12}$ Kavli Institute for Cosmology, University of Cambridge, Madingley Road, Cambridge CB3 0HA, UK \\
$^{13}$ Cavendish Laboratory, University of Cambridge, 19 JJ Thomson Avenue, Cambridge CB3 0HE, UK \\
$^{14}$ Institute for Advanced Study, Tsinghua University, Beijing 100084, People’s Republic of China
}

\date{Accepted XXX. Received YYY; in original form ZZZ}

\pubyear{2025}

\begin{document}
\label{firstpage}
\pagerange{\pageref{firstpage}--\pageref{lastpage}}
\maketitle

\begin{abstract}
Star formation in galaxies is inherently complex, involving the interplay of physical processes over a hierarchy of spatial scales. In this work, we investigate the connection between global (galaxy-scale) and local (cloud-scale) star formation efficiencies (SFEs) at high redshifts ($z\gtrsim 3$), using the state-of-the-art cosmological zoom-in simulation suite \thesanzoom. We find that the galaxy-scale average SFE, $\langle \epsilon^{\rm gal}_{\rm ff} \rangle$, scales with $M_{\rm halo}^{1/3}\,(1+z)^{1/2} \sim V_{\rm vir}$, consistent with expectations from feedback-regulated models. On cloud scales, we identify giant molecular clouds (GMCs) in a broad sample of high-redshift starbursts spanning a wide range of halo masses and redshifts. Star formation in these systems is predominantly hosted by filamentary GMCs embedded in a dense and highly turbulent interstellar medium (ISM). GMCs exhibit remarkably universal properties, including mass function, size, turbulence, and surface density, regardless of the environment in which they are identified. The global gas depletion time (and the Kennicutt–Schmidt relation) is determined by the GMC mass fraction in the ISM, while the cloud-scale SFE shows little variation. In particular, we find a nearly constant gas surface density of $\Sigma_{\rm GMC} \approx 70 \msun\,\mathrm{pc}^{-2}$ across different host galaxies. Nevertheless, we identify two regimes where phases with high SFE can arise. First, stars may form efficiently in the shock fronts generated by feedback from a preceding starburst. Second, the increasing background dark matter surface density with redshift may contribute to the gravitational potential of clouds at $z \gtrsim 8$ and confine them in high-SFE phases over extended periods. 
\end{abstract}

\begin{keywords}
methods:numerical -- galaxies:high-redshift -- galaxies:star-formation -- ISM:clouds
\end{keywords}



\section{Introduction} \label{sec:intro}

Recent \textit{James Webb Space Telescope (JWST)} observations suggest a potential tension with predictions from the standard $\LCDM$ cosmology. Specifically, there appears to be an order-of-magnitude excess of massive galaxies at redshifts $z \sim 5-9$, as well as an excess in their stellar mass content~\citep[e.g.][]{Labbe2023, Xiao2024, Casey2024, Wang2025}. In addition, \textit{JWST} has identified a surprisingly large population of ultraviolet (UV)-bright galaxies at $z \gtrsim 10$ \citep[e.g.][]{Finkelstein2022,Harikane2023,Donnan2023,Robertson2024}, whose number densities exceed most theoretical expectations from pre-\textit{JWST} models~\citep[e.g.][]{Tacchella2018, Behroozi2020, Kannan2023}. 
The formation and early evolution of these cosmic giants present a major challenge to our understanding of galaxy formation. Their existence has sparked active debate in the literature and is generally interpreted in several ways, including but not limited to the following:
(1) a deviation from the standard cosmological model that allows more massive dark matter (DM) haloes to form at earlier times~\citep[e.g.][]{Klypin2021, Shen2024-ede, Padmanabhan2023, Parashari2023, Sabti2024};
(2) a substantially enhanced baryon conversion efficiency compared to that observed in the local Universe~\citep[e.g.][]{Mason2023, Dekel2023, Li2024};
(3) a top-heavy stellar initial mass function (IMF)~\citep[e.g.][]{Inayoshi2022,Yung2024,Cueto2024,Trinca2024};
(4) bursty star formation~\citep[e.g.][]{Mirocha2023,Shen2023,Sun2023,Gelli2024,Kravtsov2024, Semenov2024-turbulentsf,Semenov2024-disk}, and (5) zero
dust attenuation~\citep[e.g.][]{Nath2023}. Among all interpretations, it is theoretically favored that star formation may proceed with substantially higher efficiency under the extreme physical conditions of the high-redshift Universe~\citep[e.g.][]{Dekel2023,Li2024,Boylan-Kolchin2025,Wang2025}. While direct observational evidence remains limited, certain trends at intermediate redshifts (i.e. $z\sim1-2$) suggest elevated star formation efficiency (SFE). For instance, high-redshift galaxies often exhibit star formation characteristics akin to those of local starburst galaxies~\citep[e.g.][]{Tacconi2010}.

However, even if one adopts the prevailing view of elevated SFE in high-redshift galaxies, star formation remains inherently complex, involving a multitude of physical scales and processes~\citep[for a review, see e.g.][]{McKee2007}, such that, the precise scales and mechanisms that give rise to enhanced SFE at early times remain poorly understood. Broadly speaking, star formation can be categorized into three characteristic spatial scales, each governed by distinct physical processes and theoretical frameworks:

On the \textbf{halo scale}, SFE is commonly quantified either through the efficiency of converting baryons to stars or the stellar-to-halo mass ratio (the integrated SFE). Both formulations are widely used in empirical and analytic models of galaxy formation, serving as a bridge between baryonic and DM components~\citep[e.g.][]{Behroozi2013, Moster2013, Somerville2015, Wechsler2018}. Observations in the local Universe suggest that halo-scale SFE is generally low. At the low-mass end, star formation is strongly regulated by stellar feedback, including supernovae (SN), stellar winds, cosmic rays, photoheating, and radiation pressure~\citep[e.g.][]{Hopkins2014, Muratov2015, Chan2018, Agertz2015, Nelson2019}, while at the high-mass end, feedback from active galactic nuclei (AGN) has been suggested as a key mechanism that may play a dominant role in suppressing star formation~\citep[e.g.][]{DiMatteo2005, Croton2006, Bower2006, Weinberger2017}.

On the \textbf{galactic (kpc) scale}, star formation is commonly described by the Kennicutt–Schmidt (KS) relation~\citep{Kennicutt1998}, which correlates the star formation rate (SFR) and the gas surface density over kpc patches of the ISM. The efficiency of star formation at this scale is typically quantified by the gas depletion time, while a dimensionless form can be obtained by normalizing it with a relevant timescale, such as the free-fall time or dynamical time. Numerous studies have demonstrated that, on average, only a few percent of a galaxy’s gas mass is converted into stars per galactic free-fall time~\citep[e.g.][]{Leroy2008, Genzel2010, Daddi2010, Krumholz2012,Tacchella2020}. 
This low efficiency is commonly attributed to stellar feedback continuously disrupting star-forming regions and driving turbulence -- a process known as self-regulation~\citep[e.g.][]{Mac_Low2004,Ostriker2011, Krumholz2012}, with magnetic fields potentially contributing as an additional source of support against gravitational collapse~\citep[e.g.][]{Krumholz2019, Federrath2015, Seifried2011, Commercon2011}.


Finally, the \textbf{cold and dense ISM} fragments into individual, self-gravitating cloud structures (i.e. giant molecular clouds, GMCs) on parsec scales, which are considered the primary fuel and initial conditions for star formation~\citep[for a review, see e.g.][]{Schinnerer2024}. 
The collective formation, dispersal, and star-forming activity of GMCs give rise to the galactic scale KS relation~\citep{Krumholz2012, Ostriker2010, Faucher2013}, highlighting the importance of accurately characterizing the cloud-scale SFE. In observations of local GMCs, the SFE is typically defined as the fraction of gas converted into stars within one free-fall time~\citep{Krumholz2007}, referred to as the instantaneous SFE per free-fall time, $\epsilon^{\rm GMC}_{\rm ff}$. This quantity exhibits significant scatter -- often exceeding 0.3 dex -- due to both variations in observational tracers and the inherently stochastic nature of GMC evolution~\citep[e.g.][]{Murray2011, Lee2016, Vutisalchavakul2016, Evans2014, Sun2023}.
On the theoretical side, cloud-scale SFE in simulations is also sensitive to numerical treatments and cloud identification methodologies, such that related analyses have yet to converge on a consistent picture~\citep[e.g.][]{Dobbs2015, Jeffreson2018,Grudic2018,Fotopoulou2023,Ni2025}. Taken together, assessing whether star formation is locally efficient remains a particularly challenging and unsettled question.

We are now in a position to revisit which stage (or the bridge between which two scales) may be responsible for enhancing the star formation efficiency. In previous work (\citealt{Shen2025}; S25, hereafter), we examined both halo-scale and galaxy-scale SFE and their connection at high redshift, where we found only a mild redshift evolution in the instantaneous halo-scale SFE for low-mass haloes, along with a universal KS relation for neutral gas. However, it is essential to connect galaxy-scale SFE to the cloud scale, especially since several key differences at high redshift may fundamentally alter the properties of GMCs (e.g. mass function, size, morphology, lifetime, star formation efficiency, and spatial distribution), and their mass fractions in the ISM. 
For example, unlike the relatively discrete, gravitationally bound clouds found in nearby galaxies with low surface densities, GMCs at high redshift are often embedded within gas-rich environments characterized by elevated background surface densities. As a result, there is no phase transition at their edges to decouple them from the turbulence in the ambient ISM~\citep{Dekel2009, Ceverino2010, Krumholz2012}. In metal-poor dwarf galaxies, which are common in the early Universe, GMCs also tend to adopt more filamentary morphologies~\citep{Shi2020, Fotopoulou2023}.
Moreover, the increasing matter surface density at high redshift may help GMCs withstand stellar feedback, allowing them to sustain star formation for longer periods~\citep[e.g.][]{Grudic2018, Menon2024, Boylan-Kolchin2025}. In addition, the clumpy, irregular morphology and the large turbulence of their host galaxies could influence the formation and evolution of GMCs~\citep[e.g.][]{Chevance2020, Hopkins2023}.

Due to observational limitations, small-scale structures at such high redshifts ($z\gtrsim3$) remain largely unresolved, even with the most advanced telescopes currently available (e.g. \textit{JWST}, ALMA;~\citealt{DessaugesZavadsky2019,Swinbank2015,Sharda2018}). However, with the rapid development of state-of-the-art galaxy formation simulations over the past decades, several numerical models have achieved remarkable agreement with recent high-redshift observations~\citep[e.g.][]{Feng2016,Ceverino2017,Ma2018,Vogelsberger2020highz,Shen2020,Shen2022,Shen2024size,Lovell2021,Pallottini2022,Kannan2022}, and some are now capable of resolving individual star-forming regions~\citep[e.g.][]{Wang2015, Hopkins2018feedback, Hopkins2023fire3, Li2020, Li2022, Agertz2020, Agertz2021, Gutcke2021, ReinaCampos2022, Nobels2023, Wibking2023,Zhao2024}. These modern simulations thus provide essential laboratories for studying local star formation in realistic galactic environments. 

In this paper, as a follow-up to S25, we utilize the newly developed \thesanzoom simulation suite~\citep{Kannan2025} to extend the analysis of multi-SFE, with a particular focus on the connection between galaxy-scale and cloud-scale star formation. 
This paper is organized as follows: In Section~\ref{sec:sim}, we introduce the basic setup of the simulations and the various physical models involved. In Section~\ref{sec:multi}, we quantify the multi-scale SFE of galaxies and show the analyses on the galaxy-scale average SFE and its dependence on halo mass and redshift. In Section~\ref{sec:local}, we examine the properties of local star-forming regions across different host galaxies. In Section~\ref{sec:connect}, we bridge the global and local star formation with GMC mass fraction, and examine the dominant factors affecting the galaxy-scale SFE. In Sections~\ref{sec:discuss} and \ref{sec:conclusion}, we present our discussions and conclusions.

Throughout the paper, we assume the cosmological parameters from \citet{Planck2016} (obtained from their TT,TE,EE+lowP+lensing+BAO+JLA+H0 dataset), with $H_0=67.74\,\kms/\Mpc$, $\Omega_{\rm m}=0.3089$, $\Omega_{\Lambda}=0.6911$, $\Omega_{\rm b}=0.0486$, $\sigma_8=0.8159$, and $n_{\rm s}=0.9667$. 

\section{Simulations}\label{sec:sim}

\subsection{Simulation details and model variations} 
\label{subsec:sim}

The analysis presented in this work is based on the \thesanzoom simulation suite. The simulation campaign is designed to provide a realistic simulation counterpart to the extensive array of \textit{JWST} observations of high-redshift galaxies, and has already yielded several early results~\citep[e.g.][]{McClymont2025-MSscatter,McClymont2025-size,Zier2025-reion,Zier2025-pop3,Shen2025}. An overview of the suite and its underlying methodology is provided in \citet{Kannan2025}. 

In brief, \thesanzoom is a high-resolution zoom-in extension of the \thesan project~\citep{Kannan2022thesan, Garaldi2022, Smith2022, Garaldi2024}, which were large-volume ($\sim100\, \rm cMpc$) simulations using the IllustrisTNG galaxy formation model~\citep{Pillepich2017, Springel2017, Nelson2017} and on-the-fly radiative transfer (RT).
The zoom-in simulations are built upon the successful experiments of the \textsc{Smuggle} galaxy formation model~\citep{Marinacci2019}, which includes radiative cooling and heating (down to $\sim 10\,{\rm K}$), star formation, and explicit stellar feedback from radiation, stellar winds, and SNe. In addition, on-the-fly RT~\citep{Kannan2019} is included through the \textsc{Arepo-rt} code~\citep{Kannan2019,Zier2024-gpu} with explicit modeling of the non-equilibrium thermochemistry between radiation, gas, and dust~\citep{Kannan2020}.

In particular, unlike large-volume simulations that adopt an effective equation-of-state model for star formation and feedback~\citep[e.g.][]{Springel2003, Vogelsberger2013, Vogelsberger2014, Vogelsberger2014nature, Pillepich2017}, \thesanzoom employs a set of physical criteria for star formation as well as an explicit and comprehensive stellar feedback model. Star formation occurs in dense (with a minimum threshold of $n_{\rm H} > 10\, \mathrm{cm}^{-3}$), self-gravitating~\citep{Hopkins2013}, and Jeans-unstable ~\citep[$L_{\rm J}=\sqrt{\pi c^2_{\rm s}/G\rho}<\Delta x$, where $\Delta x$ is the cell size;][]{Truelove1997} gas via
\begin{equation}
    \dot\rho_{\star} = \epsilon^{\rm cl}_{\rm ff}\frac{\rho_{\rm gas}}{t_{\rm ff}} \, ,
    \label{eq:cell-level_SFE}
\end{equation}
where $t_{\rm ff}=\sqrt{3\pi/32G\rho}$ is the cell-level free-fall timescale, and the cell-level star formation efficiency per free-fall time $\epsilon^{\rm cl}_{\rm ff}$ is set to 100\% in the fiducial runs. Stellar populations are represented by collisionless particles, which are stochastically spawned from eligible gas cells~\citep{Springel2003, Vogelsberger2014,Vogelsberger2020}.

Stellar feedback is modeled from several channels, including photoionization, radiation pressure, stellar winds, and SN feedback. Stellar radiation is modeled self-consistently, with photons emitted by young stars contributing to local heating and ionization. The luminosity and spectral energy distribution of stars, as a function of stellar age and metallicity, are taken from the Binary Population and Spectral Synthesis (BPASS) models~\citep{Eldridge2017}, and are tracked by the RT scheme. Photoionization, radiation pressure, and photoelectric heating are handled by the non-equilibrium thermochemical network~\citep{Kannan2020, Kannan2021}. The mass, momentum, and energy injection rates from stellar winds are computed using the analytic prescriptions in \citet{Hopkins2018feedback, Hopkins2023fire3}, based on stellar evolution tracks from \textsc{Starburst99}~\citep{Leitherer1999}. SNe feedback is implemented by injecting both thermal energy and momentum into the surrounding gas. The SN rate for each stellar particle is computed assuming a ~\citet{Chabrier2003} IMF. Each SN event injects a canonical $10^{51}\,\mathrm{erg}$ of energy into the ISM within a coupling radius, which is capped at 2 physical kpc in the fiducial runs. To account for unresolved Sedov–Taylor phases of the blast wave, additional momentum injection is applied in the form of a terminal momentum boost~\citep{Hopkins2018feedback, Marinacci2019}.

In addition, an empirical early stellar feedback (ESF) model is implemented by applying a momentum injection rate of $1000\, \mathrm{km\,s^{-1}\,Myr^{-1}}$ per unit stellar mass formed during the first 5 Myr after star formation. This component is intended to improve agreement with observed stellar mass–halo mass relations at high redshift~\citep[e.g.][]{Tacchella2018, Behroozi2019}, and may effectively compensate for missing physical processes in the simulation, such as cosmic rays~\citep[e.g.][]{Pakmor2016b,Buck2020,Hopkins2020-cr}, magnetic fields~\citep[e.g.][]{Marinacci2016,Hopkins2020-cr}, Lyman-$\alpha$ radiation pressure~\citep[e.g.][]{Smith2017, Kimm2018, Nebrin2025} or other numerical uncertainties.

The \thesanzoom\ suite includes three zoom levels, corresponding to spatial resolution improvements by factors of 4, 8, and 16, labeled as ``4x'', “8x”, and “16x”, respectively. These correspond to median baryonic mass resolutions of
$9.09 \times 10^3 \msun,\,1.14 \times 10^3\, \msun,\, \text{and}\,1.42 \times 10^2\, \msun$. The full suite incorporates several model variations, and here we describe those most relevant to this study. The noESF runs remove the additional ESF mentioned above. 
The varSFE runs modify the local star formation efficiency per free-fall time from a fixed value to a density-dependent one. Specifically, for gas cells eligible for star formation under the aforementioned criteria, $\epsilon_{\rm ff}^{\rm cl}$ starts at 1\% at the threshold density of $n_{\rm H} = 10\, \mathrm{cm}^{-3}$ and increases linearly with gas density, reaching a maximum of 100\% for $n_{\rm H} \geq 10^3\, \mathrm{cm}^{-3}$. 
These variations allow us to assess the robustness of our results and isolate the impact of different feedback and star-formation prescriptions on both global- and cloud-scale star-formation efficiency.

\subsection{Starbursts in \thesanzoom}
\begin{figure}
    \centering
    \includegraphics[width=1\linewidth]{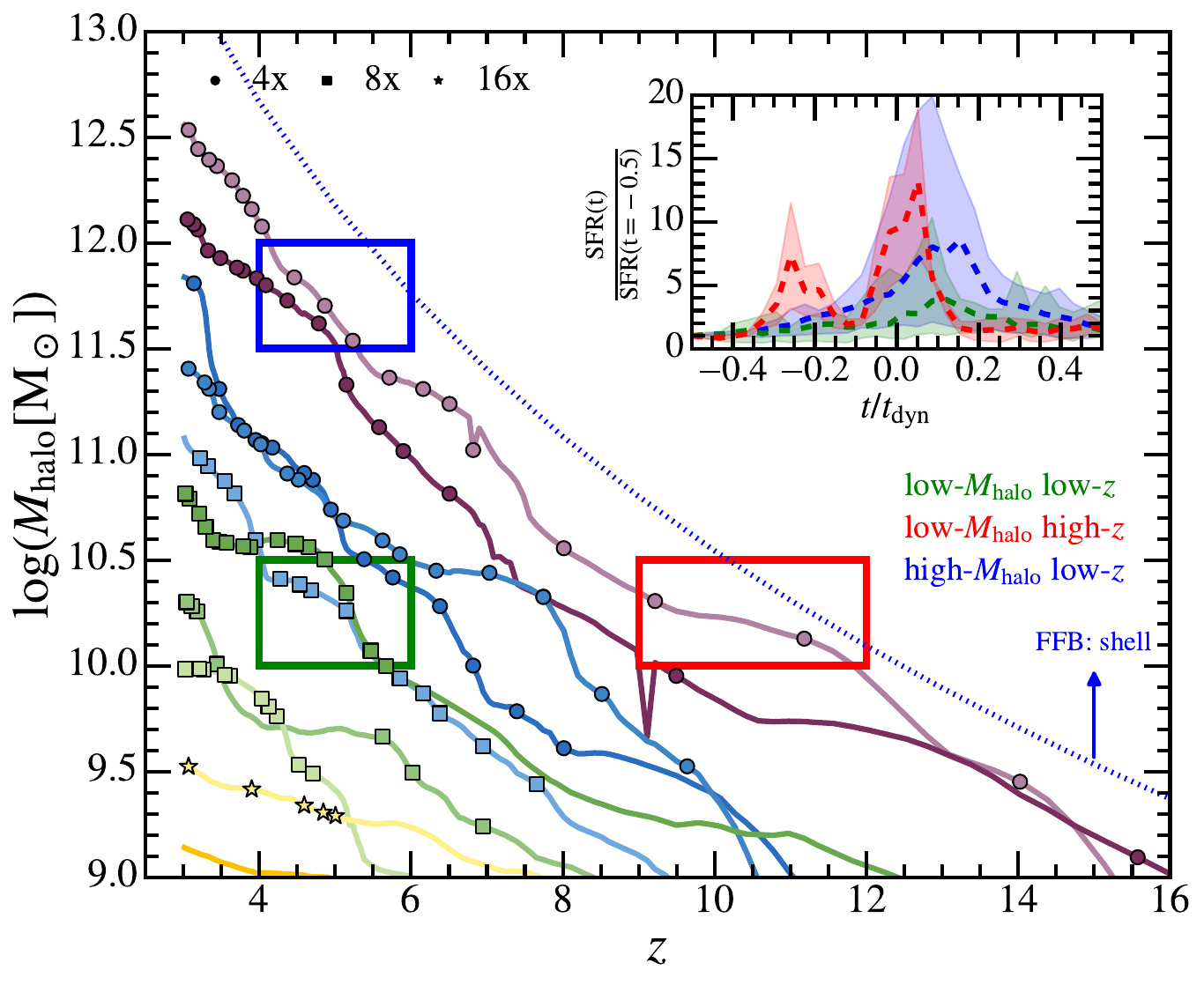}
    \caption{Examples of selected starbursts from the main target galaxies in \thesanzoom. Their dynamical evolution is shown with lines, and the starburst phases are marked by discrete data points. Starburst segments are defined as periods exhibiting a local peak in SFR within a duration of one dynamical time, $t_{\rm dyn}$. The marker shape indicates the highest resolution available for each target (circles for 4x, squares for 8x, and stars for 16x). Due to the limited dynamical range of the \thesanzoom simulations, no starbursts in our sample fully satisfy the conditions for the FFB regime (blue dashed lines).
    In the inset panel, we show the normalized SFHs, $\mathrm{SFR}(t/t_{\rm dyn})/\mathrm{SFR}(t/t_{\rm dyn}=-0.5)$, for starbursts in three different regimes to illustrate the selection criteria. The dashed lines represent the median normalized SFH in each bin, while the shaded regions indicate the $1\sigma$ galaxy-to-galaxy scatter.}
    \label{fig:data_sample}
\end{figure}

The simulation outputs are saved from $z = 16$ down to $z = 3$ in 189 snapshots, with a time cadence of $\sim 10\,\mathrm{Myr}$. Each snapshot records the full properties of gas, DM, and stellar particles. DM haloes are identified using the friends-of-friends (FOF) algorithm~\citep{Davis1985,Springel2005} with a linking length of 0.2 times the mean inter-particle separation. 
Merger trees that track the progenitors of subhaloes over time are constructed with the SUBFIND-HBT algorithm~\citep{Springel2021}. In this work, we do not restrict our analysis to the main target galaxies in \citet{Kannan2025}, but rather focus on the starbursts of all central galaxies within the zoom-in regions, where central galaxies are defined as the most massive galaxy in a given DM halo (i.e., the main subhalo identified by SUBFIND-HBT).
Starbursts are identified as episodes featuring a local peak in the SFR within a dynamical timescale ($t_{\rm dyn} \equiv R_{\rm vir} / V_{\rm vir}$). By default, we require the peak SFR to exceed twice the median SFR within the $t_{\rm dyn}$ window. However, this threshold is relaxed for low-mass haloes in the 16x run, where few galaxies satisfy such a criterion. For galaxies that meet this condition, we define the starburst phase as a period of duration $t_{\rm dyn}$ centered on the snapshot of peak SFR. For merger-induced starbursts, we follow only the main progenitor to avoid double-counting. This selection approximately ensures that the identified galaxies are in phases of active star formation, enabling fair comparisons in the subsequent analysis.

In Fig.~\ref{fig:data_sample}, we show examples of selected starbursts from the main target galaxies in \thesanzoom. We display their dynamical evolution over time, with the starburst phase identified by discrete data points, each representing the central snapshot in one dynamical time interval. The shape of the marker indicates the highest resolution available for each target (circles for 4x, squares for 8x, and stars for 16x).
In the inset panel, we present the normalized star formation histories (SFHs) of selected starbursts across three representative dynamical intervals (e.g., high-redshift low-mass, low-redshift low-mass, and low-redshift high-mass haloes). In addition to the examples shown, we include a comprehensive sample of 500 starbursts drawn from all central galaxies across all redshifts, halo masses, and resolutions.
We compare these starbursts to the feedback-free regime~\citep[FFB;][]{Dekel2023}, in which stellar feedback is suppressed due to low metallicity, and galaxies undergo rapid gravitational collapse before the onset of supernovae, potentially reaching unity-level SFE. As the \thesanzoom galaxies do not form well-defined disks, we compare specifically to the shell-configuration scenario in the FFB framework, where the galaxy is formed in a self-shielded shell-like geometry.
However, because of the limited dynamical range of \thesanzoom, none of the starbursts in our sample fully satisfy the FFB criterion.


\section{Multi-scale star-formation efficiency}\label{sec:multi}
\subsection{Star formation efficiencies across different scales}\label{subsec:multiSFE}

\begin{figure*}
    \centering
    \includegraphics[width=1\linewidth]{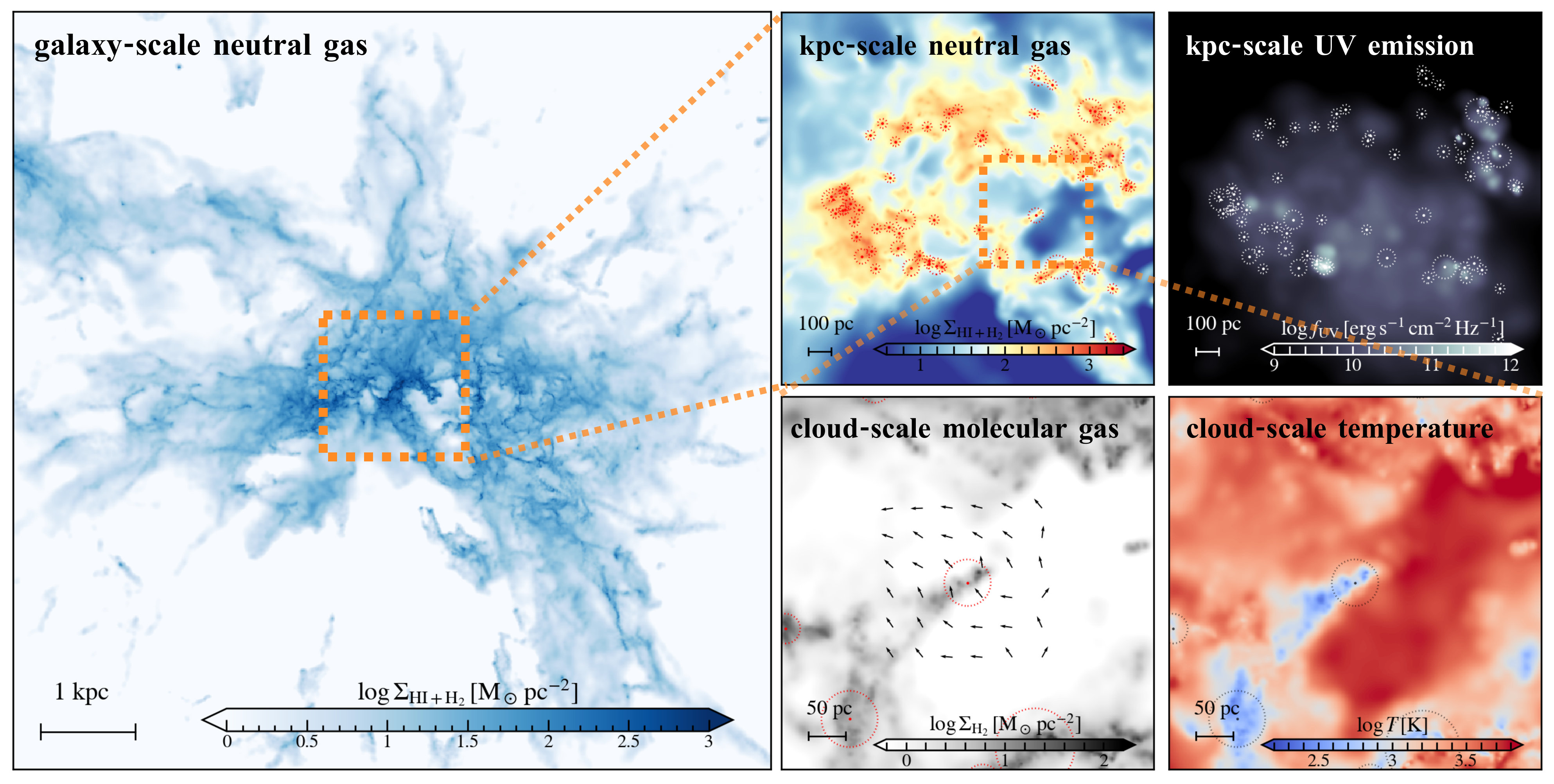}
    \caption{A visualization of a \thesanzoom galaxy (``m11.1'') at $ z \simeq 6$ across multiple spatial scales. The left panel shows the surface density of neutral and molecular gas across the entire galaxy within a field of view of $0.5\,R_{\mathrm{vir}}$, corresponding to the spatial scale at which the global SFE is measured. The orange square marks the central kpc-scale region, which is further examined in the top right panels, where the KS relation is expected to hold. There, we show the neutral gas surface density hosting the GMCs (indicated by dashed circles) and the stellar UV emission tracing young stellar objects (YSOs). The radius of each circle corresponds to the effective radius of the GMC, as defined in Eq.~\ref{eq:gmc_r}. The bottom panel provides a zoomed-in view of a representative GMC with $\alpha \simeq 3$, showing the relative velocity field of the surrounding gas and its mean temperature. {\sc CloudPhinder} successfully identifies cold, dense ISM structures consistent with GMCs.
}
    \label{fig:visulization}
\end{figure*}

Star formation within galaxies is a multiscale process that spans a wide range of physical scales. In this context, it is particularly informative to examine how star formation efficiency varies with spatial scale and to explore the transitions between different regimes. Table~\ref{tab:definition} summarizes the definitions of SFE adopted at each scale in this study, while Fig.~\ref{fig:visulization} provides a visual overview of the spatial regimes over which SFE is quantified. 

\begin{table*}
    \caption{Definitions and measurements of global and local properties in this paper.}
    \centering
    \addtolength{\tabcolsep}{-0.2pt}
    \def\arraystretch{1.2}
    \begin{tabularx}{\linewidth}{lX}
        \hline
        Definitions & Description \\
        \hline
        \hline
        Timescales & \\
        $t_{\rm H}\equiv 1/H(z)$ & Hubble time\\
        $t_{\rm dyn}\equiv R_{\rm vir}/V_{\rm vir}$ & Dynamical crossing time of the halo \\
        $t_{\rm dep}\equiv M_{\rm gas}/{\rm SFR}$ & Depletion time of an assembly of gas  with mass $M_{\rm gas}$ \\
        $t_{\rm ff} \equiv \sqrt{3\,\pi/32\,G\,\rho}$ & Free-fall time of gas at density $\rho$  \\
        ${\langle t_{\rm ff}\rangle} \equiv \langle{ t_{\rm ff}}^{-1}\rangle_{\rm m}^{-1}$ & Mass-weighted harmonic average free-fall timescale \vspace{.1cm} \\
        \hline
        Star formation efficiency & \\
        $\epsilon^*_{\rm halo}\equiv {\rm SFR}/(f_{\rm b}\,\dot M_{\rm halo})$ & Halo-scale instantaneous SFE (Eq.~\ref{eq:haloSFE}) \\
        $\langle\epsilon^{\rm gal}_{\rm ff}\rangle\equiv{\rm SFR}\times \langle t^{\rm gal}_{\rm ff}\rangle/{M_{\rm gas}}$  & Galaxy-scale average SFE per free-fall time (Eq.~\ref{eq:glbSFE}) \\
        $\langle \epsilon^{\rm gal}_{\rm ff}\rangle(n_{\rm crit}) \equiv {{\rm SFR}_{{n>n_{\rm crit}}}}\times \langle t^{\rm gal}_{\rm ff}\rangle_{n>n_{\rm crit}}/M_{{\rm gas},\,n>n_{\rm crit}}$  & Galaxy-scale average SFE per free-fall time with a density threshold (Eq.~\ref{eq:multiSEF}) \\
        $\epsilon^{\rm GMC}_{\rm ff} \equiv {\rm SFR}\times t^{\rm GMC}_{\rm ff}/{M_{\rm gas}}$ & GMC-scale SFE per free-fall time  (Eq.~\ref{eq:gmc_sfe_ff})\\
        $\epsilon^{\rm GMC}_{\rm int} \equiv M_{\rm \star}(t=\infty)/M_{\rm gas}(t=0)$ & GMC-scale integrated SFE (Eq.~\ref{eq:gmc_sfe_int}) \\
        $\epsilon_{\rm ff}^{\rm cl}\equiv {\rm SFR}\times t_{\rm ff}/M_{\rm gas}$ & Cell-level SFE per free-fall time, which is set to be unity in the fiducial runs (Eq.~\ref{eq:cell-level_SFE})
        \vspace{.15cm} \\
        \hline
    \end{tabularx}
    \def\arraystretch{1.2}
    \label{tab:definition}
\end{table*}

A subtle approach to studying SFE across different scales is to compute the average SFE of the ISM in different phases. This assumes that, as the physical scale gradually narrows toward the star-forming regions within the ISM, the average density of the ISM increases, such that a higher density threshold captures progressively smaller-scale ISM clumps.
Here we define a multi-scale star formation efficiency as
\begin{equation}
    \epsilon_{\rm ff}(n_{\rm crit}) \equiv \frac{{{\rm SFR}_{{n>n_{\rm crit}}}}\times \langle t_{\rm ff}\rangle_{n>n_{\rm crit}}}{M_{\rm gas,\,n>n_{\rm crit}}} \, ,
    \label{eq:multiSEF}
\end{equation}
where ${\langle t_{\rm ff}\rangle} \equiv \langle{1/t_{{\rm ff},i}}\rangle_{\rm m}^{-1}$ is the harmonic mean of the free-fall time weighted by mass. The expression depends on the phase of the gas tracer used, which is taken to be neutral gas in this work.

In fiducial runs of \thesanzoom, star-forming gas cells collapse into stars in one free-fall time with cell-level SFE $\epsilon^{\rm cl}_{\rm ff}=1$. Incorporating ${\rm SFR}=\epsilon^{\rm cl}_{\rm ff}m/{t_{\rm ff}}$ and $\langle t_{\rm ff}\rangle$ into Eq.~(\ref{eq:multiSEF}), one can obtain
\begin{equation}
    \epsilon_{\rm ff}(n_{\rm crit})=\frac{\sum_{{\rm sf},n>n_{\rm crit}}m^{\rm HI+H_2}_i/t_{{\rm ff},i}}{\sum_{n>n_{\rm crit}}m^{\rm HI+H_2}_i/t_{{\rm ff},i}} \, ,
\label{eq:multiSEF2}
\end{equation}
where the denominator sums over gas cells exceeding the density threshold, and the numerator includes only those that are star-forming. This is roughly consistent with the form in \citet{Semenov2018}, but it is influenced by two factors: the fraction of gas in the star-forming phase and the difference in free-fall time due to the clumping of star-forming gas.

\begin{figure*}
    \centering
    \includegraphics[width=\linewidth]{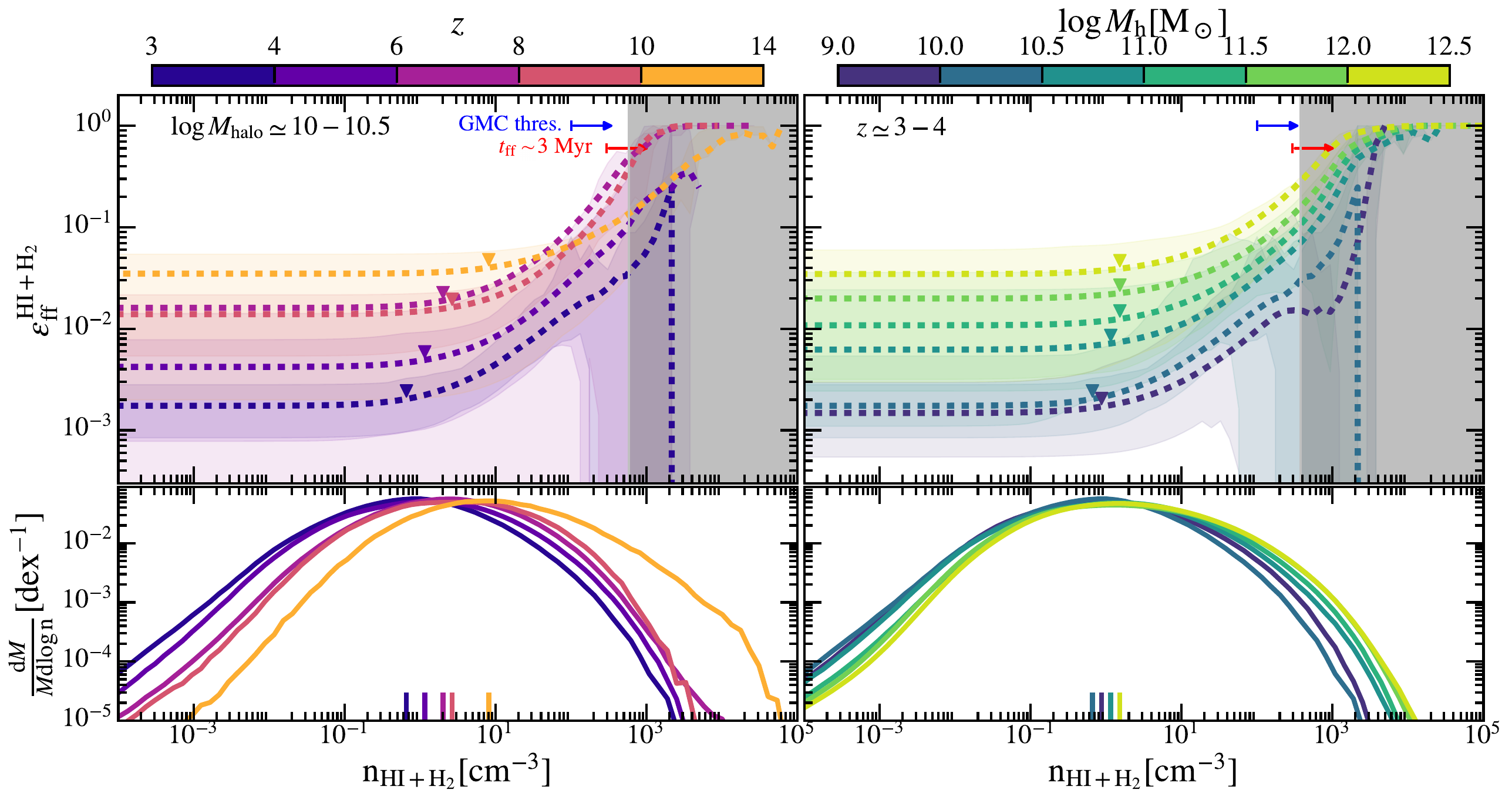}
    \caption{\textit{Top panels:} Multi-scale SFE manifested in the SFE versus the critical number density of \thesanzoom galaxies from selected redshift and halo mass bins. For each bin, we combine snapshots within $t_{\rm dyn}$ around the starburst peak and integrate all cool gas cells ($T<10^5 \K$) to measure the SFE of gas above a certain number density. The dashed lines represent the mass-weighted average multi-scale SFE within each bin, while the triangles mark the point where 50\% of gas mass is enclosed. The shaded contours illustrate the snapshot-to-snapshot variation, enclosing the $10^{\rm th}$–$90^{\rm th}$ percentile range of the multi-scale SFE values during the corresponding starburst periods.
    The gray shaded region on the right corresponds to densities above the median density of star-forming gas. Since the cell-level SFE for these star-forming gas is hardcoded to unity, this regime serves as a reference for the unresolved limit. Results from all bins trace a similar pattern where the SFE remains steady due to self-regulation, then gradually increases as self-regulation begins to fail, until it reaches a numerical limit and approaches unity as only star-forming gas cells are included. \textit{Lower panels:} Average ISM density distribution of galaxies in each bin. The mean density for each bin is indicated by a vertical line at the bottom, color-coded consistently with the corresponding distribution. Each follows a log-normal profile. The median density rises with increasing redshift and exhibits a modest upward trend with halo mass.
    }
    \label{fig:multiSFE}
\end{figure*}

In Fig.~\ref{fig:multiSFE}, we show the multi-scale SFE of the main target galaxies as a function of redshift and halo mass. For this series of analyses, we focus on the gas within the central region of radius $0.5\,R_{\rm vir}$ of each galaxy, excluding gas with temperatures above $10^5\,\mathrm{K}$. This serves as our working definition of the ISM in galaxies.
Starbursts are binned according to the redshift at which their SFR peaks and the mass of their host halo. For each bin, we combine gas cells from all included starbursts across snapshots spanning $t_{\rm dyn}$, yielding a result equivalent to a mass-weighted average.
The shaded regions indicate the $10^{\rm th}$–$90^{\rm th}$ percentile range of the corresponding function across all snapshots within each bin. In general, results from all bins show a similar pattern that SFE remains steady due to self-regulation, then gradually increases as self-regulation begins to fail at higher densities, until it reaches the numerical limit where cell-level SFE is hardcoded to unity. In the figure, we show the median number density of star-forming gas cells as a reference for the unresolved limit.

For the redshift dependence, we present results at a fixed halo mass range of $10^{10} \lesssim M_{\rm h} / \rm M_{\odot} \lesssim 10^{10.5}$ in the fiducial L4 run. The global SFE rises with increasing redshift, though it exhibits significant snapshot-to-snapshot scatter. However, the shape of the function does not display a strong redshift evolution, indicating a universal star formation efficiency at small scales.
We do not find evidence that self-regulation breaks down earlier at higher redshift, as might be expected in the FFB scenario~\citep{Dekel2023, Li2024}. Instead, we observe a systematic shift along the density axis, which can be attributed to differences in the ISM density distributions across redshifts.
As shown in the inset panel, each distribution follows a log-normal profile, consistent with the turbulence-dominated theory for star-forming regions~\citep[e.g.][]{Vazquez1994, Padoan1997, Federrath2008}. As the redshift increases, the distribution shifts toward higher densities, particularly at the low-density end, where gas density roughly scales with the critical density of the Universe $\rho_{\rm crit}(z)$. Meanwhile, a larger fraction of gas reaches star-forming densities at higher redshift, enchancing the global SFE.

The average global SFE increases with halo mass when considering a fixed redshift range of $3 \lesssim  z \lesssim 4$. At the high-density end, the rising trend remains nearly self-similar, except for the lowest halo mass bin, where some starbursts show star formation dominated by lower-density gas ($n_{\rm H}<10^3\cm^{-3}$).
While the median gas density shows only a weak dependence on halo mass, the high-density tail becomes more prominent in higher-mass haloes, contributing to elevated global SFE. This trend is consistent with the findings of S25, where more massive haloes host galaxies with higher effective surface densities, leading to an increased GMC mass fraction in the ISM, as will be discussed in the Sect.~\ref{sec:connect}.

Density is not the only criterion for star formation in \thesanzoom. The simulation additionally requires that star-forming gas cells satisfy a Jeans instability condition, namely that the thermal Jeans length is smaller than the cell size, $L_{\rm J} = \sqrt{\pi c_{\rm s}^2 / G\rho} < \Delta x$. This implies that the multi-scale SFE can be reformulated as a function of both density and temperature, or more generally, as a function of the distance to the star formation threshold. In Fig.~\ref{fig:multiSFE_lj} of the Appendix, we present the multi-scale SFE as a function of $x \equiv \Delta x / L_{\rm J}$, where similar trends are observed.

\subsection{Global star formation efficiency}\label{subsec:glbSFE}

\begin{figure*}
    \centering
    \includegraphics[width=1\linewidth]{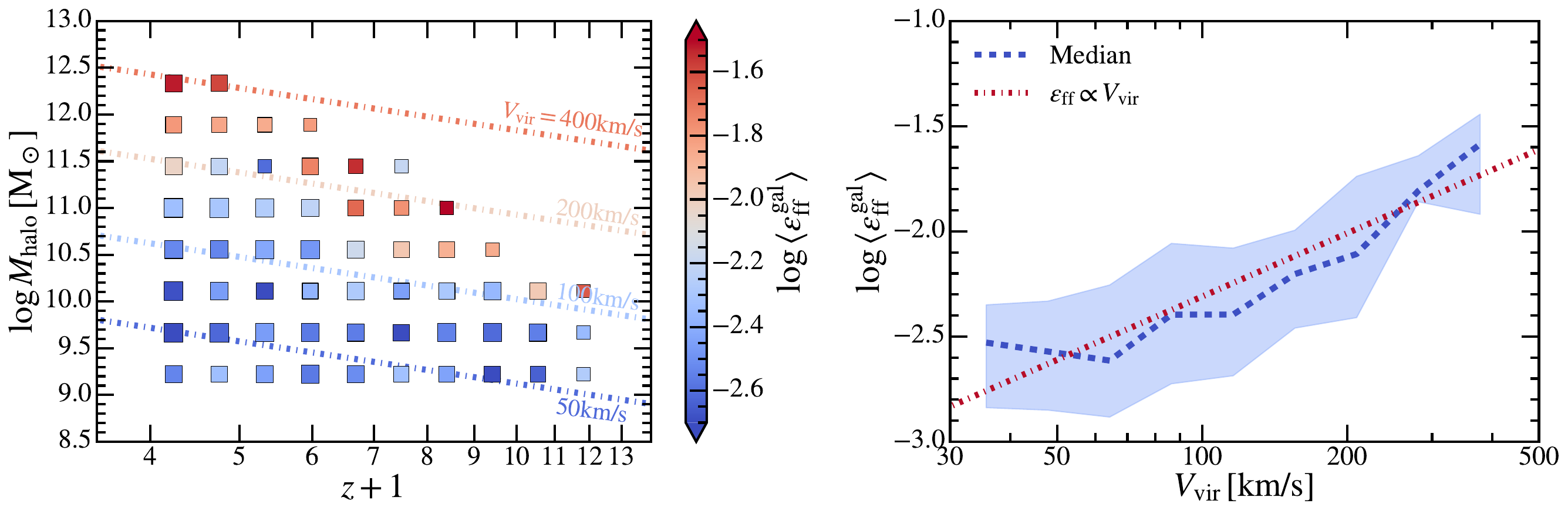}
    \caption{\textit{Left panel:} Distribution of starburst samples in the redshift–halo mass plane, color-coded by the galaxy-scale average SFE. For each starburst, we compute the median SFE within $t_{\rm dyn}$ and use it as the representative value. The color of each square denotes the median SFE within that bin, while the square size reflects the number of starbursts in the bin. Dashed lines indicate analytical predictions derived from the right panel for four reference values of virial velocity. 
    \textit{Right panel:} Galaxy-scale average SFE as a function of virial velocity. The blue dashed line shows the median relation, with the shaded region indicating the $1\sigma$ scatter. The red dotted line marks the scaling relation, $\epsilon_{\rm ff} \propto V_{\rm vir}$. 
    Both panels illustrate that galaxy-scale SFE increases with redshift and halo mass, roughly scaling with virial velocity.} 
    
    \label{fig:glbSFE}
\end{figure*}

When all gas cells are included in the multi-scale SFE by setting $n_{\rm crit} = 0$, one obtains the global average SFE. As shown at the low-density end of Fig.~\ref{fig:multiSFE}, the global SFE increases with both redshift and halo mass. In this section, we statistically characterize the global star formation efficiency as a function of redshift and halo mass. The galaxy-scale SFE per free-fall time is defined as
\begin{equation}
    \langle\epsilon^{\rm gal}_{\rm ff}\rangle\equiv{\rm SFR_{\rm 10Myr}}\times \langle t^{\rm gal}_{\rm ff}\rangle/{M_{\rm gas}} \, ,
    \label{eq:glbSFE}
\end{equation}
where $\rm SFR_{\rm 10Myr}$ is the 10 Myr-average SFR, and $\langle t^{\rm gal}_{\rm ff}\rangle$ is the mass-weighted free fall time average over the ISM in the galaxy, which reflects its mean density. The gas selection criteria are identical to those used in the previous section for computing the multi-scale SFE.

Fig.~\ref{fig:glbSFE} presents how the galaxy-averaged SFE of starbursts varies with redshift and halo mass. In the left panel, we show the distribution of starburst samples in the redshift–halo mass plane, color-coded by the galaxy-scale average SFE. For each starburst, the median SFE within $t_{\rm dyn}$ is adopted as its representative value. The size of each square is proportional to the logarithm of the number of starbursts in the corresponding bin. In general, the SFE tends to be higher in more massive haloes and at higher redshifts.
In the right panel, we show the galaxy-scale average SFE as a function of virial velocity. Despite the large scatter, the global SFE increases with $V_{\rm vir}$, approximately following the scaling relation
\begin{equation}
    \langle \epsilon_{\rm ff}^{\rm gal} \rangle  \propto V_{\rm vir} \, \propto (1+z)^{1/2} M_{\rm h}^{1/3}.
\end{equation}

This is consistent with what has been discussed in S25 from the perspective of galactic surface density and the KS relation. S25 found a universal KS relation of $\Sigma_{\rm SFR} \propto \Sigma_{\rm HI+H_2}^2$ in \thesanzoom galaxies across a wide range of redshifts and halo masses. This implies that the global depletion time, $t_{\rm dep} = \Sigma_{\rm gas} / \Sigma_{\rm SFR}$, is primarily determined by the effective surface density,  $\Sigma_{\rm eff} \equiv \langle \Sigma_{\rm gas}^2 \rangle / \langle \Sigma_{\rm gas} \rangle$, for which $t_{\rm dep} \propto \Sigma_{\rm eff}^{-1}$. We further showed that the effective surface density in \thesanzoom galaxies generally follows a simple analytic model, with $\Sigma_{\rm gas} \propto \rho_{\rm gas}^{1/2} \sigma \propto \rho_{\rm gas}^{1/2} V_{\rm vir}$.
Given $\epsilon_{\rm ff} \equiv t_{\rm ff} / t_{\rm dep}$, the key difference in our formulation is the explicit inclusion of the free-fall time, which reflects the mean ISM density and effectively cancels out the density dependence present in the surface density scaling. As a result, the derived scaling becomes more directly sensitive to turbulence. In general, our scaling relation shows strong agreement with the $\Sigma_{\rm eff}$ measurements reported in S25.
A similar scaling can also be derived in a disk configuration using the Toomre stability criterion for a marginally stable disk, where the star formation efficiency is proportional to the turbulent gas velocity dispersion~\citep[e.g.][]{Faucher2013}, if assuming that the turbulence is driven by cold gas inflows with streaming velocity $\sim V_{\rm vir}$.


\section{Local star formation efficiency}\label{sec:local}
In this section, we investigate the properties of local star-forming regions at high redshift. The minimum resolvable size of star-forming regions varies with numerical resolution, ranging from dense clumps to GMCs, or more generally, to self-gravitating star-forming complexes.
For the analyses presented here, we adopt the 8x resolution as the fiducial run, as it resolves GMCs with sizes most comparable to those observed in both observational data and other high-resolution simulations. A detailed comparison of the properties of local star-forming regions across different resolutions is presented in the Appendix.

\subsection{GMC identification}
We use {\sc CloudPhinder}\footnote[1]{{\url{https://github.com/mikegrudic/CloudPhinder}}}~(see \citealt{Guszejnov2019} for its first application) to identify scattered regions within galaxies that are either actively forming stars or have the potential to do so. This algorithm starts from local density peaks within galaxies and identifies the largest self-gravitating structures by searching for neighboring cells satisfying a certain selection function. We refine the original algorithm in the following aspects to better adapt it to our simulation. First, in addition to gaseous structures, we incorporate young stars as an additional component of GMCs.\footnote[2]{As implied by \citet{Li2020} and \citet{Ni2025}, in simulations where $\epsilon_{\rm ff}^{\rm cl} = 1$, gas is converted into stars rapidly, such that the resulting gravitational binding of the structure can be dominated by the newly formed stellar component rather than the gas itself. To account for GMCs that are bound by young stars, we include young stars in the identified structures, which may also help trace GMCs in their late evolutionary stages.} Here, young stars are defined as stellar particles younger than 3 Myr, a timescale roughly comparable to the free-fall time of star-forming gas. The linking length for stellar particles is set to the median smoothing length of star-forming gas, which is approximately 10 pc in the 8x resolution. 
Second, target gas cells must either have a number density exceeding $100\cm^{-3}$ or a thermal Jeans length smaller than the cell size (consistent with the star formation criteria of the simulations). In addition, we limit the gas temperature to below $1000\,\mathrm{K}$ to exclude shock-heated gas that is less likely to constitute GMCs. We set a fairly tolerant criterion of the virial parameter $\alpha \leq \alpha_{\mathrm{max}}=20$\footnote[3]{Despite the large $\alpha_{\rm max}$ allowed, the majority of GMCs exhibit $\alpha < 10$, with a median around 4 (see Fig.~\ref{fig:gmc_vp_res}).}, where
\begin{equation}
\alpha = 2 (E_{\mathrm{kin}} + E_{\mathrm{th}})/|E_{\mathrm{grav}}|.
\end{equation}
Here, $E_{\mathrm{kin}}$ and $E_{\mathrm{th}}$ are the kinetic and thermal energy of the cloud, and $E_{\mathrm{grav}}$ is the gravitational potential energy. The energies are computed as follows:
\begin{align}
&E_{\mathrm{kin}} = \sum_i \frac{1}{2} \,m_i \,|\vec{v_i} - \vec{v_c}|^2, \\
&E_{\mathrm{th}} = \sum_i m_i \, u_{i}, \\
&E_{\mathrm{grav}} = -\frac{1}{2}\, \sum_{i \neq j} \frac{G\, m_i\, m_j}{|\vec{r_i} - \vec{r_j}|},
\end{align}
where $m_i$, $u_{i}$, $\vec{v_i}$, and $\vec{r_i}$ are the mass, internal energy (per unit mass), velocity, and position of the $i$-th gas cell (the same definitions apply to gas cell $j$), and $\vec{v_c}$ is the center-of-mass velocity of the cloud. 
To ensure that the identified GMCs are well resolved, we discard clouds that contain fewer than $30$ elements, approximately corresponding to the number of gas cells within one gravitational softening length.
This parameter set is similar to that of previous work~\citep[e.g.][]{Tasker2009, Grisdale2018, Fotopoulou2023, Ni2025}, and found to be most effective in capturing the majority of instantaneously star-forming gas cells.
In the Appendix~\ref{apdx:parameters}, we will discuss the impact of free parameter choices on the properties of GMCs. Overall, we find that the results are robust against variations in these parameters.

We define effective radius $R_{\mathrm{eff}}$ and velocity dispersion $\sigma_v$ of GMCs as 
\begin{equation}
\begin{split}
    &R_{\mathrm{eff}}=\sqrt{\frac{5}{3}\frac{\sum (m_i r^2_i)}{\sum m_i}} \, , \\
    &\sigma_v=\sqrt{ \frac{\sum [m_i (|\vec{v_i}-\vec{v_{\rm c}}|^2+c^2_{{\rm s},i})]}   {\sum m_i} } \, ,
\end{split}
\label{eq:gmc_r}
\end{equation}
where $c_{\mathrm{s},i}$ is the  sound speed of the $i$-th gas cell.

\subsection{Physical properties of GMCs}\label{subsec:gmc_property}
In this section, we present the statistical properties of GMCs (e.g. mass, size, and velocity dispersion) across galaxies at different redshifts and halo masses.

\begin{figure}
    \centering
    \includegraphics[width=1\linewidth]{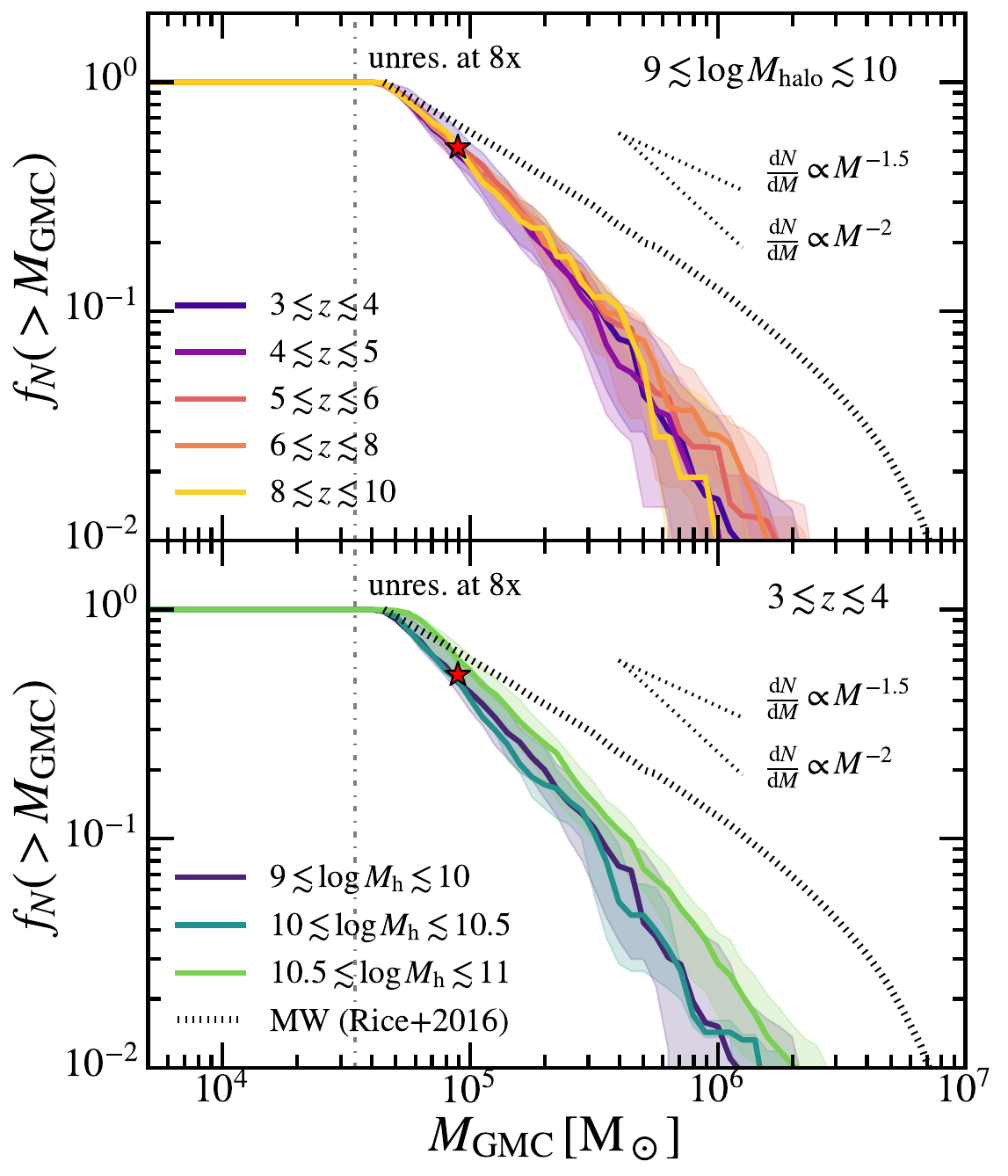}
    \caption{GMC mass functions across different redshifts and halo masses. The solid line is the median value across all starbursts with more than 50 clouds in this bin, while the shaded areas enclose the $10^{\rm th}$-$90^{\rm th}$ range. The red star indicates the median mass of the entire GMC sample, which is approximately $1 \times 10^5\,\msun$. The gray dashed lines mark our resolution limit (30 gas elements).
    Compared with observations of GMCs in the Milky Way~\citep{Rice2016}, high-mass GMCs at high redshift in \thesanzoom are noticeably less abundant, and the mass function exhibits a steeper slope. Note that the minimum mass covered by observational constraints coincides with the simulation resolution limit. } 
    \label{fig:gmc_mass}
\end{figure}

\begin{figure*}
    \includegraphics[width=1\linewidth]{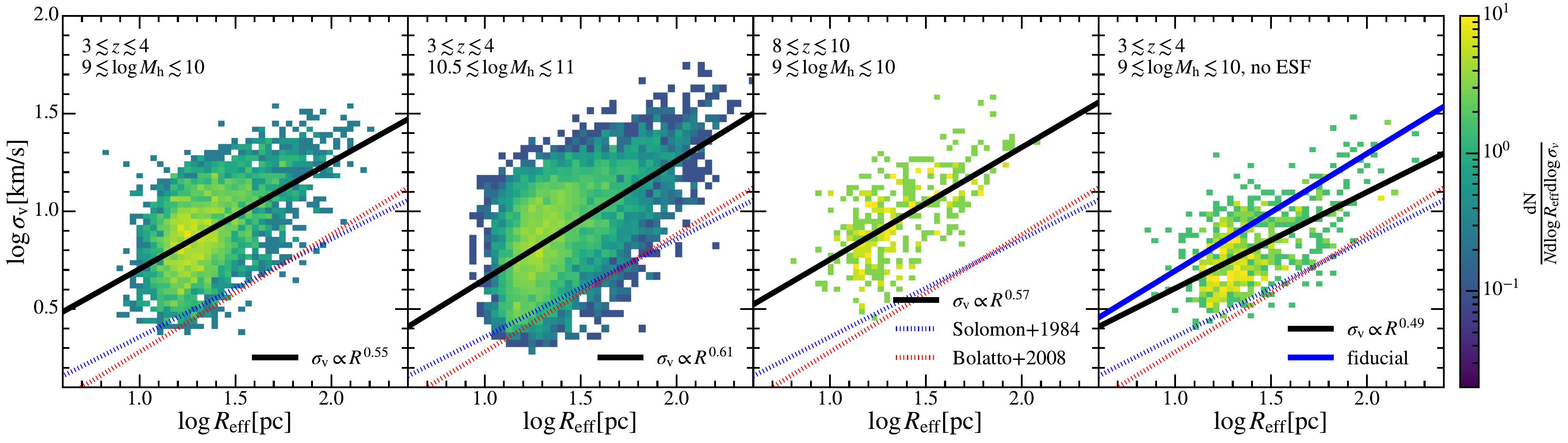}
    \caption{The radius–velocity dispersion relation of GMCs in the three different systems labeled at the top left corner in each panel. The black solid line denotes the best-fit relation of $\sigma_{\rm }\propto R_{\rm eff}^b$ for each panel. We compare it to the best-fit results based on the local observations~\citep[e.g.][]{Solomon1984, Bolatto2008}. In the rightmost panel, we further compare the results from the “no ESF” runs (black line) and the fiducial runs (blue line). Apart from the “no ESF” runs, where turbulence is systematically lower due to the removal of the additional ESF as one of its key driving sources, the slope remains largely unchanged across the other systems, with $b \simeq 0.6$.}
    \label{fig:gmc_larson}
\end{figure*}

In Fig.~\ref{fig:gmc_mass}, we present the cumulative mass distribution of GMCs and its dependence on redshift and halo mass. 
In the top panel, we show the evolution of the GMC mass function with redshift at a fixed halo mass of $10^9-10^{10}\,\rm M_\odot$. In the bottom panel, we present the GMC mass function across different halo masses at $z\simeq3-4$. Despite some slight variations, the slope of the relation $\mathrm{d}N/\mathrm{d}M \propto M^\gamma$ remains consistently around $\gamma = -2.49^{+0.13}_{-0.07}$, which is steeper than that observed for Galactic GMCs within the solar circle~\citep[e.g.][]{Rice2016}. 

The steeper slope and lack of massive GMCs may reflect a fundamental difference between star-forming regions in high-redshift galaxies and those in the local Universe. In high-redshift systems, the ISM assembles into massive clumps with surface densities on the order of $10^2\,\rm M_\odot\,\mathrm{pc}^{-2}$ (see Fig.~\ref{fig:visulization}), forming a continuous, turbulent, star-forming medium. GMCs embedded in such environments are only overdense by factors of $\sim1-10$ relative to their surroundings, in contrast to Galactic GMCs, which are typically overdense by factors of $\sim100$~\citep[$n_{\rm H}\sim100\cm^{-3}$ versus $n_{\rm H}\sim1\cm^{-3}$; e.g.][]{Bolatto2008,Krumholz2012}. As a result, there is no clear phase transition at their boundaries to decouple them from the ambient ISM turbulence~\citep[e.g.][]{Dekel2009, Ceverino2010, Krumholz2012}. 
Consequently, the number of discrete, self-gravitating high-mass GMCs observed in the local Universe decreases at high redshift, where GMCs of similar mass are more likely to be shattered by external perturbations rather than confined by self-gravity. This also helps explain both the need to adopt a relatively loose threshold on the virial parameter $\alpha_{\rm max}$ in order to capture the majority of instantaneous star formation, and the fact that the median $\alpha$ of GMCs exceeds unity, as a significant fraction of star-forming gas is spatially scattered throughout the ISM where it is confined by external pressure.
We note that our simulations do not extend to $z=0$ to explicitly tell whether this difference emerges at a particular redshift or simply reflects a feature of the adopted numerical models. 
However, it is encouraging that simulations of Milky Way–like galaxies based on similar ISM models can successfully reproduce the observed GMC mass function at low redshift~\citep[e.g.][]{Guszejnov2019, Li2020, Ni2025}, although the redshift evolution is still rather uncertain~\citep[e.g.][]{Guszejnov2019}.

In Fig.~\ref{fig:gmc_larson}, we examine the size–velocity dispersion relation of clouds, commonly referred to as Larson’s Law~\citep{Larson1981}, focusing on three representative systems: dwarf galaxies at $z\simeq8-10$, dwarf galaxies at $z\simeq3-4$, and more massive galaxies at $z\simeq3-4$. We also compare the relations in the fiducial and noESF runs in the far right panel in dwarf galaxies at $z\simeq3-4$. In observations of local GMCs, Larson’s Law describes an empirical relation of the form $\sigma_{\rm v} \propto R^b$, with a best-fit slope of $b = 0.5$ for Milky Way GMCs~\citep[blue dashed lines;][]{Solomon1984} and $b = 0.6$ for extragalactic GMCs~\citep[red dashed lines;][]{Bolatto2008}. Across different systems at high redshift in the fiducial runs, this relation appears universal and consistent with local GMCs, with all three systems exhibiting slopes of approximately $b \simeq 0.6$. However, the normalization is systematically higher than local observations, suggesting that GMCs at high redshift are more turbulent. Meanwhile, the velocity dispersion at fixed radius is lower in noESF runs, as one of the sources of turbulence (i.e., the additional ESF) is artificially removed.

\begin{figure*}
    \centering
    \includegraphics[width=0.95\linewidth]{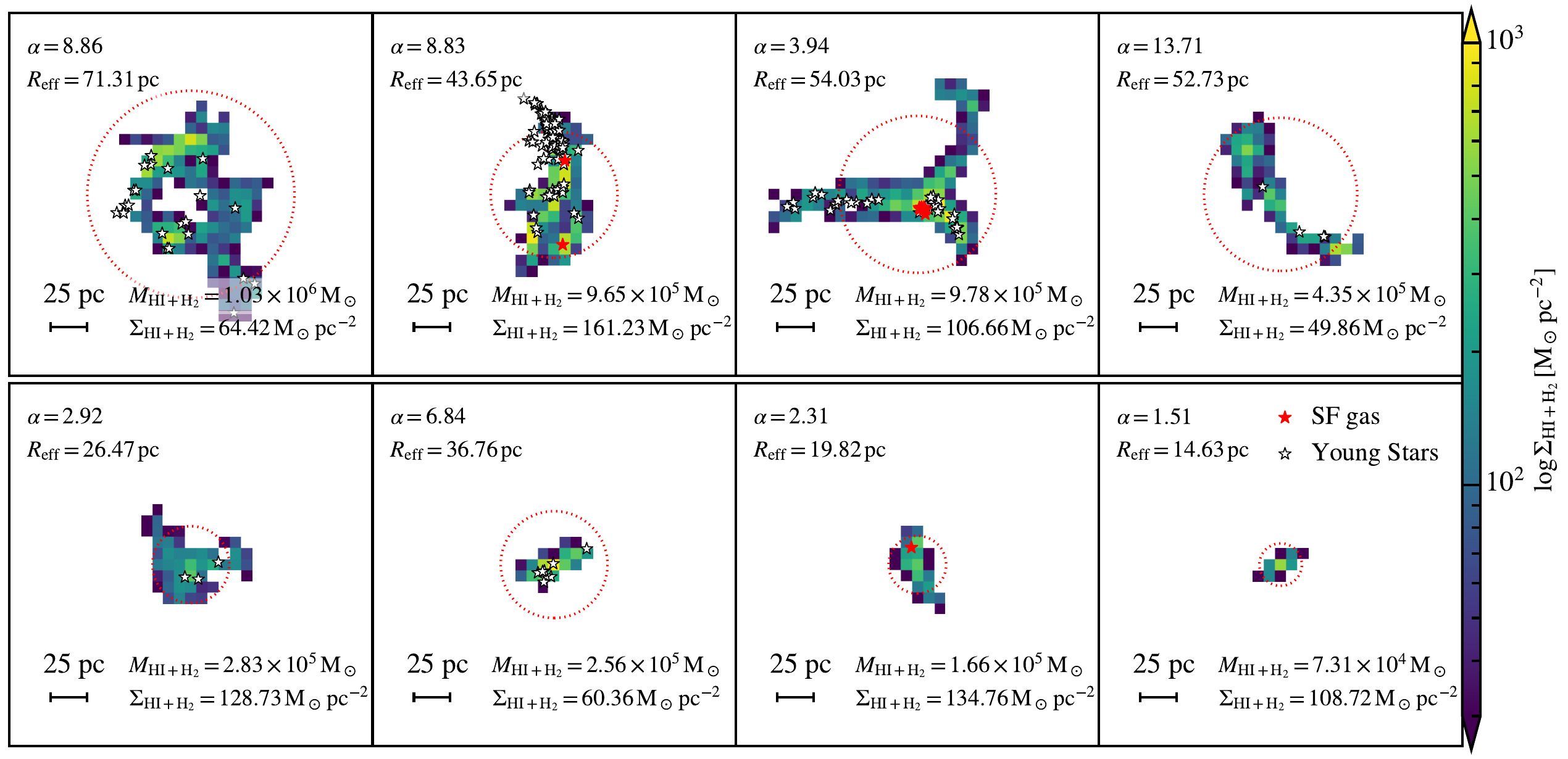}
    \caption{Visualization of eight identified GMCs in galaxy “m11.1” at $z \simeq 3$. Each cloud is centered on its center of mass. The colormap shows the gas surface density, and the red circle marks the effective radius ($R_{\rm eff}$) of each cloud. Star-forming gas cells and young stars associated with the GMCs are shown in red and white, respectively. GMCs typically exhibit filamentary and irregular morphologies. While the average surface density remains relatively low, certain local regions within the clouds can reach significantly higher surface densities.}
    \label{fig:gmc_shapes}
\end{figure*}

We now turn to the surface density of GMCs. The identified GMCs consist of both young stars and multi-phase gas cells, allowing us to define the surface density based on different baryonic components as $\Sigma = M / \pi R_{\rm eff}^2$. However, this definition may introduce some geometric bias, as the majority of GMCs in our simulation exhibit filamentary morphologies rather than well-defined spherical shapes, and the ISM has not yet settled into a disk-like configuration in \thesanzoom. 
In Fig.~\ref{fig:gmc_shapes}, we present the morphologies of eight identified GMCs. Each cloud is centered on its center of mass. For each GMC, we annotate its virial parameter, mass, size, and neutral gas surface density. The red circles indicate the effective radius, $R_{\rm eff}$.
GMCs across a wide mass range exhibit filamentary and irregular structures, likely resulting from their growth along SN-induced compressed fronts. Similar morphological features have also been observed in simulations of low-metallicity starburst galaxies~\citep{Fotopoulou2023}.

The average GMC surface density, in this context, is determined not only by the gas volume density but also by the morphology of the cloud. We have tested alternative approaches, such as smoothing the cloud structure (including the surrounding gas background) and computing the surface density using either the effective radius $R_{\rm eff}$ or the half-mass radius $R_{1/2}$. Both methods yield comparable results. We therefore conclude that our adopted definition provides a reasonable estimate of the compactness of GMCs. In addition, we measure the local DM density within a cubic region centered on each GMC, with a side length of $2 \times R_{\rm eff}$, but not smaller than 100 pc, to ensure the DM particles are properly resolved. For consistency in units, we convert this volume density into a surface density using the relation $\Sigma_{\rm DM} = \rho_{\rm DM} \times R_{\rm eff}$.

\begin{figure}
    \centering
    \includegraphics[width=1\linewidth]{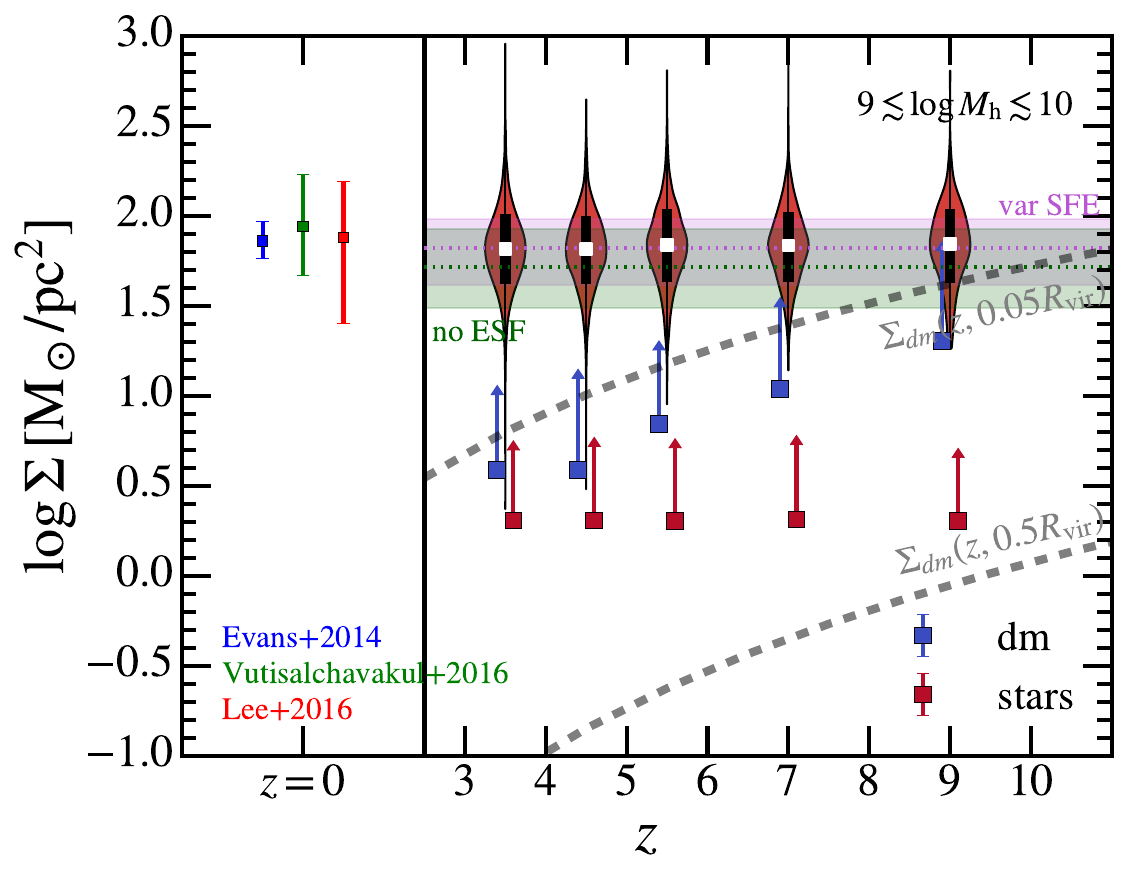}
    \caption{The surface densities of gas, stars, and DM associated with GMCs in haloes of $10^9\lesssim M_{\rm h}\lesssim10^{10}\,\rm M_{\odot}$ across different redshifts. For the gas component, the distributions are shown using violin plots, while for DM and stellar components, we display the median values along with the $84^{\rm th}$ percentiles. 
    Results are compared to local observational constraints from, e.g.~\citet{Evans2014, Vutisalchavakul2016, Lee2016}. We also overlay theoretical expectations for DM surface density at various fractions of $R_{\rm vir}$, assuming an NFW profile and a \citet{Klypin2016} concentration for haloes of $M_{\rm h} \sim 10^9 \,\rm M_\odot$. As no clear redshift dependence is found, we additionally show the median (dashed lines) and $1\sigma$ scatter (shaded regions) of gas surface densities for GMCs in the noESF and varSFE runs, colored in dark green and purple, respectively. While the median gas surface density remains $\Sigma_{\rm gas} \sim 70 \msun\,\mathrm{pc}^{-2}$ across all redshifts, the DM surface density increases with redshift following $\Sigma_{\rm DM}\propto(1+z)^3$.}
    \label{fig:gmc_sigma}
\end{figure}

In Fig.~\ref{fig:gmc_sigma}, we present the redshift evolution of the surface densities of gas, stars, and DM associated with GMCs in haloes of $10^9<M_{\rm h}/\msun<10^{10}$. For the gas component, the distribution is shown using violin plots, while for the other two components, we display the median values along with the $84^{\rm th}$ percentile. No significant redshift evolution is observed in the gas surface density, with the median remaining around $\sim 70\,\rm M_\odot \, \mathrm{pc}^{-2}$. A slight underestimation is observed compared to measurements of local GMCs~\citep[e.g.][]{Evans2014, Vutisalchavakul2016, Lee2016}, likely due to geometric effects mentioned above. Across all redshifts, the contribution of stellar surface density remains negligible. 
The DM surface density rises steadily with increasing redshift, approximately following $\Sigma_{\rm DM}\propto(1+z)^3$. We compare this trend with theoretical expectations for DM densities at various fractions of $R_{\mathrm{vir}}$. The reference surface densities are computed as $\Sigma = \rho \times R_{\rm crit}$, assuming an NFW profile~\citep{Navarro1997} and adopting the mass-concentration relation from \citet{Klypin2016}. Here, $R_{\rm crit}$ is set to 100 pc, which is roughly the same order of magnitude as the diameter of GMCs. The DM surface density of GMCs closely tracks the rise of the ambient DM density, with DM-dominated GMCs potentially emerging at $z \gtrsim  8$.

Would any numerical prescription determine the gas surface density of GMCs? To address this, we compare the gas surface densities of GMCs across different model variants. In the varSFE runs, both the median surface density and its scatter closely match those of the fiducial runs. This is expected, as the modification to the cell-level SFE has limited impact at 8x resolution. The majority of star formation occurs in the densest gas cells, where the number density exceeds the threshold required to reach the maximum cell-level SFE ($n_{\rm H} = 1000\, \mathrm{cm}^{-3}$).
On the other hand, the Jeans instability criterion at 8x corresponds to a minimum number density of approximately $100\, \mathrm{cm}^{-3}$, where the cell-level SFE already increases to $\sim10\%$. In addition, star-forming gas at lower densities typically appears in isolation, located in more diffuse regions where the associated GMCs are too small to be resolved.
As a result, the effective cell-level SFE within resolved GMCs remains essentially unchanged between the varSFE and fiducial runs.

In the noESF runs, however, the median gas surface density slightly decreases to $\sim 51 \,\rm M_\odot\,\mathrm{pc}^{-2}$, contrary to the naive expectation that suppressing early feedback would result in denser gas via unchecked collapse. This reduction may suggest that GMC formation at high redshift is governed by the global turbulence spectrum rather than by self-gravity. In a gravity-dominated regime, where dense clumps form in isolation, weakening stellar feedback allows GMCs to collapse further and reach higher surface densities~\citep{Li2020, Ni2025}. On the other hand, in a turbulence-dominated regime (more prevalent at high redshifts), GMCs are ``mixed'' with the ambient turbulent ISM. Removing a source of turbulence, such as the additional ESF, leads to geometrically more diffuse GMCs with lower surface densities. However, this affects only the overall morphology of GMCs, while gas in the dense cores of them in fact exhibits higher physical densities (see the lowest panel of Fig.~\ref{fig:ks_connection}).

\subsection{Cloud-scale star formation efficiency}
We now turn to the analysis of cloud-scale star formation efficiency in high-redshift galaxies.
Several flavours of SFE exist in the literature, motivated either by observational convenience or theoretical frameworks. In this study, we focus primarily on the instantaneous SFE per free-fall time $\epsilon_{\rm ff}$~\citep{Krumholz2005, Padoan2011, Hennebelle2011, Federrath2012}, that can be robustly derived from the \thesanzoom simulations. This is defined as the fraction of gas converted into stars per free-fall time
\begin{equation}
    \epsilon_{\rm ff}^{\rm GMC}=\frac{{\rm SFR}\times t_{\rm ff}}{M_{\rm gas}} \, ,
    \label{eq:gmc_sfe_ff}
\end{equation}
where ${t_{\rm ff}} \equiv \langle{ t_{{\rm ff,},i}}^{-1}\rangle_{\rm m}^{-1}$, and the $\rm SFR$ in this work is calculated using the sum of initial mass of young stars normalized by 3 Myr. 
This quantity reflects the current level of star formation activity within the cloud. Numerous previous studies have shown that $\epsilon_{\rm ff}$ exhibits a rise-and-fall behavior with large temporal fluctuations~\citep[e.g.][]{Grudic2018,Ni2025} over the course of a GMC’s lifetime, from the accretion of cold neutral gas to eventual disruption by stellar feedback. Nevertheless, its time-averaged value typically aligns with the observed level of approximately $1-3\%$~\citep{Grudic2018,Grudic2019}.

\begin{figure}
    \centering
    \includegraphics[width=1\linewidth]{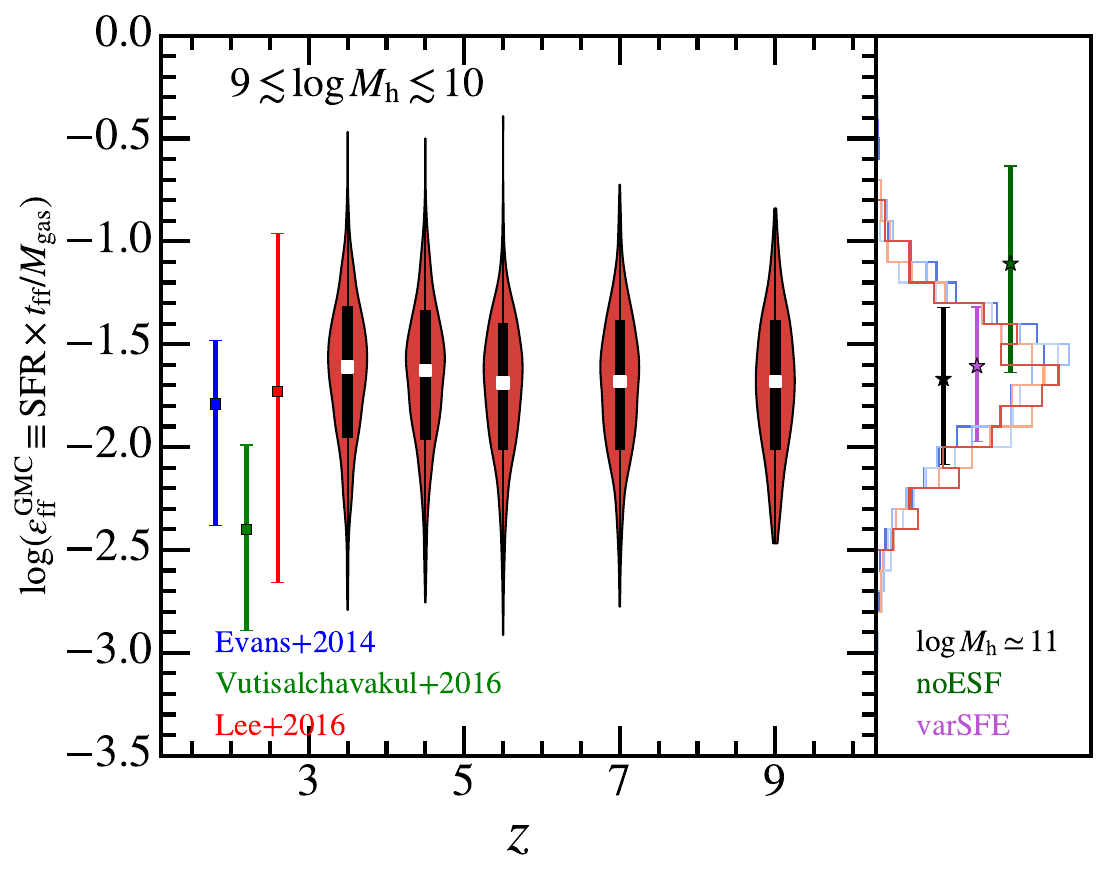}

    \caption{Redshift evolution of the instantaneous star formation efficiency of GMCs. For comparison, we include measurements of MW GMCs at $z=0$~\citep[e.g.][]{Evans2014, Vutisalchavakul2016, Lee2016}, using the same color coding as in Fig.~\ref{fig:gmc_sigma}. The right panel shows the distribution of $\epsilon_{\rm ff}^{\rm GMC}$, with redder colors corresponding to higher redshifts, and includes a comparison with results from different numerical models (dark green and purple) as well as from high-mass haloes (black) at $z \simeq 3-4$. While no significant redshift evolution is found, turning off the additional ESF increases the average SFE by a factor of $\sim 3$.}
    \label{fig:gmc_sfe}
\end{figure}

In Fig.~\ref{fig:gmc_sfe}, we present the distributions of instantaneous star formation efficiency of GMCs at different redshifts. For comparison, we include measurements of MW GMCs~\citep[e.g.][]{Evans2014, Vutisalchavakul2016, Lee2016}, using the same color coding as in Fig.~\ref{fig:gmc_sigma}.
No significant redshift evolution is observed. The median value is approximately 3\%, slightly higher than that from observations of local GMCs.
As our sample consists of a population of GMCs captured at different evolutionary phases, the measurements exhibit large scatter~\citep{Grudic2019}. Nevertheless, the universal distribution of $\epsilon_{\rm ff}^{\rm GMC}$ suggests that the overall star formation efficiency of GMC populations is largely insensitive to redshift, consistent with the trends identified in Sect.~\ref{subsec:multiSFE}. While that of varSFE runs stays consistent with the fiducial runs, the only factor that affects the instantaneous SFE is the additional ESF, as the median value in noESF runs increase by a factor of $2-3$.

\begin{figure}
    \includegraphics[width=1\linewidth]{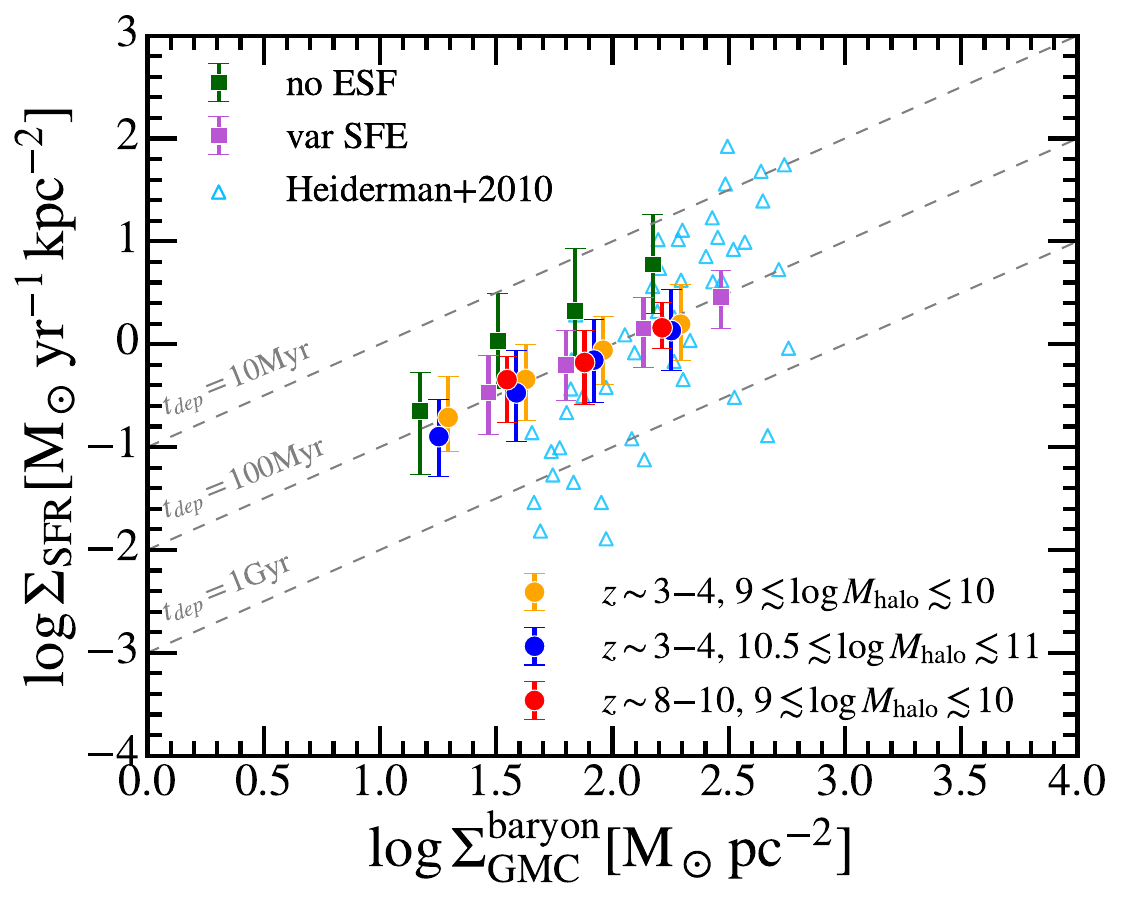}
    \caption{The relation between surface density and star formation rate surface density for the three representative systems and two model variants. We compare them to the local observations~\citep{Heiderman2010}. GMCs across different systems in the fiducial runs exhibit a similar depletion time of approximately 100 Myr, while the SFR at a fixed surface density in the noESF runs is higher by a factor of $\sim 3$.}
    \label{fig:gmc_ks}
\end{figure}

Alternatively, one can directly measure the gas depletion time of GMCs. In Fig.~\ref{fig:gmc_ks}, we present the relation between surface density and star formation rate surface density for different systems and model variants. This relation is equivalent to plotting SFR versus GMC mass. We find a universal KS-type relation at the GMC scale, with a best-fit scaling of $\Sigma_{\rm SFR} \propto \Sigma_{\rm GMC}$, that is, a depletion time of approximately 100 Myr. At the high surface density end, the results are in good agreement with local GMC observations~\citep{Heiderman2010}, while at the low surface density end, the SFRs are systematically higher than those observed locally. The noESF runs exhibit elevated SFRs in the high-surface-density regime, exceeding the fiducial case by a factor of 3, while little difference is observed at low surface densities, where the paucity of young stars limits their impact on star formation.

Another commonly reported SFE in numerical simulations of star-forming clouds~\citep[e.g.][]{McKee2007, Grudic2018, Ni2025} is the integrated SFE $\epsilon_{\rm int}$, defined as the fraction of the initial gas mass converted into stars over the entire lifetime of a cloud
\begin{equation}
    \epsilon^{\rm GMC}_{\rm int} = M_{\rm \star}(t=\infty)/M_{\rm gas}(t=0) \, ,
    \label{eq:gmc_sfe_int}
\end{equation}
where $M_{\rm \star}(t=\infty)$ is the total mass in stars formed and $M_{\rm gas}(t=0)$ is the initial gas mass\footnote[3]{For a more realistic case where the gas mass may increase during the evolution of a GMC, one may instead use the maximum baryonic mass attained throughout its lifetime as the denominator.}.
Measuring this quantity requires knowledge of the mass composition at both the formation and dispersal stages of the GMC, which is not directly accessible in observations and also not feasible to track in simulations with low output cadence. In galaxy-scale simulations, this challenge is further compounded by the highly dynamic evolution of GMC mass, as clouds frequently undergo accretion, merging, and splitting, making even the definition of an initial gas mass ambiguous. For a typical GMC of $10^5\msun$, the lifetime is on the order of 5-6 Myr~\citep{Benincasa2020, Ni2025}, shorter than the output interval of \thesanzoom of $\sim10$ Myr, rendering direct tracking impossible. Even for longer-lived ones, the mass flux between gas cells in the quasi-Lagrangian scheme and the (de)refinement of cells~\citep{Weinberger2020arepo} makes detailed tracking of GMCs challenging without additional foresight, such as the use of on-the-fly cloud identification or tracer particle techniques. Therefore, we do not measure $\epsilon_{\rm int}^{\rm GMC}$ directly but instead infer it from the distribution of instantaneous SFE, as will be discussed later in Sect~\ref{subsec:gmc_dm}.

\section{Connecting the global and local \\ star formation}\label{sec:connect}

With the cloud-scale results in hand, we can now move on to understanding multi-scale star formation efficiency.
As discussed in e.g. \citet{Faucher2013}, the total mass formed in a time duration of $\Delta t$ can be expressed in the following two different ways
\begin{align}
    M_{\star} &= \dfrac{\Delta t}{t_{\rm dep}}\, M_{\rm gas} \hspace{2.75cm}   [{\rm galactic\,\, view}] \nonumber \\
    M_{\star} &= \left(\dfrac{\Delta t}{t^{\rm GMC}_{\rm life}}\right)\, M_{\rm gas}\, f_{\rm GMC}\, \epsilon_{\rm int}^{\rm GMC} \hspace{0.6cm}   [{\rm GMC\,\, view}]
\end{align}
where $t^{\rm GMC}_{\rm life}$ is the GMC lifetime, $f_{\rm GMC}$ is the fraction of gas mass in GMCs.
Equating the two expressions, we obtain 
\begin{equation}
    t_{\rm dep} = (t^{\rm GMC}_{\rm life}/\epsilon^{\rm GMC}_{\rm int})/ f_{\rm GMC} \, .
    \label{eq:bridge}
\end{equation}

Since $\epsilon^{\rm GMC}_{\rm int}$ is not directly accessible in \thesanzoom, we instead seek alternative proxies within this relation. To zeroth order, the formula holds if one replaces $t^{\rm GMC}_{\rm life}$ and $\epsilon_{\rm life}^{\rm GMC}$ with $t_{\rm ff}$ and $\epsilon_{\rm ff}$ via 
\begin{equation}
\begin{split}
    &\epsilon^{\rm GMC}_{\rm int}/t^{\rm GMC}_{\rm life}=\frac{\int{(\epsilon^{\rm GMC}_{\rm ff}(t)/t^{\rm GMC}_{\rm ff}(t))\times M_{\rm gas}(t){\rm d}t}}{{M_{\rm gas}(t=0)}\int{{\rm d}t}} \\
    &\hspace{1.6cm}\simeq\langle \epsilon^{\rm GMC}_{\rm ff}/t^{\rm GMC}_{\rm ff}\rangle_{\rm m} \, .
\end{split}
\label{eq:int2ff}
\end{equation}

In practice, we find that measuring $\langle \epsilon^{\rm GMC}_{\rm ff} \rangle_{\rm m} / \langle t^{\rm GMC}_{\rm ff} \rangle_{\rm m}$ and $\langle \epsilon^{\rm GMC}_{\rm ff} / t^{\rm GMC}_{\rm ff} \rangle_{\rm m}$ yields similar values. 
Thus we rewrite Eq.~(\ref{eq:bridge}) and obtain
\begin{equation}
    t_{\rm dep} = \langle t^{\rm GMC}_{\rm ff} \rangle / \left(\langle \epsilon^{\rm GMC}_{\rm ff} \rangle \, f_{\rm GMC}\right) \, ,
    \label{eq:ks}
\end{equation}
which reflects that the global depletion time is therefore determined by the competition between $\epsilon^{\rm GMC}_{\rm ff}$, $t^{\rm GMC}_{\rm ff}$ and $f_{\rm GMC}$.

\begin{figure}
    \centering
    \includegraphics[width=1\linewidth]{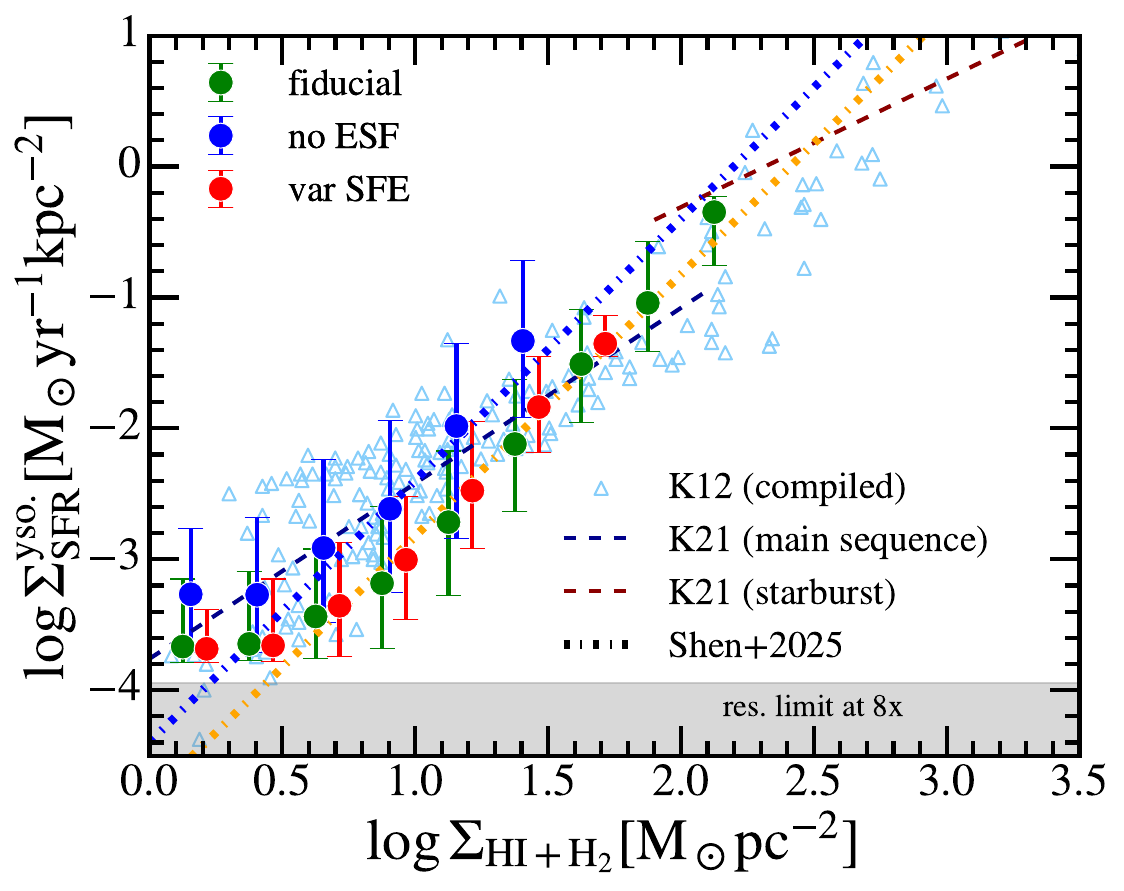}
    \caption{KS relations of neutral gas in \thesanzoom simulations with different physics variations. 
    We compare the KS relation measured in simulations to the local observed ones of main-sequence and starburst galaxies from \citet{Kennicutt2021}, individual data points compiled in \citet{Kennicutt2012}.
    We also compare the results with the analytic model presented in S25, using the feedback momentum injection rates $P_\star/m_\star$ adopted in the fiducial and “no ESF” runs, respectively. We find that a smaller geometrical factor, $f \simeq 0.15$, is required due to the absence of contamination from projection effects. The shaded region shows the regime where SFR surface density is not properly resolved; i.e. only one young stellar particle in a pixel. Before reaching the numerical floor, the KS relations follow $\Sigma_{\rm SFR}\sim \Sigma^{2}_{\rm HI+H_2}$, while the normalization increases after removing the additional ESF.}
    \label{fig:kpc_ks}
\end{figure}

We begin by briefly revisiting the global KS relation studied in S25, which follows a simple power-law form, $\Sigma_{\rm SFR} \propto \Sigma_{\rm HI+H_2}^2$. This can be interpreted as the consequence of the balance between momentum injection from stellar feedback and turbulent dissipation in the ISM. The gas depletion time is found as
\begin{equation}
    t_{\rm dep} \simeq 0.5\,\Gyr\,\left(\frac{f}{0.3}\right)\,\left(\frac{P_\star/m_\star}{3000\,{\rm km\,s^{-1}}}\right)\,\left(\frac{\Sigma_{\rm gas}}{100\,{\rm M_\odot\,pc^{-2}}}\right) \, ,
\end{equation}
where $f$ is an order-unity fudge factor to encapsulate order-unity constants in the derivation and the unaccounted geometrical effects, which is found to be 0.3 in S25, and the total momentum injection from stellar feedback per unit stellar mass formed, $P_\star/m_\star$ is roughly $8000\,{\rm km\,s^{-1}}$ in fiducial runs and $3000\,{\rm km\,s^{-1}}$ in noESF runs. In Fig.~\ref{fig:kpc_ks}, we revisit this analysis by dividing the central $0.5\ R_{\rm vir}$ region of each galaxy into cubic cells with a side length of 1 kpc, and calculating the SFR from the initial mass of young stellar objects (YSOs) younger than 10 Myr within each cube.
A key difference over S25 is that our method is largely free from projection effects, which can otherwise obscure intrinsic gas-SFR correlations. Overall, our measurements are in excellent agreement with the results of S25, with only a slightly smaller $f\simeq0.15$, likely due to the absence of contamination from projection.

\begin{figure}
    \centering
    \includegraphics[width=1\linewidth]{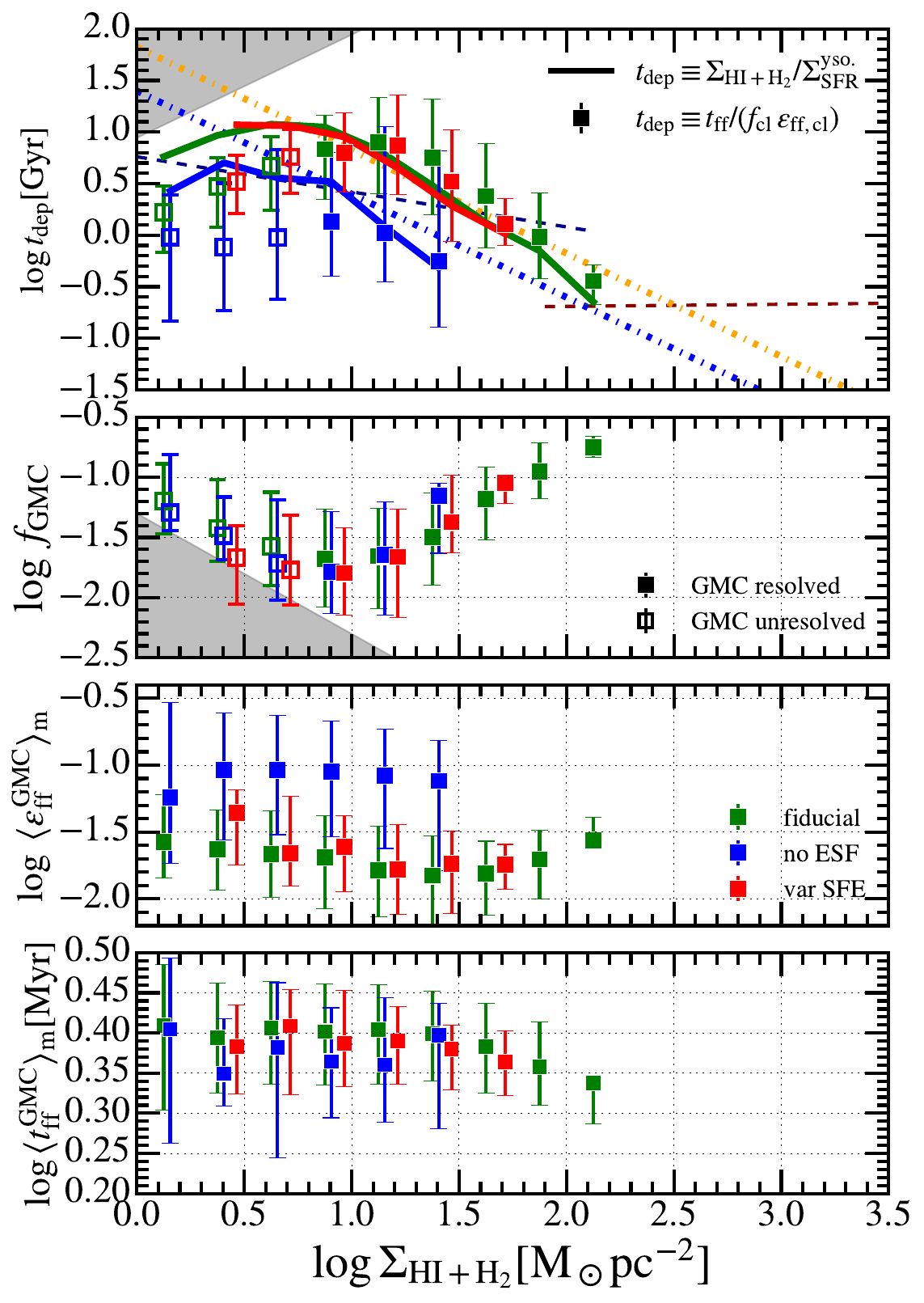}
    \caption{Microscopic decomposition of the KS relation. \textit{Top panel:} Comparison between the measured kpc-scale depletion time (solid lines) and the reconstructed values (data points with error bars), derived from the three components in Eq.~(\ref{eq:bridge}). The reconstructed values are computed only from cubes containing resolved GMCs, while the measured values include all cubes that host YSOs. 
    Different model variants are shown in different colors. Hollow data points indicate cubes that contain only a single GMC of typical mass ($ 10^5\,\msun$), where the GMC mass fraction is considered unresolved, and the shaded region marks the regime where the SFR surface density is unresolved. Overall, the two estimates show good agreement. 
    \textit{Lower panels:} Variations of the three components; i.e. the GMC mass fraction ($f_{\rm GMC}$), the mass-weighted instantaneous cloud-scale SFE ($\langle \epsilon_{\rm ff}^{\rm GMC} \rangle_{\rm m}$), and the mass-weighted harmonic mean free-fall time ($\langle t_{\rm ff} \rangle_{\rm m}$), as a function of neutral gas surface density. These quantities are measured using all identified GMCs, including non-star-forming ones, whose centers of mass lie within the selected 1-kpc cubes. Among the three components, $f_{\rm GMC}$ is the dominant factor in determining the global depletion time, while $\langle \epsilon_{\rm ff}^{\rm GMC} \rangle_{\rm m}$ exhibits a non-monotonic transition, reflecting the non-linear interplay between feedback and gravitational collapse.
}
    \label{fig:ks_connection}
\end{figure}

We now turn to a microscopic decomposition of the global SFE. To do this, we select the cubic cells identified above that contain resolved GMCs and compute their mass fraction, mass-weighted SFE, and free-fall time. Since no significant redshift and halo mass dependence is found, we combine gas cells from all galaxies in the sample. The surface density of each cube is defined as the neutral gas mass within the volume, divided by $1\,\mathrm{kpc}^2$. In calculating GMC-related quantities, we also include non-star-forming GMCs to ensure completeness. In Fig.~\ref{fig:ks_connection}, we show how the kpc-scale depletion time and its three constituent components vary as functions of surface density. In the top panel, we compare the depletion time directly measured in the simulation to that reconstructed from the three components using the formulation in Eq.~(\ref{eq:ks}). The two estimates are in good agreement, until they hit the resolution limit, where only a single GMC of typical mass $10^5\,\msun$ is present within the cube.

In the lower panels, we compare the variation strength of the three components. The most prominent change is seen in $f_{\rm GMC}$, which rises steadily from about 2\% in the first resolved patches to nearly 20\% at the high surface density end. Across all three runs, as the environmental surface density increases, the GMCs that form tend to become progressively denser, as indicated by a mild decline in $t_{\rm ff}$.

In contrast, $\epsilon_{\rm ff}$ exhibits modest fluctuations within a factor of 2, in both the fiducial and varSFE runs. Such a subtle but important fluctuation may indicate the emergence and gradual weakening of self-regulation. In the most diffuse regions, where only a single GMC is present, the absence of external feedback allows the cloud to form stars with relatively higher efficiency. As the surface density increases, the number of GMCs grows, leading to stronger cumulative feedback from neighboring regions. This external feedback, via self-regulation, acts to suppress $\epsilon_{\rm ff}$~\citep{Semenov2018}. However, at sufficiently high surface densities, gravitational binding becomes strong enough to resist feedback, resulting in a rebound of the average $\epsilon_{\rm ff}$. This also explains why fluctuations in $\epsilon_{\rm ff}$ are significantly reduced in the noESF runs. 

Furthermore, we find that GMCs in the noESF runs are statistically indistinguishable from those in the fiducial runs in terms of $f_{\rm GMC}$. The internal density structure reflected from $t_{\rm ff}$ shows larger scatter, as more dense GMCs are found in low-density patches compared to the fiducial runs. The observed increase in global depletion time in the noESF runs is therefore driven primarily by an enhancement in local GMC SFE, rather than changes in GMC abundance.

\section{Discussions}\label{sec:discuss}
\subsection{A case study of starburst near FFB regime}

\begin{figure*}
    \includegraphics[width=1\linewidth]{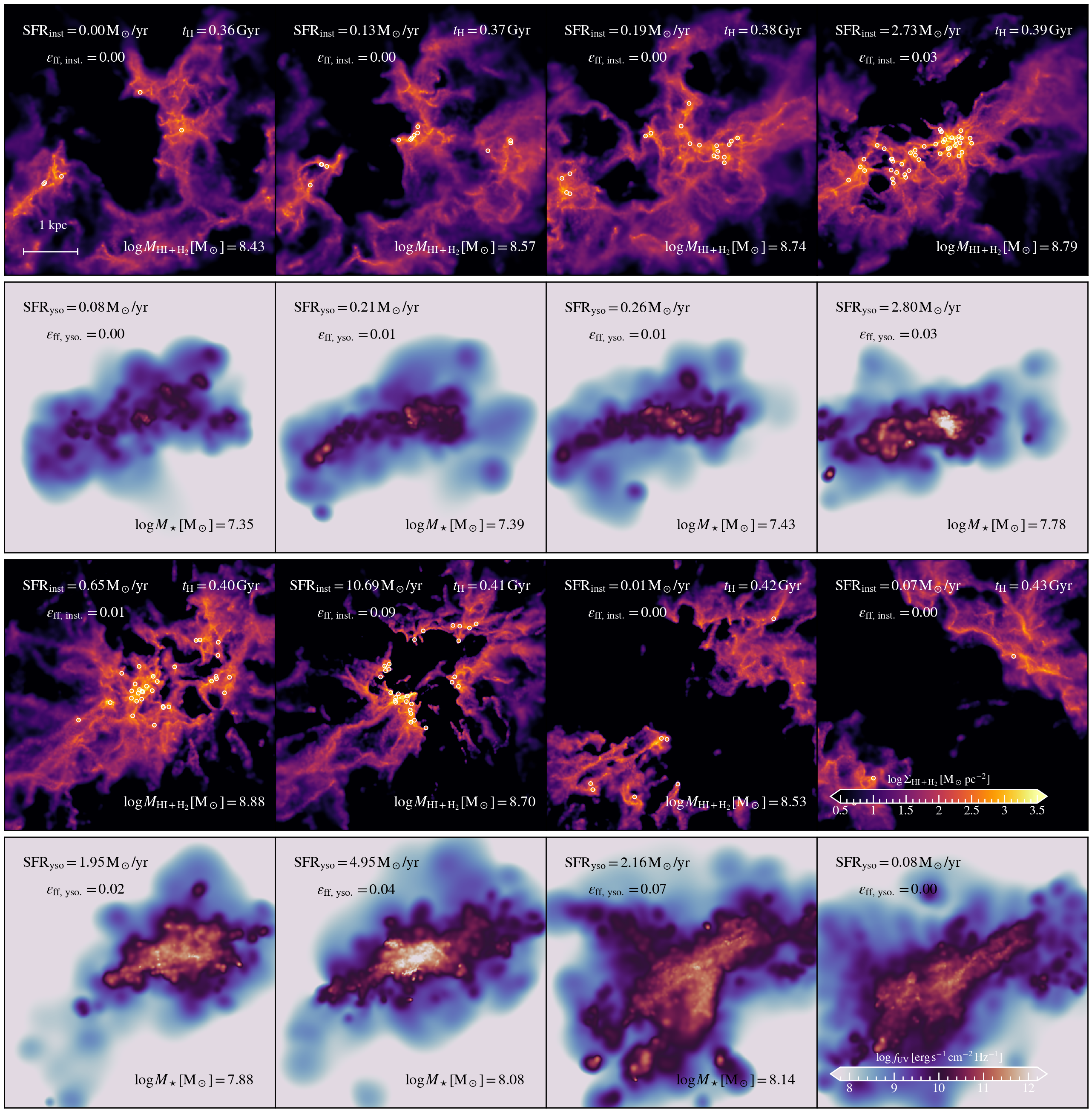}
    \caption{Duty cycle of a starburst galaxy (“m12.6”) at $z \sim 11$ with $\log M_{\rm halo} \sim 10.2$. An 80 Myr timeline is shown from top to bottom. For each snapshot, we display the projected gas surface density (top panels) and stellar UV emission (bottom panels). The instantaneous and 10 Myr-averaged SFR and SFE are also calculated. Identified GMCs are marked with white circles.
    This starburst exhibits two distinct star formation peaks: the first is driven by rapid gas accumulation, while the second is triggered by the formation of numerous GMCs along SN-induced compressed fronts, indicating positive feedback from SNe.}
    \label{fig:ffb}
\end{figure*}

\begin{figure}
\includegraphics[width=1\linewidth]{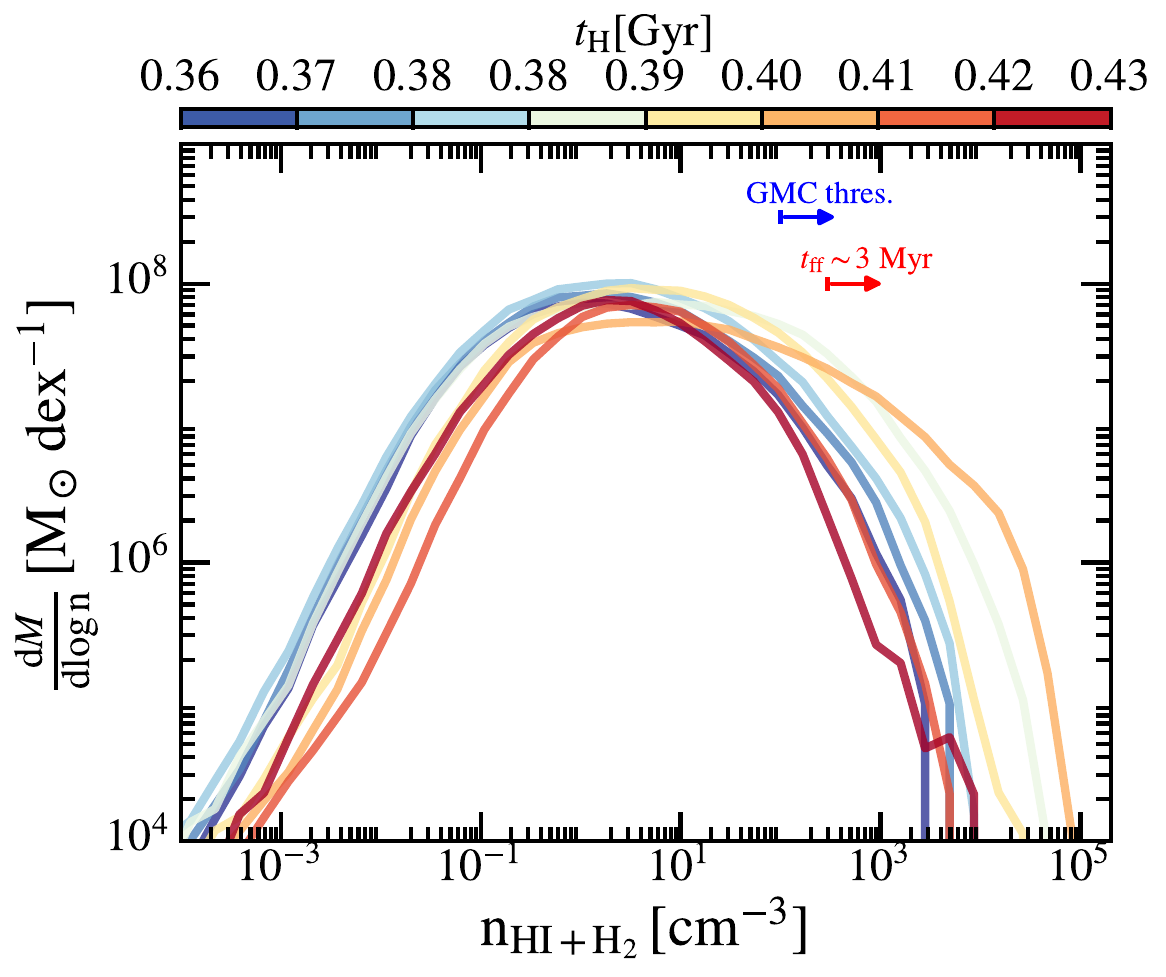}
    \caption{ISM density distribution over time during the starburst duty cycle. The timestamp corresponding to each row is indicated in the top color bar, where the color transitions from blue to red represent progressively later times. Throughout the starburst, the main body of the distribution and the median density remain largely unchanged. In contrast, the high-density tail, which closely traces the SFR, first rises and then declines. }
    \label{fig:ffb_ism}
\end{figure}

\begin{figure}
\includegraphics[width=1\linewidth]{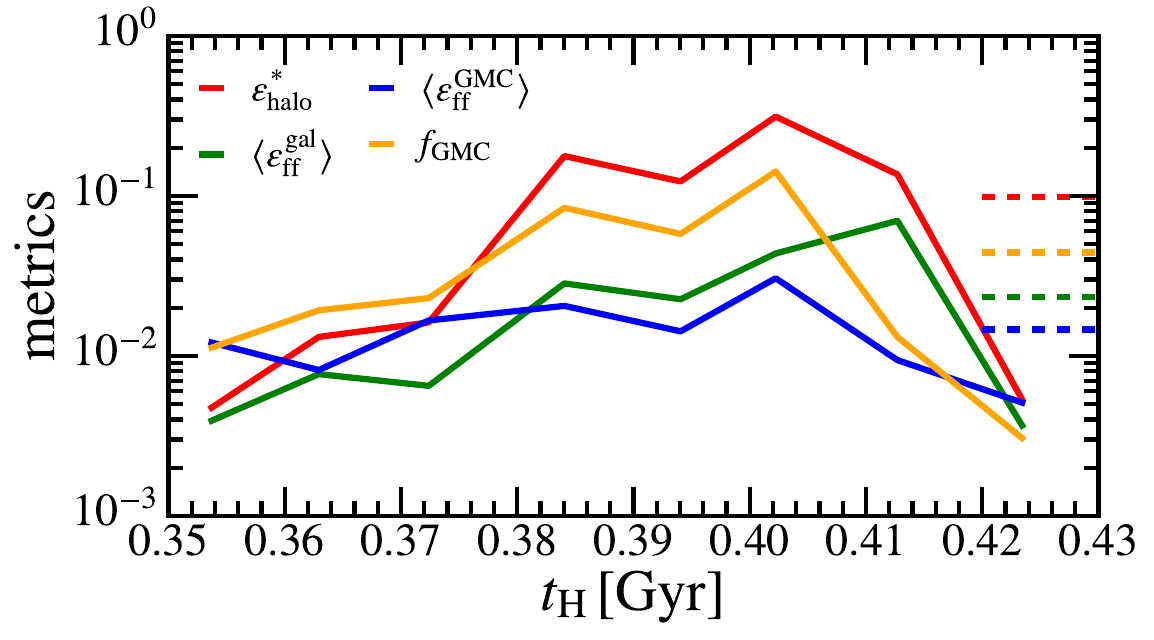}
    \caption{Evolution of star formation efficiency at different scales and the GMC mass fraction across time. The short dotted lines on the right edge indicate the time-averaged values of each quantity. All four metrics evolve in a coordinated manner, with the cloud-scale SFE showing the least variation.}
    \label{fig:ffb_sfe}
\end{figure}

At all examined scales, we find no direct evidence of feedback-free starbursts, likely because the halo mass range probed by \thesanzoom does not satisfy the conditions required for this scenario. Among all zoom-in regions in the simulation suite, we identify one galaxy with $M_{\rm h} \sim 10^{10} \, \msun$ at $z \sim 11$ in the 4x fiducial run that comes closest to the FFB regime. We investigate its evolution over one dynamical time to explore why order-unity global SFE was not realized. It is worth noting that some absolute properties of GMCs at this resolution may differ from those in the 8x runs, as will be discussed in the Appendix~\ref{apdx:resolution}. However, the key trends reported above are found to be universal across all resolutions.

In Fig.~\ref{fig:ffb}, we show the evolution of gas and stellar distribution around a single galaxy during its starburst, focusing on the central $2\times0.5\,R_{\rm vir}$ region, with a temporal cadence of approximately 10 Myr. For each snapshot, the instantaneous SFR in gas cells and the 10 Myr-averaged SFR are displayed in the top right corner. The galaxy-scale SFE is computed using Eq.~(\ref{eq:glbSFE}), and the spatial distribution of identified GMCs is marked with white circles. This starburst captures a sequence in which star formation is triggered by strong gas inflows, followed by intense stellar feedback that initially induces a second burst of star formation, and subsequently disperses most of the gas, leading to the quenching of further star formation.

In Fig.~\ref{fig:ffb_ism}, we track the evolution of the ISM density distribution over the starburst duty cycle. The timestamp corresponding to each row is indicated in the inset color bar, where the color transitions from blue to red represent progressively later times. The main body of the distribution follows a log-normal shape, with only mild variation in its median density across time. A high-density tail emerges, which directly reflects the level of instantaneous SFE. Even in the snapshot where the galaxy reaches its maximum density, the highest instantaneous SFE achieved is only 7\%, and only about 30\% of the gas has a free-fall time shorter than 3 Myr. The mean density corresponds to a free-fall time of approximately 5 Myr, which exceeds the 3 Myr timescale for SNe onset. This mismatch in timescales is one of the reasons why a feedback-free burst fails to develop. We note that the inclusion of the additional ESF in the fiducial runs may also, by construction, inhibit the emergence of FFB.

In Fig.~\ref{fig:ffb_sfe}, we quantify the SFE on three characteristic spatial scales. The halo-scale SFE is defined as
\begin{equation}
    \epsilon^*_{\rm halo}\equiv {\rm SFR}/(f_{\rm b}\,\dot M_{\rm halo}) \, ,
    \label{eq:haloSFE}
\end{equation}
where $\dot{M}_{\rm halo}$ is measured as the change in halo mass over a duration of $t_{\rm dyn}$, and SFR is averaged over 10 Myr. The galaxy-scale SFE is defined in Eq.~(\ref{eq:glbSFE}). The GMC-scale SFE is computed as the mass-weighted $\epsilon_{\rm ff}^{\rm GMC}$ of all GMCs identified in each snapshot. We also track the mass fraction of GMCs across snapshots. The time-averaged quantities over the full duty cycle are indicated on the left side of the panel.
Throughout the duty cycle, the maximum halo-scale SFE reached by this starburst is 30\%. All three SFE measures exhibit a coherent temporal evolution. 

At first glance, the variation in GMC-scale SFE may appear inconsistent with the universal trend reported earlier. However, this discrepancy may arise because in the case of a strong, short-lived starburst, the GMC population can form over a narrow time window and not reach equilibrium. That is, the fluctuations in GMC-scale SFE reflect the fact that the entire GMC population is caught in the same evolutionary phase. This may offer a cloud-scale explanation for why star formation appears more bursty in lower-mass haloes in \thesanzoom~\citep[][Shen et al. in prep.]{McClymont2025-MSscatter}, as variations in cloud-scale SFE could amplify burstiness in the star formation histories.

It's also interesting that this starburst exhibits two distinct peaks in SFR and SFE. While the first peak is likely driven by gas accretion, the second appears to be triggered by strong, simultaneous SNe, as a large number of GMCs emerge along SNe-compressed fronts. Rather than quenching star formation, the SNe act as positive feedback~\citep[e.g.][]{Greif2007, Cosentino_2022}, compressing the gas and triggering a second episode of star formation\footnote[4]{\thesanzoom does not adopt a converging flow criterion~\citep{Hopkins2013} for star formation. This could potentially promote the compressive or triggered mode of star formation.} that contributes a comparable amount of stellar mass. This feature is inline with the shell configuration proposed in the FFB scenario~\citep{Dekel2023}, where the interaction between cold supersonic gas inflow and stellar winds from a previous generation of stars drives shock compression, thereby triggering subsequent starburst. However, the prevalence of such positive feedback, as well as the conditions under which it operates, remain open questions that warrant further investigation.


\subsection{DM-dominated GMCs at $z\gtrsim 8$}\label{subsec:gmc_dm}

\begin{figure}
    \centering
    \includegraphics[width=1\linewidth]{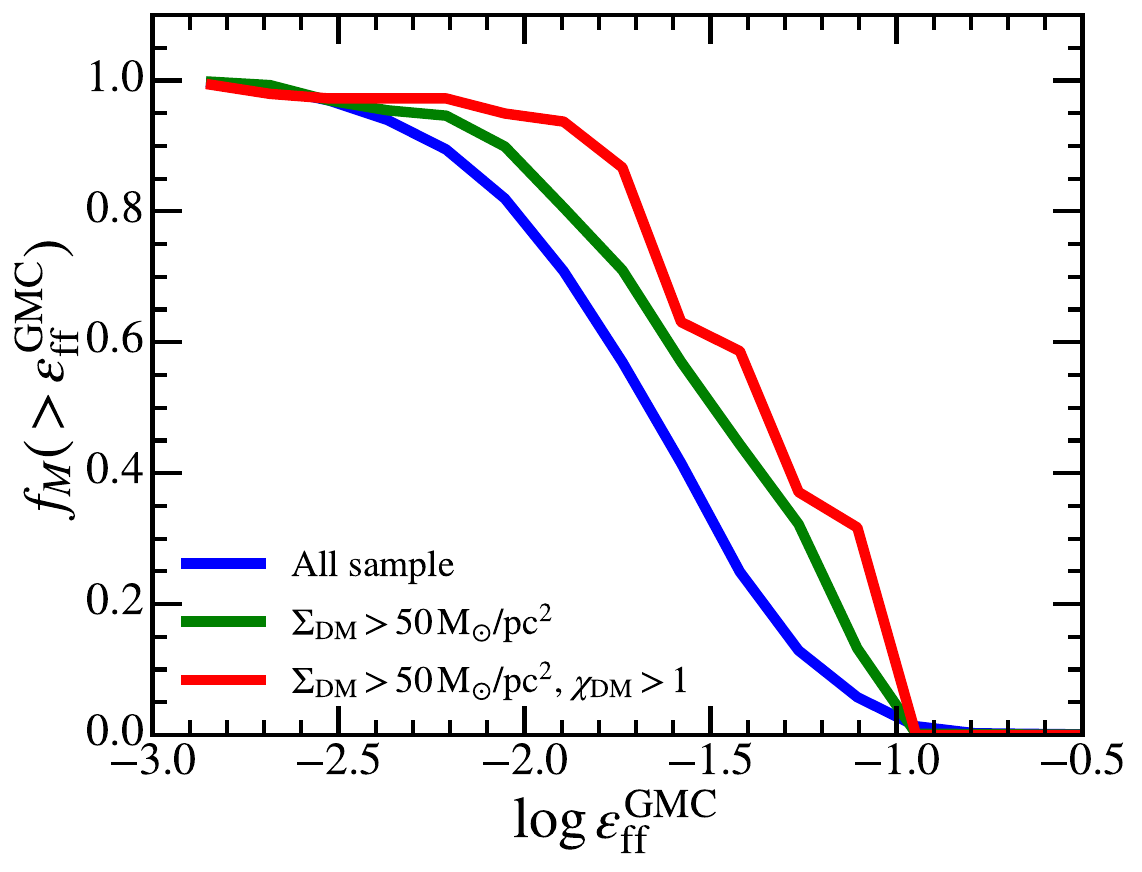}
    \caption{Cumulative mass-weighted distribution of $\epsilon_{\rm ff}^{\rm GMC}$ for DM-rich (green), DM-bounded (red), and regular (blue) star-forming GMCs. The GMC sample is compiled from all selected starbursts in the L8 run. DM-rich and DM-bounded GMCs exhibit systematically higher SFE compared to regular ones.}
    \label{fig:gmc_dm_dom}
\end{figure}

As reported in Section~\ref{subsec:gmc_property}, the DM surface density of GMCs increases with redshift. The increasing DM surface density may indicate the exceptional nature of certain GMC properties in high-redshift galaxies. For instance, \citet{Grudic2018} demonstrated a positive correlation between the integrated star formation efficiency and the surface density of clouds, 
\begin{equation}
    \epsilon_{\rm int}^{\rm GMC}=\left(1+\frac{\Sigma_{\rm fb}}{\Sigma_{\rm tot}}\right)^{-1},
\end{equation}
where $\Sigma_{\rm fb}$ is the critical feedback density, and $\Sigma_{\rm tot}$ is the total surface density of all mass components contributing to the gravitational potential. This is derived by equating the feedback force acting on the gas with gravitational forces, assuming that this balance marks the termination of star formation within a GMC~\citep{Murray2010, Grudic2018, Li2019}. In this context, DM, acting as an additional source of gravity, may enhance the integrated star formation efficiency at fixed baryonic surface density — the DM-induced feedback failure scenario proposed by \citet{Boylan-Kolchin2025}.
However, deriving the integrated star formation efficiency requires tracing the full GMC life cycle to capture its evolution, which is not feasible in \thesanzoom due to the limited output cadence of 10 Myr. 

Though we cannot measure $\epsilon_{\rm int}^{\rm GMC}$ that is directly predicted by the feedback-failure scenario, the distribution of instantaneous SFE may shed light on it. The underlying idea is that, assuming a given population of GMCs follows a similar evolutionary pathway, the distribution of their instantaneous SFE reflects the relative amount of time these clouds spend in a particular phase of star formation~\citep{Grudic2018}. In Fig.~\ref{fig:gmc_dm_dom}, we present the mass-weighted distribution of $\epsilon^{\rm GMC}_{\rm ff}$, for three populations: DM-rich, DM-bounded, and regular star-forming GMCs. We define DM-rich GMCs as those with DM surface densities exceeding $50\, \rm M_\odot\, \mathrm{pc}^{-2}$. The binding strength between GMCs and the surrounding DM is quantified using a coupling factor defined as
\begin{equation}
    \chi_{\rm DM}=\sigma_{{\rm GMC},\,v}/\sigma_{{\rm DM},\,v} \, .
    \label{eq:dm_couple}
\end{equation}
Here, $\sigma_{{\rm GMC},\,v}$ is the velocity dispersion of GMCs, and $\sigma_{{\rm DM},\,v}$ is the average relative velocity of nearby DM particles (within the effective radius of the GMC) with respect to the GMC, computed as
\begin{equation}
    \sigma_{{\rm DM},\,v}=\sqrt{ \frac{\sum_{\,r_{{\rm DM},i}<R_{\rm eff}} m_{{\rm DM},i} |\vec{v_{{\rm DM},i}}-\vec{v_{\rm c}}|^2}   {\sum_{\,r_{{\rm DM},i}<R_{\rm eff}} m_{{\rm DM},i}} }\, .
\end{equation}
where $m_{{\rm DM},i}$, $\vec{v}_{{\rm DM},i}$, and $r_{{\rm DM},i}$ are the mass, velocity, and distance to GMC center of the $i$-th DM particle.
A GMC is considered DM-bounded if it satisfies both conditions: (1) $\chi_{\rm DM} > 1$ and (2) is classified as DM-rich.

We find that both DM-rich and DM-bounded GMCs exhibit SFE distributions skewed toward higher values (with a peak at $\epsilon_{\rm ff}\sim10\%$) compared to the regular population. This may indicate that these clouds remain in high-SFE phases for a longer fraction of their lifetimes, potentially due to enhanced gravitational support from the DM background. 

However, several important caveats warrant particular attention. First, only a small subset (50 out of 356) of the DM-rich clouds are truly gravitationally bound to the surrounding DM, as indicated by their coupling factors. This implies that most DM-rich clouds are likely just passing through regions of high DM density, rather than being genuinely anchored by it. The absence of a clear difference in the SFE distributions between DM-bounded and DM-rich GMCs may reflect the fact that the ambient DM density is already high in these environments, making the distinction between bound and unbound clouds less dynamically relevant. Second, the DM surface density of GMCs, given their relatively constant physical sizes, is primarily determined by the local DM volume density, which depends on redshift and, to a lesser extent, on the radial position of the GMC within the halo~\citep[as the DM density gradient is generally mild in the central regions where most GMCs reside;][]{Navarro1997}. As shown in Fig.~\ref{fig:gmc_sigma}, if the gas surface density remains roughly constant toward higher redshift, it is likely that the majority of GMCs would become DM dominated within the central $0.05\,R_{\rm vir}$ regions at $z \gtrsim 10$. 

For the feedback-failure scenario, the total surface densities of the identified GMCs remain below the critical threshold of $\sim 3000 \, \rm M_\odot \, \mathrm{pc}^{-2}$~\citep[or even larger if galactic shear is considered;][]{Ni2025}, which is typically required to suppress stellar feedback effectively~\citep{Grudic2018}. As a result, we are unable to directly test the validity of this scenario with the current suite of simulations. 
However, the scaling relation suggests that in a halo with a concentration $c\sim10$, achieving a DM surface density of $1000\,\rm M_\odot\,pc^{-2}$ would require the GMC to reside within $\sim 0.01\,R_{\rm vir}$ at $z \sim 20$, implying it is extremely difficult for DM alone to suppress feedback entirely on the cloud scale. 

We note that our prediction differs from that of \citet{Boylan-Kolchin2025}, who argued that in a MW–mass halo at $z \sim 10$, the DM surface density may already be sufficient for gravitational forces to overcome stellar feedback for most of the gas. This discrepancy arises because their estimate of the gravitational potential treats the galaxy (or halo) as a whole, such that the force at radius $r$ within the halo depends on the total enclosed mass within $r$, thus incorporating a global dependence on halo mass. In contrast, our analysis focuses on the local gravitational potential at the cloud scale, which depends solely on the local DM density. Consequently, our model generally predicts that even higher redshifts are required to reach a similar level of gravitational confinement. Nonetheless, the two results are not in conflict: while in our framework the star formation efficiency of individual GMCs is determined by the local gravitational potential, the global potential considered by \citet{Boylan-Kolchin2025} may still play a role by determining the mass fraction of GMCs in the ISM.

While we cannot draw definitive conclusions about its applicability at even higher redshifts, our findings still highlight the importance of incorporating the DM background when interpreting the formation and evolution of GMCs in the early Universe.

\section{Conclusion}\label{sec:conclusion}
In this work, we have investigated star formation efficiency on sub-galactic scales, with a particular focus on the properties of GMCs, using the \thesanzoom simulations. \thesanzoom is a suite of radiation-hydrodynamic simulations, which incorporates explicit models for star formation, multi-channel stellar feedback in a resolved multiphase ISM, and a non-equilibrium thermochemistry module that captures the complex interplay between radiation, gas, and dust. This provides a reliable framework for studying small-scale ISM structures in a cosmological context. We analyze the evolution of GMC properties in starbursts across redshift and under different model variants, and examine how these connect to galaxy-scale star formation. Our key findings are summarized below.

\begin{itemize}
    \item \textbf{Galaxy-scale SFE and its transition to smaller scales:} We examine the galaxy-scale SFE in \thesanzoom starbursts, defined as the ratio of the free-fall time averaged over all neutral ISM to the global depletion time. The resulting SFE follows a scaling relation of $\langle \epsilon_{\rm ff}^{\rm gal} \rangle \propto M_{\rm halo}^{1/3}(1+z)^{1/2}$ (Fig.~\ref{fig:glbSFE}), consistent with theoretical expectations for feedback-regulated models. Using a multi-scale SFE framework, we trace the evolution of the SFE across different gas phases, capturing the transition from efficient self-regulation to its eventual breakdown and the onset of numerical saturation (Fig.~\ref{fig:multiSFE}). Despite the variation in global SFE and slight redshift-dependent shifts in ISM density distribution, star formation in dense gas appears largely universal. For systems approaching the FFB regime, no clear transitional behavior is seen. 

    \item \textbf{GMC properties at high redshift:} We identify and characterize GMCs in high-redshift galaxies. Star formation in these systems is predominantly hosted by filamentary GMCs interacting with the surrounding ISM. The properties of GMCs, including mass function, size, and velocity dispersion, are remarkably universal across halo masses and redshifts. The GMC mass function follows a robust power-law slope of $\mathrm{d}N/\mathrm{d}M \propto M^{-2.5}$, with only mild dependence on redshift and halo mass at the high-mass end (Fig.~\ref{fig:gmc_mass}). At a given resolution, GMCs exhibit an approximately constant gas surface density, while the DM surface density increases with redshift, potentially leading to the emergence of DM–dominated clouds (Fig.~\ref{fig:gmc_sigma}). Interestingly, when the additional ESF is disabled, the gas surface density of GMCs decreases, likely because the reduced level of turbulence in the environment alters the balance between turbulent support and gravitational collapse.

    \item \textbf{GMC-scale SFE and its connection to global star formation:} The GMC-scale SFE shows little dependence on redshift or halo mass (Fig.~\ref{fig:gmc_sfe}). In fiducial runs, the median SFE per free-fall time is around 2-3\%, with a corresponding depletion time of approximately 100 Myr. When additional ESF is removed, the SFE increases to 6\%. We revisit the KS relation by decomposing the kpc-scale depletion time into three components: $\epsilon_{\rm ff}^{\rm GMC},\,f_{\rm GMC}$, and $t_{\rm ff}^{\rm GMC}$ (Fig.~\ref{fig:ks_connection}). We find that the GMC mass fraction $f_{\rm GMC}$ is the primary driver of variations in global depletion time, although modest fluctuations exist in the other two components. The GMC-scale SFE exhibits a non-monotonic dependence on ambient surface density that initially declines with increasing density due to stronger self-regulation, then rises again in high-density environments where feedback is increasingly suppressed. This suggests that self-regulation becomes less effective either when GMCs are sparse or when the surrounding surface density increases to some extreme values.

    \item \textbf{Implications for the feedback-free/failure scenarios:} While we do not find direct evidence for the feedback-free or feedback-failure scenarios, we observe several indirect trends that may be relevant. For example, although \thesanzoom galaxies do not fully reach the extreme conditions assumed in the FFB scenario, the simulation still naturally reproduces the shell-like configuration predicted by FFB, which may help sustain elevated SFE at both halo and galaxy scales (Fig.~\ref{fig:ffb}). We also identify a population of DM–dominated GMCs, many of which reside in high-SFE phases, potentially contributing to an elevated $\epsilon_{\rm int}^{\rm GMC}$ (Fig.~\ref{fig:gmc_dm_dom}). Based on current scaling relations, we predict that GMCs located within the central $0.05\,R_{\rm vir}$ regions at $z \gtrsim 10$ are likely to be dominated by DM. However, relying solely on DM gravitational confinement to fully suppress stellar feedback appears to be extremely challenging, even at $z \gtrsim 20$, although this assessment remains tentative and requires further investigation. 
\end{itemize}

In summary, we find that the bulk properties of GMCs are remarkably universal across different systems and highlight the potential impact of DM on GMC physics at extremely high redshifts. However, our current analysis lacks a dynamical description of individual GMCs, such as their lifetimes, merger rates, and integrated SFE. In addition, a systematic comparison across different numerical recipes remains unexplored. Therefore, the evolution of GMCs at high redshift warrants further investigation using simulations with higher output frequency, tracer particles, and a broader range of model variants.
Besides, an alternative and more observationally accessible approach to understanding small-scale star formation is to examine the final products of these GMCs; i.e. dense stellar clumps and potential progenitors of globular clusters, which we plan to investigate in future work.

\section*{Acknowledgements}
XS acknowledges the support of the NASA theory grant JWST-AR-04814. RK acknowledges support of the Natural Sciences and Engineering Research Council of Canada (NSERC) through a Discovery Grant and a Discovery Launch Supplement (funding reference numbers RGPIN-2024-06222 and DGECR-2024-00144) and York University's Global Research Excellence Initiative. LH acknowledges support from the Simons Foundation collaboration, ``Learning the Universe''.


\section*{Data availability}
All simulation data, including snapshots, group catalogs, and merger trees will be made publicly available in the near future. Data will be distributed via \url{https://www.thesan-project.com/thesan-zoom/}. Before the public data release, the data underlying this article can be shared upon reasonable request to the corresponding author(s).




\begin{thebibliography}{}
\makeatletter
\relax
\def\mn@urlcharsother{\let\do\@makeother \do\$\do\&\do\#\do\^\do\_\do\%\do\~}
\def\mn@doi{\begingroup\mn@urlcharsother \@ifnextchar [ {\mn@doi@}
  {\mn@doi@[]}}
\def\mn@doi@[#1]#2{\def\@tempa{#1}\ifx\@tempa\@empty \href
  {http://dx.doi.org/#2} {doi:#2}\else \href {http://dx.doi.org/#2} {#1}\fi
  \endgroup}
\def\mn@eprint#1#2{\mn@eprint@#1:#2::\@nil}
\def\mn@eprint@arXiv#1{\href {http://arxiv.org/abs/#1} {{\tt arXiv:#1}}}
\def\mn@eprint@dblp#1{\href {http://dblp.uni-trier.de/rec/bibtex/#1.xml}
  {dblp:#1}}
\def\mn@eprint@#1:#2:#3:#4\@nil{\def\@tempa {#1}\def\@tempb {#2}\def\@tempc
  {#3}\ifx \@tempc \@empty \let \@tempc \@tempb \let \@tempb \@tempa \fi \ifx
  \@tempb \@empty \def\@tempb {arXiv}\fi \@ifundefined
  {mn@eprint@\@tempb}{\@tempb:\@tempc}{\expandafter \expandafter \csname
  mn@eprint@\@tempb\endcsname \expandafter{\@tempc}}}

\bibitem[\protect\citeauthoryear{Agertz \& Kravtsov}{Agertz \&
  Kravtsov}{2015}]{Agertz2015}
Agertz O.,  Kravtsov A.~V.,  2015, \mn@doi [The Astrophysical Journal]
  {10.1088/0004-637x/804/1/18}, 804, 18

\bibitem[\protect\citeauthoryear{{Agertz} et~al.,}{{Agertz}
  et~al.}{2020}]{Agertz2020}
{Agertz} O.,  et~al., 2020, \mn@doi [\mnras] {10.1093/mnras/stz3053}, \href
  {https://ui.adsabs.harvard.edu/abs/2020MNRAS.491.1656A} {491, 1656}

\bibitem[\protect\citeauthoryear{{Agertz} et~al.,}{{Agertz}
  et~al.}{2021}]{Agertz2021}
{Agertz} O.,  et~al., 2021, \mn@doi [\mnras] {10.1093/mnras/stab322}, \href
  {https://ui.adsabs.harvard.edu/abs/2021MNRAS.503.5826A} {503, 5826}

\bibitem[\protect\citeauthoryear{Behroozi, Wechsler  \& Conroy}{Behroozi
  et~al.}{2013}]{Behroozi2013}
Behroozi P.~S.,  Wechsler R.~H.,   Conroy C.,  2013, \mn@doi [The Astrophysical
  Journal] {10.1088/0004-637X/770/1/57}, 770, 57

\bibitem[\protect\citeauthoryear{{Behroozi}, {Wechsler}, {Hearin}  \&
  {Conroy}}{{Behroozi} et~al.}{2019}]{Behroozi2019}
{Behroozi} P.,  {Wechsler} R.~H.,  {Hearin} A.~P.,   {Conroy} C.,  2019,
  \mn@doi [\mnras] {10.1093/mnras/stz1182}, \href
  {https://ui.adsabs.harvard.edu/abs/2019MNRAS.488.3143B} {488, 3143}

\bibitem[\protect\citeauthoryear{{Behroozi} et~al.,}{{Behroozi}
  et~al.}{2020}]{Behroozi2020}
{Behroozi} P.,  et~al., 2020, \mn@doi [\mnras] {10.1093/mnras/staa3164}, \href
  {https://ui.adsabs.harvard.edu/abs/2020MNRAS.499.5702B} {499, 5702}

\bibitem[\protect\citeauthoryear{{Benincasa} et~al.,}{{Benincasa}
  et~al.}{2020}]{Benincasa2020}
{Benincasa} S.~M.,  et~al., 2020, \mn@doi [\mnras] {10.1093/mnras/staa2116},
  \href {https://ui.adsabs.harvard.edu/abs/2020MNRAS.497.3993B} {497, 3993}

\bibitem[\protect\citeauthoryear{Bolatto, Leroy, Rosolowsky, Walter  \&
  Blitz}{Bolatto et~al.}{2008}]{Bolatto2008}
Bolatto A.~D.,  Leroy A.~K.,  Rosolowsky E.,  Walter F.,   Blitz L.,  2008,
  Resolved Properties of Extragalactic Giant Molecular Clouds (\mn@eprint
  {arXiv} {0810.2726}), \url {https://arxiv.org/abs/0810.2726}

\bibitem[\protect\citeauthoryear{Bower, Benson, Malbon, Helly, Frenk, Baugh,
  Cole  \& Lacey}{Bower et~al.}{2006}]{Bower2006}
Bower R.~G.,  Benson A.~J.,  Malbon R.,  Helly J.~C.,  Frenk C.~S.,  Baugh
  C.~M.,  Cole S.,   Lacey C.~G.,  2006, \mn@doi [Monthly Notices of the Royal
  Astronomical Society] {10.1111/j.1365-2966.2006.10519.x}, 370, 645

\bibitem[\protect\citeauthoryear{Boylan-Kolchin}{Boylan-Kolchin}{2025}]{Boylan-Kolchin2025}
Boylan-Kolchin M.,  2025, \mn@doi [Monthly Notices of the Royal Astronomical
  Society] {10.1093/mnras/staa0000}, 000, 1

\bibitem[\protect\citeauthoryear{{Buck}, {Pfrommer}, {Pakmor}, {Grand}  \&
  {Springel}}{{Buck} et~al.}{2020}]{Buck2020}
{Buck} T.,  {Pfrommer} C.,  {Pakmor} R.,  {Grand} R. J.~J.,   {Springel} V.,
  2020, \mn@doi [\mnras] {10.1093/mnras/staa1960}, \href
  {https://ui.adsabs.harvard.edu/abs/2020MNRAS.497.1712B} {497, 1712}

\bibitem[\protect\citeauthoryear{{Casey} et~al.,}{{Casey}
  et~al.}{2024}]{Casey2024}
{Casey} C.~M.,  et~al., 2024, \mn@doi [\apj] {10.3847/1538-4357/ad2075}, \href
  {https://ui.adsabs.harvard.edu/abs/2024ApJ...965...98C} {965, 98}

\bibitem[\protect\citeauthoryear{Ceverino, Dekel  \& Bournaud}{Ceverino
  et~al.}{2010}]{Ceverino2010}
Ceverino D.,  Dekel A.,   Bournaud F.,  2010, \mn@doi [Monthly Notices of the
  Royal Astronomical Society] {10.1111/j.1365-2966.2010.16433.x}, 404, 2151

\bibitem[\protect\citeauthoryear{{Ceverino}, {Glover}  \& {Klessen}}{{Ceverino}
  et~al.}{2017}]{Ceverino2017}
{Ceverino} D.,  {Glover} S. C.~O.,   {Klessen} R.~S.,  2017, \mn@doi [\mnras]
  {10.1093/mnras/stx1386}, \href
  {https://ui.adsabs.harvard.edu/abs/2017MNRAS.470.2791C} {470, 2791}

\bibitem[\protect\citeauthoryear{{Chabrier}}{{Chabrier}}{2003}]{Chabrier2003}
{Chabrier} G.,  2003, \mn@doi [\pasp] {10.1086/376392}, \href
  {https://ui.adsabs.harvard.edu/abs/2003PASP..115..763C} {115, 763}

\bibitem[\protect\citeauthoryear{Chan, Kereš, Wetzel, Hopkins,
  Faucher-Giguère, El-Badry  \& Hafen}{Chan et~al.}{2018}]{Chan2018}
Chan T.~L.,  Kereš D.,  Wetzel A.~R.,  Hopkins P.~F.,  Faucher-Giguère C.-A.,
   El-Badry K.,   Hafen Z.,  2018, \mn@doi [Monthly Notices of the Royal
  Astronomical Society] {10.1093/mnras/sty1024}, 478, 906

\bibitem[\protect\citeauthoryear{Chevance et~al.,}{Chevance
  et~al.}{2020}]{Chevance2020}
Chevance M.,  et~al., 2020, \mn@doi [Space Science Reviews]
  {10.1007/s11214-020-00674-x}, 216, 50

\bibitem[\protect\citeauthoryear{Commerçon, Hennebelle  \& Henning}{Commerçon
  et~al.}{2011}]{Commercon2011}
Commerçon B.,  Hennebelle P.,   Henning T.,  2011, \mn@doi [The Astrophysical
  Journal Letters] {10.1088/2041-8205/742/1/L9}, 742, L9

\bibitem[\protect\citeauthoryear{Cosentino et~al.,}{Cosentino
  et~al.}{2022}]{Cosentino_2022}
Cosentino G.,  et~al., 2022, \mn@doi [Monthly Notices of the Royal Astronomical
  Society] {10.1093/mnras/stac070}, 511, 953–963

\bibitem[\protect\citeauthoryear{Croton et~al.,}{Croton
  et~al.}{2006}]{Croton2006}
Croton D.~J.,  et~al., 2006, \mn@doi [Monthly Notices of the Royal Astronomical
  Society] {10.1111/j.1365-2966.2005.09675.x}, 365, 11

\bibitem[\protect\citeauthoryear{Cueto, Hutter, Dayal, Gottlöber, Heintz,
  Mason, Trebitsch  \& Yepes}{Cueto et~al.}{2024}]{Cueto2024}
Cueto E.~R.,  Hutter A.,  Dayal P.,  Gottlöber S.,  Heintz K.~E.,  Mason C.,
  Trebitsch M.,   Yepes G.,  2024, \mn@doi [Astronomy \&amp; Astrophysics]
  {10.1051/0004-6361/202349017}, 686, A138

\bibitem[\protect\citeauthoryear{Daddi, Elbaz, Walter  \& et al.}{Daddi
  et~al.}{2010}]{Daddi2010}
Daddi E.,  Elbaz D.,  Walter F.,   et al. 2010, \mn@doi [The Astrophysical
  Journal Letters] {10.1088/2041-8205/714/1/L118}, 714, L118

\bibitem[\protect\citeauthoryear{{Davis}, {Efstathiou}, {Frenk}  \&
  {White}}{{Davis} et~al.}{1985}]{Davis1985}
{Davis} M.,  {Efstathiou} G.,  {Frenk} C.~S.,   {White} S.~D.~M.,  1985,
  \mn@doi [\apj] {10.1086/163168}, \href
  {https://ui.adsabs.harvard.edu/abs/1985ApJ...292..371D} {292, 371}

\bibitem[\protect\citeauthoryear{Dekel et~al.,}{Dekel et~al.}{2009}]{Dekel2009}
Dekel A.,  et~al., 2009, \mn@doi [Nature] {10.1038/nature07648}, 457, 451

\bibitem[\protect\citeauthoryear{{Dekel}, {Sarkar}, {Birnboim}, {Mandelker}  \&
  {Li}}{{Dekel} et~al.}{2023}]{Dekel2023}
{Dekel} A.,  {Sarkar} K.~C.,  {Birnboim} Y.,  {Mandelker} N.,   {Li} Z.,  2023,
  \mn@doi [\mnras] {10.1093/mnras/stad1557}, \href
  {https://ui.adsabs.harvard.edu/abs/2023MNRAS.523.3201D} {523, 3201}

\bibitem[\protect\citeauthoryear{Dessauges-Zavadsky et~al.,}{Dessauges-Zavadsky
  et~al.}{2019}]{DessaugesZavadsky2019}
Dessauges-Zavadsky M.,  et~al., 2019, \mn@doi [Nature Astronomy]
  {10.1038/s41550-019-0874-0}, 3, 1115–1121

\bibitem[\protect\citeauthoryear{{Di Matteo}, {Springel}  \& {Hernquist}}{{Di
  Matteo} et~al.}{2005}]{DiMatteo2005}
{Di Matteo} T.,  {Springel} V.,   {Hernquist} L.,  2005, \mn@doi [\nat]
  {10.1038/nature03335}, \href
  {https://ui.adsabs.harvard.edu/abs/2005Natur.433..604D} {433, 604}

\bibitem[\protect\citeauthoryear{{Dobbs}, {Pringle}  \&
  {Duarte-Cabral}}{{Dobbs} et~al.}{2015}]{Dobbs2015}
{Dobbs} C.~L.,  {Pringle} J.~E.,   {Duarte-Cabral} A.,  2015, \mn@doi [\mnras]
  {10.1093/mnras/stu2319}, \href
  {https://ui.adsabs.harvard.edu/abs/2015MNRAS.446.3608D} {446, 3608}

\bibitem[\protect\citeauthoryear{Donnan, McLeod, McLure, Dunlop, Carnall,
  Cullen  \& Magee}{Donnan et~al.}{2023}]{Donnan2023}
Donnan C.~T.,  McLeod D.~J.,  McLure R.~J.,  Dunlop J.~S.,  Carnall A.~C.,
  Cullen F.,   Magee D.,  2023, \mn@doi [Monthly Notices of the Royal
  Astronomical Society] {10.1093/mnras/stad471}, 520, 4554

\bibitem[\protect\citeauthoryear{{Eldridge}, {Stanway}, {Xiao}, {McClelland},
  {Taylor}, {Ng}, {Greis}  \& {Bray}}{{Eldridge} et~al.}{2017}]{Eldridge2017}
{Eldridge} J.~J.,  {Stanway} E.~R.,  {Xiao} L.,  {McClelland} L.~A.~S.,
  {Taylor} G.,  {Ng} M.,  {Greis} S.~M.~L.,   {Bray} J.~C.,  2017, \mn@doi
  [\pasa] {10.1017/pasa.2017.51}, \href
  {https://ui.adsabs.harvard.edu/abs/2017PASA...34...58E} {34, e058}

\bibitem[\protect\citeauthoryear{Evans, Heiderman  \& Vutisalchavakul}{Evans
  et~al.}{2014}]{Evans2014}
Evans N.~J.,  Heiderman A.,   Vutisalchavakul N.,  2014, \mn@doi [The
  Astrophysical Journal] {10.1088/0004-637x/782/2/114}, 782, 114

\bibitem[\protect\citeauthoryear{Faucher-Giguère, Quataert  \&
  Hopkins}{Faucher-Giguère et~al.}{2013}]{Faucher2013}
Faucher-Giguère C.-A.,  Quataert E.,   Hopkins P.~F.,  2013, \mn@doi [Monthly
  Notices of the Royal Astronomical Society] {10.1093/mnras/stt829}, 433, 1970

\bibitem[\protect\citeauthoryear{Federrath}{Federrath}{2015}]{Federrath2015}
Federrath C.,  2015, \mn@doi [Monthly Notices of the Royal Astronomical
  Society] {10.1093/mnras/stv941}, 450, 4035

\bibitem[\protect\citeauthoryear{Federrath \& Klessen}{Federrath \&
  Klessen}{2012}]{Federrath2012}
Federrath C.,  Klessen R.~S.,  2012, \mn@doi [The Astrophysical Journal]
  {10.1088/0004-637X/761/2/156}, 761, 156

\bibitem[\protect\citeauthoryear{Federrath, Klessen  \& Schmidt}{Federrath
  et~al.}{2008}]{Federrath2008}
Federrath C.,  Klessen R.~S.,   Schmidt W.,  2008, \mn@doi [The Astrophysical
  Journal] {10.1086/595280}, 688, L79

\bibitem[\protect\citeauthoryear{{Feng}, {Di-Matteo}, {Croft}, {Bird},
  {Battaglia}  \& {Wilkins}}{{Feng} et~al.}{2016}]{Feng2016}
{Feng} Y.,  {Di-Matteo} T.,  {Croft} R.~A.,  {Bird} S.,  {Battaglia} N.,
  {Wilkins} S.,  2016, \mn@doi [\mnras] {10.1093/mnras/stv2484}, \href
  {https://ui.adsabs.harvard.edu/abs/2016MNRAS.455.2778F} {455, 2778}

\bibitem[\protect\citeauthoryear{Finkelstein et~al.,}{Finkelstein
  et~al.}{2022}]{Finkelstein2022}
Finkelstein S.~L.,  et~al., 2022, \mn@doi [The Astrophysical Journal Letters]
  {10.3847/2041-8213/ac966e}, 940, L55

\bibitem[\protect\citeauthoryear{Fotopoulou}{Fotopoulou}{2023}]{Fotopoulou2023}
Fotopoulou K.,  2023, PhD thesis, Ludwig-Maximilians-Universität München,
  \url {https://edoc.ub.uni-muenchen.de/31182/1/Fotopoulou_Konstantina.pdf}

\bibitem[\protect\citeauthoryear{{Garaldi}, {Kannan}, {Smith}, {Springel},
  {Pakmor}, {Vogelsberger}  \& {Hernquist}}{{Garaldi}
  et~al.}{2022}]{Garaldi2022}
{Garaldi} E.,  {Kannan} R.,  {Smith} A.,  {Springel} V.,  {Pakmor} R.,
  {Vogelsberger} M.,   {Hernquist} L.,  2022, \mn@doi [\mnras]
  {10.1093/mnras/stac257}, \href
  {https://ui.adsabs.harvard.edu/abs/2022MNRAS.512.4909G} {512, 4909}

\bibitem[\protect\citeauthoryear{{Garaldi} et~al.,}{{Garaldi}
  et~al.}{2024}]{Garaldi2024}
{Garaldi} E.,  et~al., 2024, \mn@doi [\mnras] {10.1093/mnras/stae839}, \href
  {https://ui.adsabs.harvard.edu/abs/2024MNRAS.530.3765G} {530, 3765}

\bibitem[\protect\citeauthoryear{Gelli, Mason  \& Hayward}{Gelli
  et~al.}{2024}]{Gelli2024}
Gelli V.,  Mason C.,   Hayward C.~C.,  2024, The impact of mass-dependent
  stochasticity at cosmic dawn (\mn@eprint {arXiv} {2405.13108}), \url
  {https://arxiv.org/abs/2405.13108}

\bibitem[\protect\citeauthoryear{Genzel, Tacconi, Gracia-Carpio  \& et
  al.}{Genzel et~al.}{2010}]{Genzel2010}
Genzel R.,  Tacconi L.~J.,  Gracia-Carpio J.,   et al. 2010, \mn@doi [Monthly
  Notices of the Royal Astronomical Society]
  {10.1111/j.1365-2966.2010.17078.x}, 407, 2091

\bibitem[\protect\citeauthoryear{Greif, Johnson, Bromm  \& Klessen}{Greif
  et~al.}{2007}]{Greif2007}
Greif T.~H.,  Johnson J.~L.,  Bromm V.,   Klessen R.~S.,  2007, \mn@doi [The
  Astrophysical Journal] {10.1086/522028}, 670, 1–14

\bibitem[\protect\citeauthoryear{{Grisdale}, {Agertz}, {Renaud}  \&
  {Romeo}}{{Grisdale} et~al.}{2018}]{Grisdale2018}
{Grisdale} K.,  {Agertz} O.,  {Renaud} F.,   {Romeo} A.~B.,  2018, \mn@doi
  [\mnras] {10.1093/mnras/sty1595}, \href
  {https://ui.adsabs.harvard.edu/abs/2018MNRAS.479.3167G} {479, 3167}

\bibitem[\protect\citeauthoryear{Grudić, Hopkins, Faucher-Giguère, Quataert,
  Murray  \& Kereš}{Grudić et~al.}{2018}]{Grudic2018}
Grudić M.~Y.,  Hopkins P.~F.,  Faucher-Giguère C.-A.,  Quataert E.,  Murray
  N.,   Kereš D.,  2018, \mn@doi [Monthly Notices of the Royal Astronomical
  Society] {10.1093/mnras/sty035}, 475, 3511–3528

\bibitem[\protect\citeauthoryear{Grudić, Hopkins, Lee, Murray,
  Faucher-Giguère  \& Johnson}{Grudić et~al.}{2019}]{Grudic2019}
Grudić M.~Y.,  Hopkins P.~F.,  Lee E.~J.,  Murray N.,  Faucher-Giguère C.-A.,
    Johnson L.~C.,  2019, \mn@doi [Monthly Notices of the Royal Astronomical
  Society] {10.1093/mnras/stz1758}, 488, 1501–1518

\bibitem[\protect\citeauthoryear{Guszejnov, Grudić, Offner, Boylan-Kolchin,
  Faucher-Gigère, Wetzel, Benincasa  \& Loebman}{Guszejnov
  et~al.}{2019}]{Guszejnov2019}
Guszejnov D.,  Grudić M.~Y.,  Offner S. S.~R.,  Boylan-Kolchin M.,
  Faucher-Gigère C.-A.,  Wetzel A.,  Benincasa S.~M.,   Loebman S.,  2019,
  \mn@doi [Monthly Notices of the Royal Astronomical Society]
  {10.1093/mnras/stz3527}, 492, 488

\bibitem[\protect\citeauthoryear{{Gutcke}, {Pakmor}, {Naab}  \&
  {Springel}}{{Gutcke} et~al.}{2021}]{Gutcke2021}
{Gutcke} T.~A.,  {Pakmor} R.,  {Naab} T.,   {Springel} V.,  2021, \mn@doi
  [\mnras] {10.1093/mnras/staa3875}, \href
  {https://ui.adsabs.harvard.edu/abs/2021MNRAS.501.5597G} {501, 5597}

\bibitem[\protect\citeauthoryear{{Harikane} et~al.,}{{Harikane}
  et~al.}{2023}]{Harikane2023}
{Harikane} Y.,  et~al., 2023, \mn@doi [\apjs] {10.3847/1538-4365/acaaa9}, \href
  {https://ui.adsabs.harvard.edu/abs/2023ApJS..265....5H} {265, 5}

\bibitem[\protect\citeauthoryear{Heiderman, Evans, Allen, Huard  \&
  Heyer}{Heiderman et~al.}{2010}]{Heiderman2010}
Heiderman A.,  Evans N.~J.,  Allen L.~E.,  Huard T.,   Heyer M.,  2010, \mn@doi
  [The Astrophysical Journal] {10.1088/0004-637x/723/2/1019}, 723, 1019–1037

\bibitem[\protect\citeauthoryear{Hennebelle \& Chabrier}{Hennebelle \&
  Chabrier}{2011}]{Hennebelle2011}
Hennebelle P.,  Chabrier G.,  2011, \mn@doi [The Astrophysical Journal]
  {10.1088/2041-8205/743/2/L29}, 743, L29

\bibitem[\protect\citeauthoryear{{Hopkins}, {Narayanan}  \& {Murray}}{{Hopkins}
  et~al.}{2013}]{Hopkins2013}
{Hopkins} P.~F.,  {Narayanan} D.,   {Murray} N.,  2013, \mn@doi [\mnras]
  {10.1093/mnras/stt723}, \href
  {https://ui.adsabs.harvard.edu/abs/2013MNRAS.432.2647H} {432, 2647}

\bibitem[\protect\citeauthoryear{Hopkins, Kereš, Oñorbe, Faucher-Giguère,
  Quataert, Murray  \& Bullock}{Hopkins et~al.}{2014}]{Hopkins2014}
Hopkins P.~F.,  Kereš D.,  Oñorbe J.,  Faucher-Giguère C.-A.,  Quataert E.,
  Murray N.,   Bullock J.~S.,  2014, \mn@doi [Monthly Notices of the Royal
  Astronomical Society] {10.1093/mnras/stu1738}, 445, 581

\bibitem[\protect\citeauthoryear{{Hopkins} et~al.,}{{Hopkins}
  et~al.}{2018}]{Hopkins2018feedback}
{Hopkins} P.~F.,  et~al., 2018, \mn@doi [\mnras] {10.1093/mnras/sty674}, \href
  {https://ui.adsabs.harvard.edu/abs/2018MNRAS.477.1578H} {477, 1578}

\bibitem[\protect\citeauthoryear{{Hopkins} et~al.,}{{Hopkins}
  et~al.}{2020}]{Hopkins2020-cr}
{Hopkins} P.~F.,  et~al., 2020, \mn@doi [\mnras] {10.1093/mnras/stz3321}, \href
  {https://ui.adsabs.harvard.edu/abs/2020MNRAS.492.3465H} {492, 3465}

\bibitem[\protect\citeauthoryear{{Hopkins} et~al.,}{{Hopkins}
  et~al.}{2023a}]{Hopkins2023fire3}
{Hopkins} P.~F.,  et~al., 2023a, \mn@doi [\mnras] {10.1093/mnras/stac3489},
  \href {https://ui.adsabs.harvard.edu/abs/2023MNRAS.519.3154H} {519, 3154}

\bibitem[\protect\citeauthoryear{Hopkins et~al.,}{Hopkins
  et~al.}{2023b}]{Hopkins2023}
Hopkins P.~F.,  et~al., 2023b, \mn@doi [Monthly Notices of the Royal
  Astronomical Society] {10.1093/mnras/stad1902}, 525, 2241–2286

\bibitem[\protect\citeauthoryear{{Inayoshi}, {Harikane}, {Inoue}, {Li}  \&
  {Ho}}{{Inayoshi} et~al.}{2022}]{Inayoshi2022}
{Inayoshi} K.,  {Harikane} Y.,  {Inoue} A.~K.,  {Li} W.,   {Ho} L.~C.,  2022,
  \mn@doi [\apjl] {10.3847/2041-8213/ac9310}, \href
  {https://ui.adsabs.harvard.edu/abs/2022ApJ...938L..10I} {938, L10}

\bibitem[\protect\citeauthoryear{{Jeffreson} \& {Kruijssen}}{{Jeffreson} \&
  {Kruijssen}}{2018}]{Jeffreson2018}
{Jeffreson} S. M.~R.,  {Kruijssen} J.~M.~D.,  2018, \mn@doi [\mnras]
  {10.1093/mnras/sty594}, \href
  {https://ui.adsabs.harvard.edu/abs/2018MNRAS.476.3688J} {476, 3688}

\bibitem[\protect\citeauthoryear{{Kannan}, {Vogelsberger}, {Marinacci},
  {McKinnon}, {Pakmor}  \& {Springel}}{{Kannan} et~al.}{2019}]{Kannan2019}
{Kannan} R.,  {Vogelsberger} M.,  {Marinacci} F.,  {McKinnon} R.,  {Pakmor} R.,
    {Springel} V.,  2019, \mn@doi [\mnras] {10.1093/mnras/stz287}, \href
  {https://ui.adsabs.harvard.edu/abs/2019MNRAS.485..117K} {485, 117}

\bibitem[\protect\citeauthoryear{{Kannan}, {Marinacci}, {Vogelsberger},
  {Sales}, {Torrey}, {Springel}  \& {Hernquist}}{{Kannan}
  et~al.}{2020}]{Kannan2020}
{Kannan} R.,  {Marinacci} F.,  {Vogelsberger} M.,  {Sales} L.~V.,  {Torrey} P.,
   {Springel} V.,   {Hernquist} L.,  2020, \mn@doi [\mnras]
  {10.1093/mnras/staa3249}, \href
  {https://ui.adsabs.harvard.edu/abs/2020MNRAS.499.5732K} {499, 5732}

\bibitem[\protect\citeauthoryear{{Kannan}, {Vogelsberger}, {Marinacci},
  {Sales}, {Torrey}  \& {Hernquist}}{{Kannan} et~al.}{2021}]{Kannan2021}
{Kannan} R.,  {Vogelsberger} M.,  {Marinacci} F.,  {Sales} L.~V.,  {Torrey} P.,
    {Hernquist} L.,  2021, \mn@doi [\mnras] {10.1093/mnras/stab416}, \href
  {https://ui.adsabs.harvard.edu/abs/2021MNRAS.503..336K} {503, 336}

\bibitem[\protect\citeauthoryear{{Kannan}, {Garaldi}, {Smith}, {Pakmor},
  {Springel}, {Vogelsberger}  \& {Hernquist}}{{Kannan}
  et~al.}{2022a}]{Kannan2022thesan}
{Kannan} R.,  {Garaldi} E.,  {Smith} A.,  {Pakmor} R.,  {Springel} V.,
  {Vogelsberger} M.,   {Hernquist} L.,  2022a, \mn@doi [\mnras]
  {10.1093/mnras/stab3710}, \href
  {https://ui.adsabs.harvard.edu/abs/2022MNRAS.511.4005K} {511, 4005}

\bibitem[\protect\citeauthoryear{{Kannan}, {Smith}, {Garaldi}, {Shen},
  {Vogelsberger}, {Pakmor}, {Springel}  \& {Hernquist}}{{Kannan}
  et~al.}{2022b}]{Kannan2022}
{Kannan} R.,  {Smith} A.,  {Garaldi} E.,  {Shen} X.,  {Vogelsberger} M.,
  {Pakmor} R.,  {Springel} V.,   {Hernquist} L.,  2022b, \mn@doi [\mnras]
  {10.1093/mnras/stac1557}, \href
  {https://ui.adsabs.harvard.edu/abs/2022MNRAS.514.3857K} {514, 3857}

\bibitem[\protect\citeauthoryear{{Kannan} et~al.,}{{Kannan}
  et~al.}{2023}]{Kannan2023}
{Kannan} R.,  et~al., 2023, \mn@doi [\mnras] {10.1093/mnras/stac3743}, \href
  {https://ui.adsabs.harvard.edu/abs/2023MNRAS.524.2594K} {524, 2594}

\bibitem[\protect\citeauthoryear{{Kannan} et~al.,}{{Kannan}
  et~al.}{2025}]{Kannan2025}
{Kannan} R.,  et~al., 2025, arXiv e-prints, \href
  {https://ui.adsabs.harvard.edu/abs/2025arXiv250220437K} {p. arXiv:2502.20437}

\bibitem[\protect\citeauthoryear{Kennicutt}{Kennicutt}{1998}]{Kennicutt1998}
Kennicutt R.~C.,  1998, \mn@doi [The Astrophysical Journal] {10.1086/305588},
  498, 541

\bibitem[\protect\citeauthoryear{{Kennicutt} \& {De Los Reyes}}{{Kennicutt} \&
  {De Los Reyes}}{2021}]{Kennicutt2021}
{Kennicutt} Robert~C. J.,  {De Los Reyes} M. A.~C.,  2021, \mn@doi [\apj]
  {10.3847/1538-4357/abd3a2}, \href
  {https://ui.adsabs.harvard.edu/abs/2021ApJ...908...61K} {908, 61}

\bibitem[\protect\citeauthoryear{{Kennicutt} \& {Evans}}{{Kennicutt} \&
  {Evans}}{2012}]{Kennicutt2012}
{Kennicutt} R.~C.,  {Evans} N.~J.,  2012, \mn@doi [\araa]
  {10.1146/annurev-astro-081811-125610}, \href
  {https://ui.adsabs.harvard.edu/abs/2012ARA&A..50..531K} {50, 531}

\bibitem[\protect\citeauthoryear{{Kimm}, {Haehnelt}, {Blaizot}, {Katz},
  {Michel-Dansac}, {Garel}, {Rosdahl}  \& {Teyssier}}{{Kimm}
  et~al.}{2018}]{Kimm2018}
{Kimm} T.,  {Haehnelt} M.,  {Blaizot} J.,  {Katz} H.,  {Michel-Dansac} L.,
  {Garel} T.,  {Rosdahl} J.,   {Teyssier} R.,  2018, \mn@doi [\mnras]
  {10.1093/mnras/sty126}, \href
  {https://ui.adsabs.harvard.edu/abs/2018MNRAS.475.4617K} {475, 4617}

\bibitem[\protect\citeauthoryear{{Klypin}, {Yepes}, {Gottl{\"o}ber}, {Prada}
  \& {He{\ss}}}{{Klypin} et~al.}{2016}]{Klypin2016}
{Klypin} A.,  {Yepes} G.,  {Gottl{\"o}ber} S.,  {Prada} F.,   {He{\ss}} S.,
  2016, \mn@doi [\mnras] {10.1093/mnras/stw248}, \href
  {https://ui.adsabs.harvard.edu/abs/2016MNRAS.457.4340K} {457, 4340}

\bibitem[\protect\citeauthoryear{Klypin et~al.,}{Klypin
  et~al.}{2021}]{Klypin2021}
Klypin A.,  et~al., 2021, \mn@doi [Monthly Notices of the Royal Astronomical
  Society] {10.1093/mnras/stab769}, 504, 769–781

\bibitem[\protect\citeauthoryear{Kravtsov \& Belokurov}{Kravtsov \&
  Belokurov}{2024}]{Kravtsov2024}
Kravtsov A.,  Belokurov V.,  2024, Stochastic star formation and the abundance
  of $z>10$ UV-bright galaxies (\mn@eprint {arXiv} {2405.04578}), \url
  {https://arxiv.org/abs/2405.04578}

\bibitem[\protect\citeauthoryear{Krumholz \& Federrath}{Krumholz \&
  Federrath}{2019}]{Krumholz2019}
Krumholz M.~R.,  Federrath C.,  2019, \mn@doi [Frontiers in Astronomy and Space
  Sciences] {10.3389/fspas.2019.00007}, 6, 7

\bibitem[\protect\citeauthoryear{Krumholz \& McKee}{Krumholz \&
  McKee}{2005}]{Krumholz2005}
Krumholz M.~R.,  McKee C.~F.,  2005, \mn@doi [The Astrophysical Journal]
  {10.1086/431734}, 630, 250

\bibitem[\protect\citeauthoryear{Krumholz \& Tan}{Krumholz \&
  Tan}{2007}]{Krumholz2007}
Krumholz M.~R.,  Tan J.~C.,  2007, \mn@doi [The Astrophysical Journal]
  {10.1086/509862}, 654, 304

\bibitem[\protect\citeauthoryear{Krumholz, Dekel  \& McKee}{Krumholz
  et~al.}{2012}]{Krumholz2012}
Krumholz M.~R.,  Dekel A.,   McKee C.~F.,  2012, \mn@doi [The Astrophysical
  Journal] {10.1088/0004-637X/745/1/69}, 745, 69

\bibitem[\protect\citeauthoryear{{Labb{\'e}} et~al.,}{{Labb{\'e}}
  et~al.}{2023}]{Labbe2023}
{Labb{\'e}} I.,  et~al., 2023, \mn@doi [\nat] {10.1038/s41586-023-05786-2},
  \href {https://ui.adsabs.harvard.edu/abs/2023Natur.616..266L} {616, 266}

\bibitem[\protect\citeauthoryear{{Larson}}{{Larson}}{1981}]{Larson1981}
{Larson} R.~B.,  1981, \mn@doi [\mnras] {10.1093/mnras/194.4.809}, \href
  {https://ui.adsabs.harvard.edu/abs/1981MNRAS.194..809L} {194, 809}

\bibitem[\protect\citeauthoryear{Lee, Miville-Deschênes  \& Murray}{Lee
  et~al.}{2016}]{Lee2016}
Lee E.~J.,  Miville-Deschênes M.-A.,   Murray N.~W.,  2016, \mn@doi [The
  Astrophysical Journal] {10.3847/1538-4357/833/2/229}, 833, 229

\bibitem[\protect\citeauthoryear{{Leitherer} et~al.,}{{Leitherer}
  et~al.}{1999}]{Leitherer1999}
{Leitherer} C.,  et~al., 1999, \mn@doi [\apjs] {10.1086/313233}, \href
  {https://ui.adsabs.harvard.edu/abs/1999ApJS..123....3L} {123, 3}

\bibitem[\protect\citeauthoryear{Leroy, Walter, Brinks, Bigiel, de Blok, Madore
   \& Thornley}{Leroy et~al.}{2008}]{Leroy2008}
Leroy A.~K.,  Walter F.,  Brinks E.,  Bigiel F.,  de Blok W. J.~G.,  Madore B.,
    Thornley M.~D.,  2008, \mn@doi [The Astronomical Journal]
  {10.1088/0004-6256/136/6/2782}, 136, 2782

\bibitem[\protect\citeauthoryear{{Li}, {Vogelsberger}, {Marinacci}  \&
  {Gnedin}}{{Li} et~al.}{2019}]{Li2019}
{Li} H.,  {Vogelsberger} M.,  {Marinacci} F.,   {Gnedin} O.~Y.,  2019, \mn@doi
  [\mnras] {10.1093/mnras/stz1271}, \href
  {https://ui.adsabs.harvard.edu/abs/2019MNRAS.487..364L} {487, 364}

\bibitem[\protect\citeauthoryear{{Li}, Feldmann, Hopkins, Faucher-Giguère, Ma,
  Orr, Hayward  \& Wetzel}{{Li} et~al.}{2020}]{Li2020}
{Li} H.,  Feldmann R.,  Hopkins P.~F.,  Faucher-Giguère C.-A.,  Ma X.,  Orr
  M.~E.,  Hayward C.~C.,   Wetzel A.,  2020, \mn@doi [Monthly Notices of the
  Royal Astronomical Society] {10.1093/mnras/staa3123}, 499, 5862

\bibitem[\protect\citeauthoryear{{Li}, {Vogelsberger}, {Bryan}, {Marinacci},
  {Sales}  \& {Torrey}}{{Li} et~al.}{2022}]{Li2022}
{Li} H.,  {Vogelsberger} M.,  {Bryan} G.~L.,  {Marinacci} F.,  {Sales} L.~V.,
  {Torrey} P.,  2022, \mn@doi [\mnras] {10.1093/mnras/stac1136}, \href
  {https://ui.adsabs.harvard.edu/abs/2022MNRAS.514..265L} {514, 265}

\bibitem[\protect\citeauthoryear{{Li}, {Dekel}, {Sarkar}, {Aung}, {Giavalisco},
  {Mandelker}  \& {Tacchella}}{{Li} et~al.}{2024}]{Li2024}
{Li} Z.,  {Dekel} A.,  {Sarkar} K.~C.,  {Aung} H.,  {Giavalisco} M.,
  {Mandelker} N.,   {Tacchella} S.,  2024, \mn@doi [\aap]
  {10.1051/0004-6361/202348727}, \href
  {https://ui.adsabs.harvard.edu/abs/2024A&A...690A.108L} {690, A108}

\bibitem[\protect\citeauthoryear{{Lovell}, {Vijayan}, {Thomas}, {Wilkins},
  {Barnes}, {Irodotou}  \& {Roper}}{{Lovell} et~al.}{2021}]{Lovell2021}
{Lovell} C.~C.,  {Vijayan} A.~P.,  {Thomas} P.~A.,  {Wilkins} S.~M.,  {Barnes}
  D.~J.,  {Irodotou} D.,   {Roper} W.,  2021, \mn@doi [\mnras]
  {10.1093/mnras/staa3360}, \href
  {https://ui.adsabs.harvard.edu/abs/2021MNRAS.500.2127L} {500, 2127}

\bibitem[\protect\citeauthoryear{{Ma} et~al.,}{{Ma} et~al.}{2018}]{Ma2018}
{Ma} X.,  et~al., 2018, \mn@doi [\mnras] {10.1093/mnras/sty1024}, \href
  {https://ui.adsabs.harvard.edu/abs/2018MNRAS.478.1694M} {478, 1694}

\bibitem[\protect\citeauthoryear{Mac~Low \& Klessen}{Mac~Low \&
  Klessen}{2004}]{Mac_Low2004}
Mac~Low M.-M.,  Klessen R.~S.,  2004, \mn@doi [Reviews of Modern Physics]
  {10.1103/revmodphys.76.125}, 76, 125–194

\bibitem[\protect\citeauthoryear{{Marinacci} \& {Vogelsberger}}{{Marinacci} \&
  {Vogelsberger}}{2016}]{Marinacci2016}
{Marinacci} F.,  {Vogelsberger} M.,  2016, \mn@doi [\mnras]
  {10.1093/mnrasl/slv176}, \href
  {https://ui.adsabs.harvard.edu/abs/2016MNRAS.456L..69M} {456, L69}

\bibitem[\protect\citeauthoryear{{Marinacci}, {Sales}, {Vogelsberger}, {Torrey}
   \& {Springel}}{{Marinacci} et~al.}{2019}]{Marinacci2019}
{Marinacci} F.,  {Sales} L.~V.,  {Vogelsberger} M.,  {Torrey} P.,   {Springel}
  V.,  2019, \mn@doi [\mnras] {10.1093/mnras/stz2391}, \href
  {https://ui.adsabs.harvard.edu/abs/2019MNRAS.489.4233M} {489, 4233}

\bibitem[\protect\citeauthoryear{Mason, Trenti  \& Treu}{Mason
  et~al.}{2023}]{Mason2023}
Mason C.~A.,  Trenti M.,   Treu T.,  2023, \mn@doi [Monthly Notices of the
  Royal Astronomical Society] {10.1093/mnras/stad035}, 521, 497

\bibitem[\protect\citeauthoryear{{McClymont} et~al.,}{{McClymont}
  et~al.}{2025a}]{McClymont2025-MSscatter}
{McClymont} W.,  et~al., 2025a, \mn@doi [arXiv e-prints]
  {10.48550/arXiv.2503.00106}, \href
  {https://ui.adsabs.harvard.edu/abs/2025arXiv250300106M} {p. arXiv:2503.00106}

\bibitem[\protect\citeauthoryear{{McClymont} et~al.,}{{McClymont}
  et~al.}{2025b}]{McClymont2025-size}
{McClymont} W.,  et~al., 2025b, \mn@doi [arXiv e-prints]
  {10.48550/arXiv.2503.04894}, \href
  {https://ui.adsabs.harvard.edu/abs/2025arXiv250304894M} {p. arXiv:2503.04894}

\bibitem[\protect\citeauthoryear{{McKee} \& {Ostriker}}{{McKee} \&
  {Ostriker}}{2007}]{McKee2007}
{McKee} C.~F.,  {Ostriker} E.~C.,  2007, \mn@doi [\araa]
  {10.1146/annurev.astro.45.051806.110602}, \href
  {https://ui.adsabs.harvard.edu/abs/2007ARA&A..45..565M} {45, 565}

\bibitem[\protect\citeauthoryear{{Menon}, {Lancaster}, {Burkhart},
  {Somerville}, {Dekel}  \& {Krumholz}}{{Menon} et~al.}{2024}]{Menon2024}
{Menon} S.~H.,  {Lancaster} L.,  {Burkhart} B.,  {Somerville} R.~S.,  {Dekel}
  A.,   {Krumholz} M.~R.,  2024, \mn@doi [\apjl] {10.3847/2041-8213/ad462d},
  \href {https://ui.adsabs.harvard.edu/abs/2024ApJ...967L..28M} {967, L28}

\bibitem[\protect\citeauthoryear{{Mirocha} \& {Furlanetto}}{{Mirocha} \&
  {Furlanetto}}{2023}]{Mirocha2023}
{Mirocha} J.,  {Furlanetto} S.~R.,  2023, \mn@doi [\mnras]
  {10.1093/mnras/stac3578}, \href
  {https://ui.adsabs.harvard.edu/abs/2023MNRAS.519..843M} {519, 843}

\bibitem[\protect\citeauthoryear{Moster, Naab  \& White}{Moster
  et~al.}{2013}]{Moster2013}
Moster B.~P.,  Naab T.,   White S. D.~M.,  2013, \mn@doi [Monthly Notices of
  the Royal Astronomical Society] {10.1093/mnras/sts261}, 428, 3121

\bibitem[\protect\citeauthoryear{Muratov, Kereš, Faucher-Giguère, Hopkins,
  Quataert  \& Murray}{Muratov et~al.}{2015}]{Muratov2015}
Muratov A.~L.,  Kereš D.,  Faucher-Giguère C.-A.,  Hopkins P.~F.,  Quataert
  E.,   Murray N.,  2015, \mn@doi [Monthly Notices of the Royal Astronomical
  Society] {10.1093/mnras/stv2126}, 454, 2691

\bibitem[\protect\citeauthoryear{Murray}{Murray}{2011}]{Murray2011}
Murray N.,  2011, \mn@doi [The Astrophysical Journal]
  {10.1088/0004-637X/729/2/133}, 729, 133

\bibitem[\protect\citeauthoryear{Murray, Quataert  \& Thompson}{Murray
  et~al.}{2010}]{Murray2010}
Murray N.,  Quataert E.,   Thompson T.~A.,  2010, \mn@doi [The Astrophysical
  Journal] {10.1088/0004-637X/709/1/191}, 709, 191

\bibitem[\protect\citeauthoryear{Nath, Vasiliev, Drozdov  \& Shchekinov}{Nath
  et~al.}{2023}]{Nath2023}
Nath B.~B.,  Vasiliev E.~O.,  Drozdov S.~A.,   Shchekinov Y.~A.,  2023, \mn@doi
  [Monthly Notices of the Royal Astronomical Society] {10.1093/mnras/stad505},
  521, 662–667

\bibitem[\protect\citeauthoryear{Navarro, Frenk  \& White}{Navarro
  et~al.}{1997}]{Navarro1997}
Navarro J.~F.,  Frenk C.~S.,   White S. D.~M.,  1997, \mn@doi [The
  Astrophysical Journal] {10.1086/304888}, 490, 493–508

\bibitem[\protect\citeauthoryear{{Nebrin}, {Smith}, {Lorinc}, {H{\"o}rnquist},
  {Larson}, {Mellema}  \& {Giri}}{{Nebrin} et~al.}{2025}]{Nebrin2025}
{Nebrin} O.,  {Smith} A.,  {Lorinc} K.,  {H{\"o}rnquist} J.,  {Larson}
  {\r{A}}.,  {Mellema} G.,   {Giri} S.~K.,  2025, \mn@doi [\mnras]
  {10.1093/mnras/staf038}, \href
  {https://ui.adsabs.harvard.edu/abs/2025MNRAS.tmp...39N} {}

\bibitem[\protect\citeauthoryear{Nelson et~al.,}{Nelson
  et~al.}{2017}]{Nelson2017}
Nelson D.,  et~al., 2017, \mn@doi [Monthly Notices of the Royal Astronomical
  Society] {10.1093/mnras/stx3040}, 475, 624–647

\bibitem[\protect\citeauthoryear{Nelson et~al.,}{Nelson
  et~al.}{2019}]{Nelson2019}
Nelson D.,  et~al., 2019, \mn@doi [Monthly Notices of the Royal Astronomical
  Society] {10.1093/mnras/stz2306}, 490, 3234–3261

\bibitem[\protect\citeauthoryear{Ni, Li, Vogelsberger, Sales, Marinacci  \&
  Torey}{Ni et~al.}{2025}]{Ni2025}
Ni Y.,  Li H.,  Vogelsberger M.,  Sales L.~V.,  Marinacci F.,   Torey P.,
  2025, The life cycle of giant molecular clouds in simulated Milky Way-mass
  galaxies (\mn@eprint {arXiv} {2502.12256}), \url
  {https://arxiv.org/abs/2502.12256}

\bibitem[\protect\citeauthoryear{{Nobels}, {Schaye}, {Schaller}, {Ploeckinger},
  {Chaikin}  \& {Richings}}{{Nobels} et~al.}{2023}]{Nobels2023}
{Nobels} F. S.~J.,  {Schaye} J.,  {Schaller} M.,  {Ploeckinger} S.,  {Chaikin}
  E.,   {Richings} A.~J.,  2023, \mn@doi [arXiv e-prints]
  {10.48550/arXiv.2309.13750}, \href
  {https://ui.adsabs.harvard.edu/abs/2023arXiv230913750N} {p. arXiv:2309.13750}

\bibitem[\protect\citeauthoryear{Ostriker \& Shetty}{Ostriker \&
  Shetty}{2011}]{Ostriker2011}
Ostriker E.~C.,  Shetty R.,  2011, \mn@doi [The Astrophysical Journal]
  {10.1088/0004-637X/731/1/41}, 731, 41

\bibitem[\protect\citeauthoryear{Ostriker, McKee  \& Leroy}{Ostriker
  et~al.}{2010}]{Ostriker2010}
Ostriker E.~C.,  McKee C.~F.,   Leroy A.~K.,  2010, \mn@doi [The Astrophysical
  Journal] {10.1088/0004-637X/721/1/975}, 721, 975

\bibitem[\protect\citeauthoryear{Padmanabhan \& Loeb}{Padmanabhan \&
  Loeb}{2023}]{Padmanabhan2023}
Padmanabhan H.,  Loeb A.,  2023, \mn@doi [The Astrophysical Journal Letters]
  {10.3847/2041-8213/acea7a}, 953, L4

\bibitem[\protect\citeauthoryear{Padoan \& Nordlund}{Padoan \&
  Nordlund}{2011}]{Padoan2011}
Padoan P.,  Nordlund A.,  2011, \mn@doi [The Astrophysical Journal]
  {10.1088/0004-637X/730/1/40}, 730, 40

\bibitem[\protect\citeauthoryear{Padoan, Nordlund  \& Jones}{Padoan
  et~al.}{1997}]{Padoan1997}
Padoan P.,  Nordlund A.,   Jones B. J.~T.,  1997, \mn@doi [Monthly Notices of
  the Royal Astronomical Society] {10.1093/mnras/288.1.145}, 288, 145

\bibitem[\protect\citeauthoryear{{Pakmor}, {Pfrommer}, {Simpson}  \&
  {Springel}}{{Pakmor} et~al.}{2016}]{Pakmor2016b}
{Pakmor} R.,  {Pfrommer} C.,  {Simpson} C.~M.,   {Springel} V.,  2016, \mn@doi
  [\apjl] {10.3847/2041-8205/824/2/L30}, \href
  {https://ui.adsabs.harvard.edu/abs/2016ApJ...824L..30P} {824, L30}

\bibitem[\protect\citeauthoryear{{Pallottini} et~al.,}{{Pallottini}
  et~al.}{2022}]{Pallottini2022}
{Pallottini} A.,  et~al., 2022, \mn@doi [\mnras] {10.1093/mnras/stac1281},
  \href {https://ui.adsabs.harvard.edu/abs/2022MNRAS.513.5621P} {513, 5621}

\bibitem[\protect\citeauthoryear{Parashari \& Laha}{Parashari \&
  Laha}{2023}]{Parashari2023}
Parashari P.,  Laha R.,  2023, \mn@doi [Monthly Notices of the Royal
  Astronomical Society: Letters] {10.1093/mnrasl/slad107}, 526, L63–L69

\bibitem[\protect\citeauthoryear{Pillepich et~al.,}{Pillepich
  et~al.}{2017}]{Pillepich2017}
Pillepich A.,  et~al., 2017, \mn@doi [Monthly Notices of the Royal Astronomical
  Society] {10.1093/mnras/stx2656}, 473, 4077

\bibitem[\protect\citeauthoryear{{Planck Collaboration} et~al.,}{{Planck
  Collaboration} et~al.}{2016}]{Planck2016}
{Planck Collaboration} et~al., 2016, \mn@doi [\aap]
  {10.1051/0004-6361/201525830}, \href
  {https://ui.adsabs.harvard.edu/abs/2016A&A...594A..13P} {594, A13}

\bibitem[\protect\citeauthoryear{{Reina-Campos}, {Keller}, {Kruijssen},
  {Gensior}, {Trujillo-Gomez}, {Jeffreson}, {Pfeffer}  \&
  {Sills}}{{Reina-Campos} et~al.}{2022}]{ReinaCampos2022}
{Reina-Campos} M.,  {Keller} B.~W.,  {Kruijssen} J.~M.~D.,  {Gensior} J.,
  {Trujillo-Gomez} S.,  {Jeffreson} S. M.~R.,  {Pfeffer} J.~L.,   {Sills} A.,
  2022, \mn@doi [\mnras] {10.1093/mnras/stac1934}, \href
  {https://ui.adsabs.harvard.edu/abs/2022MNRAS.517.3144R} {517, 3144}

\bibitem[\protect\citeauthoryear{Rice, Goodman, Bergin, Beaumont  \& Dame}{Rice
  et~al.}{2016}]{Rice2016}
Rice T.~S.,  Goodman A.~A.,  Bergin E.~A.,  Beaumont C.,   Dame T.~M.,  2016,
  \mn@doi [The Astrophysical Journal] {10.3847/0004-637x/822/1/52}, 822, 52

\bibitem[\protect\citeauthoryear{Robertson et~al.,}{Robertson
  et~al.}{2024}]{Robertson2024}
Robertson B.,  et~al., 2024, Earliest Galaxies in the JADES Origins Field:
  Luminosity Function and Cosmic Star-Formation Rate Density 300 Myr after the
  Big Bang (\mn@eprint {arXiv} {2312.10033}), \url
  {https://arxiv.org/abs/2312.10033}

\bibitem[\protect\citeauthoryear{Sabti, Muñoz  \& Kamionkowski}{Sabti
  et~al.}{2024}]{Sabti2024}
Sabti N.,  Muñoz J.~B.,   Kamionkowski M.,  2024, Insights from HST into
  Ultra-Massive Galaxies and Early-Universe Cosmology (\mn@eprint {arXiv}
  {2305.07049}), \url {https://arxiv.org/abs/2305.07049}

\bibitem[\protect\citeauthoryear{Schinnerer \& Leroy}{Schinnerer \&
  Leroy}{2024}]{Schinnerer2024}
Schinnerer E.,  Leroy A.~K.,  2024, Molecular Gas and the Star Formation
  Process on Cloud Scales in Nearby Galaxies (\mn@eprint {arXiv} {2403.19843}),
  \url {https://arxiv.org/abs/2403.19843}

\bibitem[\protect\citeauthoryear{Seifried, Banerjee, Klessen, Duffin  \&
  Pudritz}{Seifried et~al.}{2011}]{Seifried2011}
Seifried D.,  Banerjee R.,  Klessen R.~S.,  Duffin D.,   Pudritz R.~E.,  2011,
  \mn@doi [Monthly Notices of the Royal Astronomical Society]
  {10.1111/j.1365-2966.2011.19326.x}, 417, 1054

\bibitem[\protect\citeauthoryear{Semenov, Kravtsov  \& Gnedin}{Semenov
  et~al.}{2018}]{Semenov2018}
Semenov V.~A.,  Kravtsov A.~V.,   Gnedin N.~Y.,  2018, \mn@doi [The
  Astrophysical Journal] {10.3847/1538-4357/aac6eb}, 861, 4

\bibitem[\protect\citeauthoryear{{Semenov}, {Conroy}, {Smith}, {Puchwein}  \&
  {Hernquist}}{{Semenov} et~al.}{2024a}]{Semenov2024-disk}
{Semenov} V.~A.,  {Conroy} C.,  {Smith} A.,  {Puchwein} E.,   {Hernquist} L.,
  2024a, \mn@doi [arXiv e-prints] {10.48550/arXiv.2409.18173}, \href
  {https://ui.adsabs.harvard.edu/abs/2024arXiv240918173S} {p. arXiv:2409.18173}

\bibitem[\protect\citeauthoryear{{Semenov}, {Conroy}  \& {Hernquist}}{{Semenov}
  et~al.}{2024b}]{Semenov2024-turbulentsf}
{Semenov} V.~A.,  {Conroy} C.,   {Hernquist} L.,  2024b, \mn@doi [arXiv
  e-prints] {10.48550/arXiv.2410.09205}, \href
  {https://ui.adsabs.harvard.edu/abs/2024arXiv241009205S} {p. arXiv:2410.09205}

\bibitem[\protect\citeauthoryear{{Sharda}, {Federrath}, {da Cunha}, {Swinbank}
  \& {Dye}}{{Sharda} et~al.}{2018}]{Sharda2018}
{Sharda} P.,  {Federrath} C.,  {da Cunha} E.,  {Swinbank} A.~M.,   {Dye} S.,
  2018, \mn@doi [\mnras] {10.1093/mnras/sty886}, \href
  {https://ui.adsabs.harvard.edu/abs/2018MNRAS.477.4380S} {477, 4380}

\bibitem[\protect\citeauthoryear{Shen et~al.,}{Shen et~al.}{2020}]{Shen2020}
Shen X.,  et~al., 2020, \mn@doi [Monthly Notices of the Royal Astronomical
  Society] {10.1093/mnras/staa1423}, 495, 4747–4768

\bibitem[\protect\citeauthoryear{Shen, Vogelsberger, Nelson, Tacchella,
  Hernquist, Springel, Marinacci  \& Torrey}{Shen et~al.}{2022}]{Shen2022}
Shen X.,  Vogelsberger M.,  Nelson D.,  Tacchella S.,  Hernquist L.,  Springel
  V.,  Marinacci F.,   Torrey P.,  2022, \mn@doi [Monthly Notices of the Royal
  Astronomical Society] {10.1093/mnras/stab3794}, 510, 5560–5578

\bibitem[\protect\citeauthoryear{Shen, Vogelsberger, Boylan-Kolchin, Tacchella
  \& Kannan}{Shen et~al.}{2023}]{Shen2023}
Shen X.,  Vogelsberger M.,  Boylan-Kolchin M.,  Tacchella S.,   Kannan R.,
  2023, The impact of UV variability on the abundance of bright galaxies at $z
  \geq 9$ (\mn@eprint {arXiv} {2305.05679}), \url
  {https://arxiv.org/abs/2305.05679}

\bibitem[\protect\citeauthoryear{Shen et~al.,}{Shen
  et~al.}{2024a}]{Shen2024size}
Shen X.,  et~al., 2024a, The THESAN project: galaxy sizes during the epoch of
  reionization (\mn@eprint {arXiv} {2402.08717}), \url
  {https://arxiv.org/abs/2402.08717}

\bibitem[\protect\citeauthoryear{{Shen}, {Vogelsberger}, {Boylan-Kolchin},
  {Tacchella}  \& {Naidu}}{{Shen} et~al.}{2024b}]{Shen2024-ede}
{Shen} X.,  {Vogelsberger} M.,  {Boylan-Kolchin} M.,  {Tacchella} S.,   {Naidu}
  R.~P.,  2024b, \mn@doi [\mnras] {10.1093/mnras/stae1932}, \href
  {https://ui.adsabs.harvard.edu/abs/2024MNRAS.533.3923S} {533, 3923}

\bibitem[\protect\citeauthoryear{Shen et~al.,}{Shen et~al.}{2025}]{Shen2025}
Shen X.,  et~al., 2025, The THESAN-ZOOM project: Star-formation efficiencies in
  high-redshift galaxies (\mn@eprint {arXiv} {2503.01949}), \url
  {https://arxiv.org/abs/2503.01949}

\bibitem[\protect\citeauthoryear{Shi et~al.,}{Shi et~al.}{2020}]{Shi2020}
Shi Y.,  et~al., 2020, arXiv preprint arXiv:2002.10209

\bibitem[\protect\citeauthoryear{Shirley}{Shirley}{2015}]{Shirley2015}
Shirley Y.~L.,  2015, \mn@doi [Publications of the Astronomical Society of the
  Pacific] {10.1086/680342}, 127, 299–310

\bibitem[\protect\citeauthoryear{{Smith}, {Bromm}  \& {Loeb}}{{Smith}
  et~al.}{2017}]{Smith2017}
{Smith} A.,  {Bromm} V.,   {Loeb} A.,  2017, \mn@doi [\mnras]
  {10.1093/mnras/stw2591}, \href
  {https://ui.adsabs.harvard.edu/abs/2017MNRAS.464.2963S} {464, 2963}

\bibitem[\protect\citeauthoryear{{Smith}, {Kannan}, {Garaldi}, {Vogelsberger},
  {Pakmor}, {Springel}  \& {Hernquist}}{{Smith} et~al.}{2022}]{Smith2022}
{Smith} A.,  {Kannan} R.,  {Garaldi} E.,  {Vogelsberger} M.,  {Pakmor} R.,
  {Springel} V.,   {Hernquist} L.,  2022, \mn@doi [\mnras]
  {10.1093/mnras/stac713}, \href
  {https://ui.adsabs.harvard.edu/abs/2022MNRAS.512.3243S} {512, 3243}

\bibitem[\protect\citeauthoryear{{Solomon}, {Rivolo}, {Barrett}  \&
  {Yahil}}{{Solomon} et~al.}{1987}]{Solomon1984}
{Solomon} P.~M.,  {Rivolo} A.~R.,  {Barrett} J.,   {Yahil} A.,  1987, \mn@doi
  [Astrophysical Journal] {10.1086/165493}, \href
  {https://ui.adsabs.harvard.edu/abs/1987ApJ...319..730S} {319, 730}

\bibitem[\protect\citeauthoryear{Somerville \& Davé}{Somerville \&
  Davé}{2015}]{Somerville2015}
Somerville R.~S.,  Davé R.,  2015, \mn@doi [Annual Review of Astronomy and
  Astrophysics] {10.1146/annurev-astro-082812-140951}, 53, 51

\bibitem[\protect\citeauthoryear{{Springel} \& {Hernquist}}{{Springel} \&
  {Hernquist}}{2003}]{Springel2003}
{Springel} V.,  {Hernquist} L.,  2003, \mn@doi [\mnras]
  {10.1046/j.1365-8711.2003.06206.x}, \href
  {https://ui.adsabs.harvard.edu/abs/2003MNRAS.339..289S} {339, 289}

\bibitem[\protect\citeauthoryear{{Springel} et~al.,}{{Springel}
  et~al.}{2005}]{Springel2005}
{Springel} V.,  et~al., 2005, \mn@doi [\nat] {10.1038/nature03597}, \href
  {https://ui.adsabs.harvard.edu/abs/2005Natur.435..629S} {435, 629}

\bibitem[\protect\citeauthoryear{Springel et~al.,}{Springel
  et~al.}{2017}]{Springel2017}
Springel V.,  et~al., 2017, \mn@doi [Monthly Notices of the Royal Astronomical
  Society] {10.1093/mnras/stx3304}, 475, 676–698

\bibitem[\protect\citeauthoryear{{Springel}, {Pakmor}, {Zier}  \&
  {Reinecke}}{{Springel} et~al.}{2021}]{Springel2021}
{Springel} V.,  {Pakmor} R.,  {Zier} O.,   {Reinecke} M.,  2021, \mn@doi
  [\mnras] {10.1093/mnras/stab1855}, \href
  {https://ui.adsabs.harvard.edu/abs/2021MNRAS.506.2871S} {506, 2871}

\bibitem[\protect\citeauthoryear{{Sun} et~al.,}{{Sun} et~al.}{2023}]{Sun2023}
{Sun} J.,  et~al., 2023, \mn@doi [\apjl] {10.3847/2041-8213/acbd9c}, \href
  {https://ui.adsabs.harvard.edu/abs/2023ApJ...945L..19S} {945, L19}

\bibitem[\protect\citeauthoryear{Swinbank et~al.,}{Swinbank
  et~al.}{2015}]{Swinbank2015}
Swinbank A.~M.,  et~al., 2015, \mn@doi [The Astrophysical Journal]
  {10.1088/2041-8205/806/1/l17}, 806, L17

\bibitem[\protect\citeauthoryear{{Tacchella}, {Bose}, {Conroy}, {Eisenstein}
  \& {Johnson}}{{Tacchella} et~al.}{2018}]{Tacchella2018}
{Tacchella} S.,  {Bose} S.,  {Conroy} C.,  {Eisenstein} D.~J.,   {Johnson}
  B.~D.,  2018, \mn@doi [\apj] {10.3847/1538-4357/aae8e0}, \href
  {https://ui.adsabs.harvard.edu/abs/2018ApJ...868...92T} {868, 92}

\bibitem[\protect\citeauthoryear{Tacchella, Forbes  \& Caplar}{Tacchella
  et~al.}{2020}]{Tacchella2020}
Tacchella S.,  Forbes J.~C.,   Caplar N.,  2020, \mn@doi [Monthly Notices of
  the Royal Astronomical Society] {10.1093/mnras/staa1838}, 497, 698–725

\bibitem[\protect\citeauthoryear{Tacconi et~al.,}{Tacconi
  et~al.}{2010}]{Tacconi2010}
Tacconi L.~J.,  et~al., 2010, \mn@doi [Nature] {10.1038/nature08773}, 463,
  781–784

\bibitem[\protect\citeauthoryear{{Tasker} \& {Tan}}{{Tasker} \&
  {Tan}}{2009}]{Tasker2009}
{Tasker} E.~J.,  {Tan} J.~C.,  2009, \mn@doi [\apj]
  {10.1088/0004-637X/700/1/358}, \href
  {https://ui.adsabs.harvard.edu/abs/2009ApJ...700..358T} {700, 358}

\bibitem[\protect\citeauthoryear{Trinca, Schneider, Valiante, Graziani,
  Ferrotti, Omukai  \& Chon}{Trinca et~al.}{2024}]{Trinca2024}
Trinca A.,  Schneider R.,  Valiante R.,  Graziani L.,  Ferrotti A.,  Omukai K.,
    Chon S.,  2024, Exploring the nature of UV-bright $z \gtrsim 10$ galaxies
  detected by JWST: star formation, black hole accretion, or a non-universal
  IMF? (\mn@eprint {arXiv} {2305.04944}), \url
  {https://arxiv.org/abs/2305.04944}

\bibitem[\protect\citeauthoryear{{Truelove}, {Klein}, {McKee}, {Holliman},
  {Howell}  \& {Greenough}}{{Truelove} et~al.}{1997}]{Truelove1997}
{Truelove} J.~K.,  {Klein} R.~I.,  {McKee} C.~F.,  {Holliman} John~H. I.,
  {Howell} L.~H.,   {Greenough} J.~A.,  1997, \mn@doi [\apjl] {10.1086/310975},
  \href {https://ui.adsabs.harvard.edu/abs/1997ApJ...489L.179T} {489, L179}

\bibitem[\protect\citeauthoryear{{Vogelsberger}, Genel, Sijacki, Torrey,
  Springel  \& Hernquist}{{Vogelsberger} et~al.}{2013}]{Vogelsberger2013}
{Vogelsberger} M.,  Genel S.,  Sijacki D.,  Torrey P.,  Springel V.,
  Hernquist L.,  2013, \mn@doi [Monthly Notices of the Royal Astronomical
  Society] {10.1093/mnras/stt1789}, 436, 3031–3067

\bibitem[\protect\citeauthoryear{{Vogelsberger} et~al.,}{{Vogelsberger}
  et~al.}{2014a}]{Vogelsberger2014}
{Vogelsberger} M.,  et~al., 2014a, \mn@doi [\mnras] {10.1093/mnras/stu1536},
  \href {https://ui.adsabs.harvard.edu/abs/2014MNRAS.444.1518V} {444, 1518}

\bibitem[\protect\citeauthoryear{{Vogelsberger} et~al.,}{{Vogelsberger}
  et~al.}{2014b}]{Vogelsberger2014nature}
{Vogelsberger} M.,  et~al., 2014b, \mn@doi [Nature] {10.1038/nature13316}, 509

\bibitem[\protect\citeauthoryear{{Vogelsberger}, {Marinacci}, {Torrey}  \&
  {Puchwein}}{{Vogelsberger} et~al.}{2020a}]{Vogelsberger2020}
{Vogelsberger} M.,  {Marinacci} F.,  {Torrey} P.,   {Puchwein} E.,  2020a,
  \mn@doi [Nature Reviews Physics] {10.1038/s42254-019-0127-2}, \href
  {https://ui.adsabs.harvard.edu/abs/2020NatRP...2...42V} {2, 42}

\bibitem[\protect\citeauthoryear{{Vogelsberger} et~al.,}{{Vogelsberger}
  et~al.}{2020b}]{Vogelsberger2020highz}
{Vogelsberger} M.,  et~al., 2020b, \mn@doi [Monthly Notices of the Royal
  Astronomical Society] {10.1093/mnras/staa137}, 492, 5167–5201

\bibitem[\protect\citeauthoryear{Vutisalchavakul, Evans~II  \&
  Heyer}{Vutisalchavakul et~al.}{2016}]{Vutisalchavakul2016}
Vutisalchavakul N.,  Evans~II N.~J.,   Heyer M.,  2016, \mn@doi [The
  Astrophysical Journal] {10.3847/0004-637x/831/1/73}, 831, 73

\bibitem[\protect\citeauthoryear{Vázquez-Semadeni}{Vázquez-Semadeni}{1994}]{Vazquez1994}
Vázquez-Semadeni E.,  1994, \mn@doi [The Astrophysical Journal]
  {10.1086/173843}, 423, 681

\bibitem[\protect\citeauthoryear{{Wang}, {Dutton}, {Stinson}, {Macci{\`o}},
  {Penzo}, {Kang}, {Keller}  \& {Wadsley}}{{Wang} et~al.}{2015}]{Wang2015}
{Wang} L.,  {Dutton} A.~A.,  {Stinson} G.~S.,  {Macci{\`o}} A.~V.,  {Penzo} C.,
   {Kang} X.,  {Keller} B.~W.,   {Wadsley} J.,  2015, \mn@doi [\mnras]
  {10.1093/mnras/stv1937}, \href
  {https://ui.adsabs.harvard.edu/abs/2015MNRAS.454...83W} {454, 83}

\bibitem[\protect\citeauthoryear{Wang et~al.,}{Wang et~al.}{2025}]{Wang2025}
Wang T.,  et~al., 2025, MAssive galaxies aCRoss cOSmic time revealed by
  JWST/MIRI (MACROSS): The true number density of massive galaxies in the early
  Universe (\mn@eprint {arXiv} {2403.02399}), \url
  {https://arxiv.org/abs/2403.02399}

\bibitem[\protect\citeauthoryear{Wechsler \& Tinker}{Wechsler \&
  Tinker}{2018}]{Wechsler2018}
Wechsler R.~H.,  Tinker J.~L.,  2018, \mn@doi [Annual Review of Astronomy and
  Astrophysics] {10.1146/annurev-astro-081817-051756}, 56, 435

\bibitem[\protect\citeauthoryear{Weinberger et~al.,}{Weinberger
  et~al.}{2017}]{Weinberger2017}
Weinberger R.,  et~al., 2017, \mn@doi [Monthly Notices of the Royal
  Astronomical Society] {10.1093/mnras/stw2944}, 465, 3291

\bibitem[\protect\citeauthoryear{{Weinberger}, {Springel}  \&
  {Pakmor}}{{Weinberger} et~al.}{2020}]{Weinberger2020arepo}
{Weinberger} R.,  {Springel} V.,   {Pakmor} R.,  2020, \mn@doi [\apjs]
  {10.3847/1538-4365/ab908c}, \href
  {https://ui.adsabs.harvard.edu/abs/2020ApJS..248...32W} {248, 32}

\bibitem[\protect\citeauthoryear{{Wibking} \& {Krumholz}}{{Wibking} \&
  {Krumholz}}{2023}]{Wibking2023}
{Wibking} B.~D.,  {Krumholz} M.~R.,  2023, \mn@doi [\mnras]
  {10.1093/mnras/stac2648}, \href
  {https://ui.adsabs.harvard.edu/abs/2023MNRAS.521.5972W} {521, 5972}

\bibitem[\protect\citeauthoryear{{Xiao} et~al.,}{{Xiao}
  et~al.}{2024}]{Xiao2024}
{Xiao} M.,  et~al., 2024, \mn@doi [\nat] {10.1038/s41586-024-08094-5}, \href
  {https://ui.adsabs.harvard.edu/abs/2024Natur.635..311X} {635, 311}

\bibitem[\protect\citeauthoryear{{Yung}, {Somerville}, {Finkelstein}, {Wilkins}
   \& {Gardner}}{{Yung} et~al.}{2024}]{Yung2024}
{Yung} L.~Y.~A.,  {Somerville} R.~S.,  {Finkelstein} S.~L.,  {Wilkins} S.~M.,
  {Gardner} J.~P.,  2024, \mn@doi [\mnras] {10.1093/mnras/stad3484}, \href
  {https://ui.adsabs.harvard.edu/abs/2024MNRAS.527.5929Y} {527, 5929}

\bibitem[\protect\citeauthoryear{{Zhao}, {Pudritz}, {Pillsworth}, {Robinson}
  \& {Wadsley}}{{Zhao} et~al.}{2024}]{Zhao2024}
{Zhao} B.,  {Pudritz} R.~E.,  {Pillsworth} R.,  {Robinson} H.,   {Wadsley} J.,
  2024, \mn@doi [\apj] {10.3847/1538-4357/ad67e2}, \href
  {https://ui.adsabs.harvard.edu/abs/2024ApJ...974..240Z} {974, 240}

\bibitem[\protect\citeauthoryear{{Zier}, {Kannan}, {Smith}, {Vogelsberger}  \&
  {Verbeek}}{{Zier} et~al.}{2024}]{Zier2024-gpu}
{Zier} O.,  {Kannan} R.,  {Smith} A.,  {Vogelsberger} M.,   {Verbeek} E.,
  2024, \mn@doi [\mnras] {10.1093/mnras/stae1837}, \href
  {https://ui.adsabs.harvard.edu/abs/2024MNRAS.533..268Z} {533, 268}

\bibitem[\protect\citeauthoryear{{Zier} et~al.,}{{Zier}
  et~al.}{2025a}]{Zier2025-reion}
{Zier} O.,  et~al., 2025a, \mn@doi [arXiv e-prints]
  {10.48550/arXiv.2503.02927}, \href
  {https://ui.adsabs.harvard.edu/abs/2025arXiv250302927Z} {p. arXiv:2503.02927}

\bibitem[\protect\citeauthoryear{{Zier} et~al.,}{{Zier}
  et~al.}{2025b}]{Zier2025-pop3}
{Zier} O.,  et~al., 2025b, \mn@doi [arXiv e-prints]
  {10.48550/arXiv.2503.03806}, \href
  {https://ui.adsabs.harvard.edu/abs/2025arXiv250303806Z} {p. arXiv:2503.03806}

\makeatother
\end{thebibliography}



\appendix

\section{A temperature-dependent definition of multi-scale SFE}\label{apdx:multiSFE}

\begin{figure*}
    \centering
    \includegraphics[width=1\linewidth]{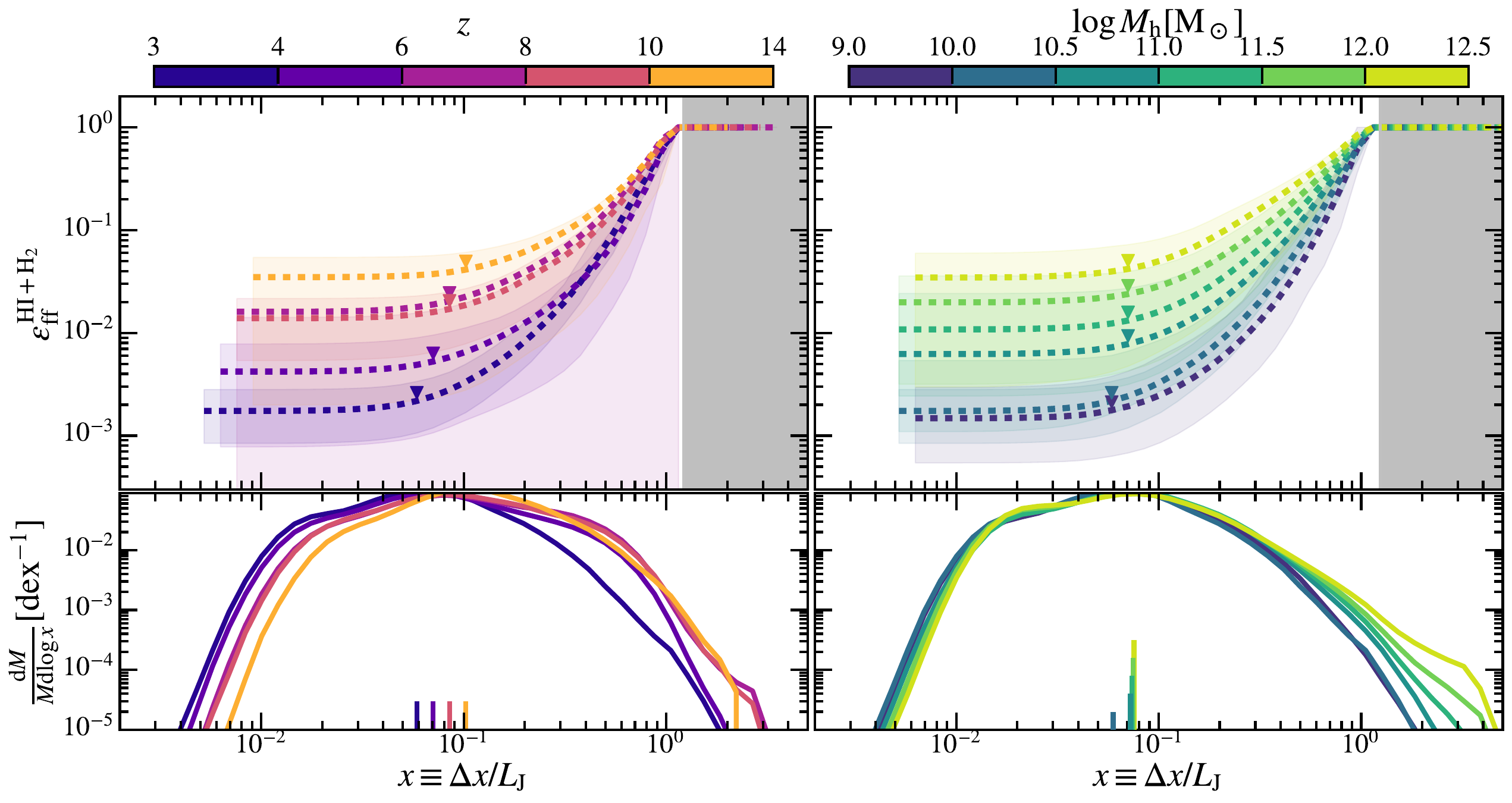}
    \caption{Similar to Fig.~\ref{fig:multiSFE}, but manifested in the SFE versus $x\equiv\Delta x/L_{\rm J}$. The gray shaded region on the right marks the regime where the Jeans length is unresolved ($\Delta x > L_{\rm J}$), and the corresponding SFEs have been numerically set to be 100\%. In the lower left panel, the short lines indicating the median $x$ values have been slightly adjusted in length to avoid visual overlap.}
    \label{fig:multiSFE_lj}
\end{figure*}

In this section, we provide a temperature-dependent definition of multi-scale SFE via,
\begin{equation}
    \epsilon_{\rm ff}(x_{\rm crit}) \equiv \frac{{{\rm SFR}_{{x>x_{\rm crit}}}}\times \langle t_{\rm ff}\rangle_{x>x_{\rm crit}}}{M_{{\rm gas},\,x>x_{\rm crit}}} \, ,
    \label{eq:multiSFE3}
\end{equation}
where $x\equiv\Delta x/L_{\rm J}$ is the distance to the Jeans unstable criteria. A larger $x$ value corresponds approximately to cooler and denser gas. In Fig.~\ref{fig:multiSFE_lj}, we present results computed using the same sample as in Fig.~\ref{fig:multiSFE}, where variations in the ISM density are effectively removed under this definition. Similarly, we find no clear dependence on redshift or halo mass in the underlying trends, with differences manifesting only in the global SFE. Self-regulation appears to weaken once the enclosed gas mass falls below approximately 50\%, and the SFE numerically approaches 100\% at $x = 1$.

\section{Impact of different parameter sets on GMC Properties}\label{apdx:parameters}
In this work, we use \textsc{CloudPhinder} to identify GMCs. The algorithm starts from local density peaks and searches for the largest gravitationally bound gaseous structures (quantified by $\alpha_{\rm max}$) above a given density threshold (quantified by $n_{\rm min}$). In the fiducial run, we adopt $n_{\rm min} = 100\,\mathrm{cm}^{-3}$ to facilitate comparisons with observations, which typically infer gas densities indirectly through emission maps of tracer molecules that have critical densities above $100\,\mathrm{cm}^{-3}$~\citep[e.g.][]{Shirley2015,Schinnerer2024}.
We have also tested different values of the gravitational boundedness threshold $\alpha_{\rm max}$ (e.g., 2, 5, and 10), and found that the distribution of identified GMCs consistently exhibits a local peak at the cutoff, reflecting the artificial influence of the imposed threshold. To mitigate this artifact, we adopt a more permissive threshold of $\alpha_{\rm max} = 20$, at which point the sharp peak becomes a weak tail rather than a prominent feature. Despite the large maximum $\alpha$ allowed, the majority of identified GMCs still have $\alpha < 10$, with a median value around $\sim 4-5$ (see Fig.~\ref{fig:gmc_vp_res}).

However, it is also informative to compare GMC properties obtained using different parameter sets. In Table~\ref{tab:pars}, we present the median properties of GMCs identified under four representative parameter choices, specifically varying the gravitational boundedness threshold ($\alpha_{\rm max} = 10$ and $20$) and the minimum density threshold ($n_{\rm min} = 50$ and $100\, \mathrm{cm}^{-3}$). The listed properties include GMC mass, size, surface density, free-fall timescale, and instantaneous SFE, as well as the fitted slopes of the mass function and Larson’s relation.

In general, adopting a lower density threshold leads to a shallower slope in the mass function but a lower median GMC mass, indicating the inclusion of a larger population of low-mass, loosely bound clouds, while the high-mass tail becomes relatively more prominent due to the relaxed selection. These GMCs also exhibit lower surface and volume densities. However, the cancellation between gas mass and free-fall time results in no significant change in the instantaneous SFE.
Varying the virial parameter threshold, on the other hand, primarily affects the slope of the Larson relation. A smaller $\alpha_{\rm max}$ value tends to exclude high-velocity-dispersion clouds, thereby lowering the overall slope.
None of the key GMC properties shows a significant dependence on redshift or halo mass across the tested parameter sets, which supports the robustness of our results.

However, even under the most permissive selection criteria (e.g. $n_{\min}=50,\,\alpha_{\rm max} > 100$), we are still unable to capture all star-forming gas cells within identified GMCs. On average, only about 80\% of star-forming gas is contained within identified GMCs. A similar phenomenon was reported in \citet{Guszejnov2019}, primarily in galactic centers where gas is predominantly dense and can form stars rapidly even when the virial criterion is not satisfied, although their star formation criteria differ from ours in allowing a fraction of dense but non-self-gravitating gas to form stars. In our case, the star-forming gas not captured by our GMC identification is instead located mainly in the outskirts of galaxies, where star-forming cells are less dense ($n_{\rm H}\simeq10^2\cm^{-3}$), and lack enough nearby dense gas to constitute a well-resolved GMC. We find that lowering the density threshold to $n_{\rm min} = 1\,\mathrm{cm}^{-3}$ results in $\sim 99\%$ of star formation being associated with identified GMCs, but at the cost of significantly overproducing small gas clumps with $n_{\rm H} < 10\,\mathrm{cm}^{-3}$ that are too diffuse to contribute meaningfully to star formation. Likewise, at higher resolution (e.g., in the 16x run), approximately 96\% of star formation is already associated with resolved GMCs.

\begin{table*}
    \caption{
    Median properties of GMCs identified under four representative parameter choices, specifically varying the gravitational boundedness threshold ($\alpha_{\rm max} = 10$ and $20$) and the minimum density threshold ($n_{\rm min} = 50$ and $100\,\mathrm{cm}^{-3}$). 
    The listed properties include GMC mass ($M_{\rm GMC}$), size ($R_{\rm eff}$), surface density ($\Sigma_{\rm gas}$), free-fall timescale ($t_{\rm ff}$), instantaneous SFE ($\epsilon_{\rm ff}^{\rm GMC}$), and the fitted slopes of the mass function ($\gamma$) and Larson’s relation ($b$).
    }

    \centering
    \def\arraystretch{1.2}
    \begin{tabularx}{0.8\linewidth}{lc|ccccccc}
        \hline
        $n_{\rm min}\,[\mathrm{cm}^{-3}]$ & $\alpha_{\rm max}$ & $M_{\rm GMC}\,[\msun]$ & $R_{\rm eff}\,[\mathrm{pc}]$ & $\Sigma_{\rm gas}\,[M_\odot\,\mathrm{pc}^{-2}]$ & $t_{\rm ff}\,[\mathrm{Myr}]$ & $\epsilon_{\rm ff}^{\rm GMC}\,[\%]$ & $\gamma$ & $b$ \\
        \hline \hline
        50  & 10 & $1.0\times10^5$ & 34.5 & 53.2 & 3.56 & 2.45 & -2.34 & 0.51 \\
        50  & 20 & $1.1\times10^5$ & 35.0 & 54.2 & 3.66 & 2.43 & -2.43 & 0.57 \\
        100 & 10 & $1.1\times10^5$ & 31.0 & 67.7 & 2.77 & 2.12 & -2.47 & 0.54 \\
        100 & 20 & $1.1\times10^5$ & 31.5 & 70.2 & 2.80 & 2.30 & -2.49 & 0.59 \\
        \hline
    \end{tabularx}
    \label{tab:pars}
\end{table*}

\section{Impact of resolutions on GMC properties}\label{apdx:resolution}

In the main text, we adopt the 8x resolution as our fiducial run for the analysis of GMC properties. In this section, we examine the similarities and differences in GMCs identified using the same set of \textsc{CloudPhinder} parameters across the three resolution levels available in the \thesanzoom suite. The linking length for stellar particles is set to the median smoothing length of star-forming gas in the corresponding resolutions. For each resolution level, we adopt a minimum threshold of 30 elements to consider a GMC as resolved, while it is also informative to examine how GMCs that would be considered resolved in a lower-resolution run behave when captured in a higher-resolution simulation. That is, 30 elements at a given resolution correspond to approximately 240 elements at the next higher resolution.

\begin{figure}
    \centering
    \includegraphics[width=1\linewidth]{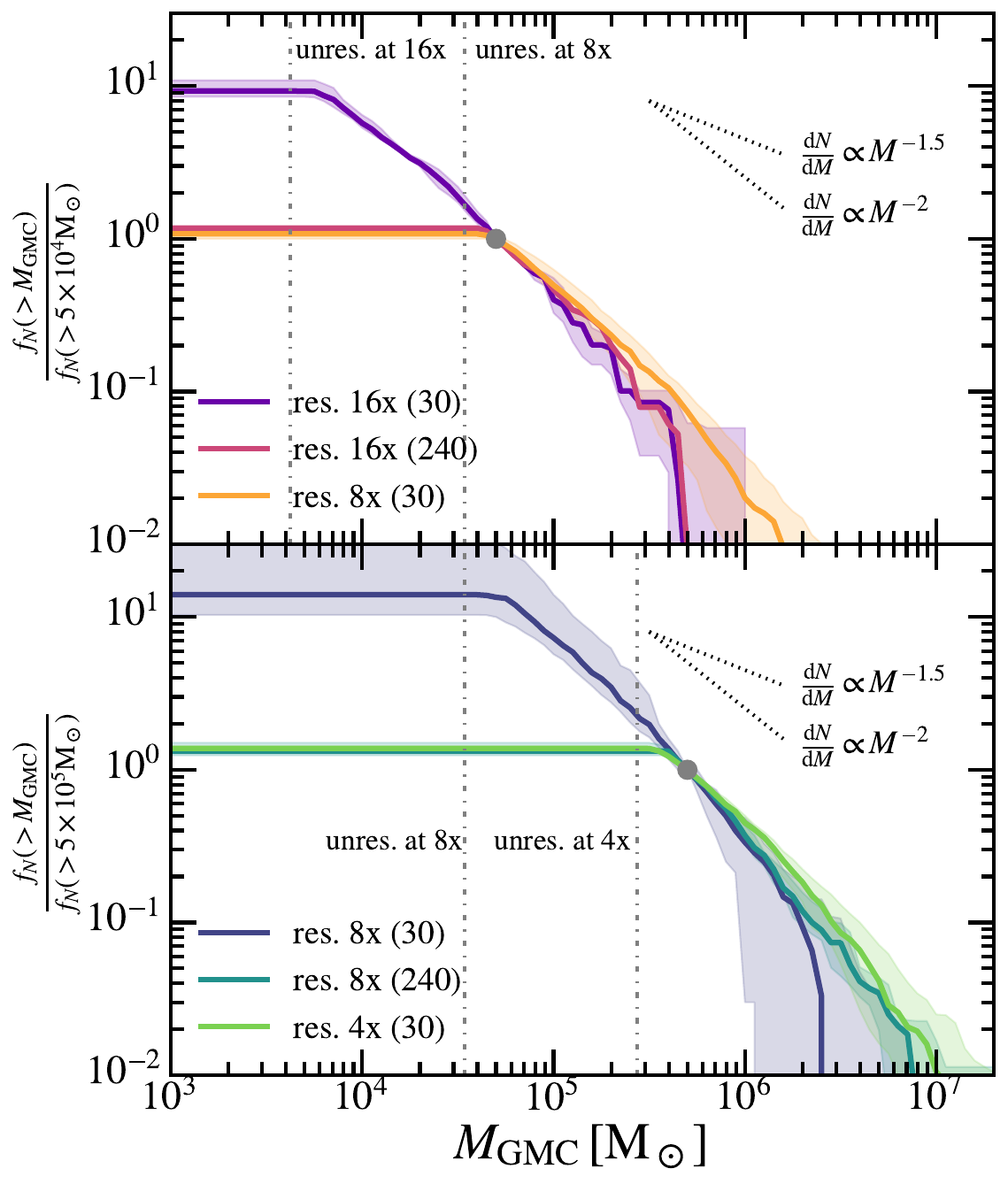}
    \caption{Similar to Fig.~\ref{fig:gmc_mass}, but a comparison across different resolution levels (4x, 8x, and 16x), with the numbers in parentheses indicating the minimum number of elements required for a GMC to be considered resolved. The mass functions are normalized by the number of GMCs that can be properly resolved in 8x ($>5 \times 10^4\,\msun$) and 4x ($>5 \times 10^5\,\msun$), respectively. The corresponding critical mass in each case is marked with gray circles, where the normalized value is by definition unity. Dashed lines indicate the minimum GMC mass that can be resolved at each resolution level, corresponding to 30 times the median baryonic mass. We also compare the properties of GMCs identified in higher-resolution runs that would be considered resolved in lower-resolution runs.}
    \label{fig:gmc_mf_res}
\end{figure}

\begin{figure}
    \centering
    \includegraphics[width=1\linewidth]{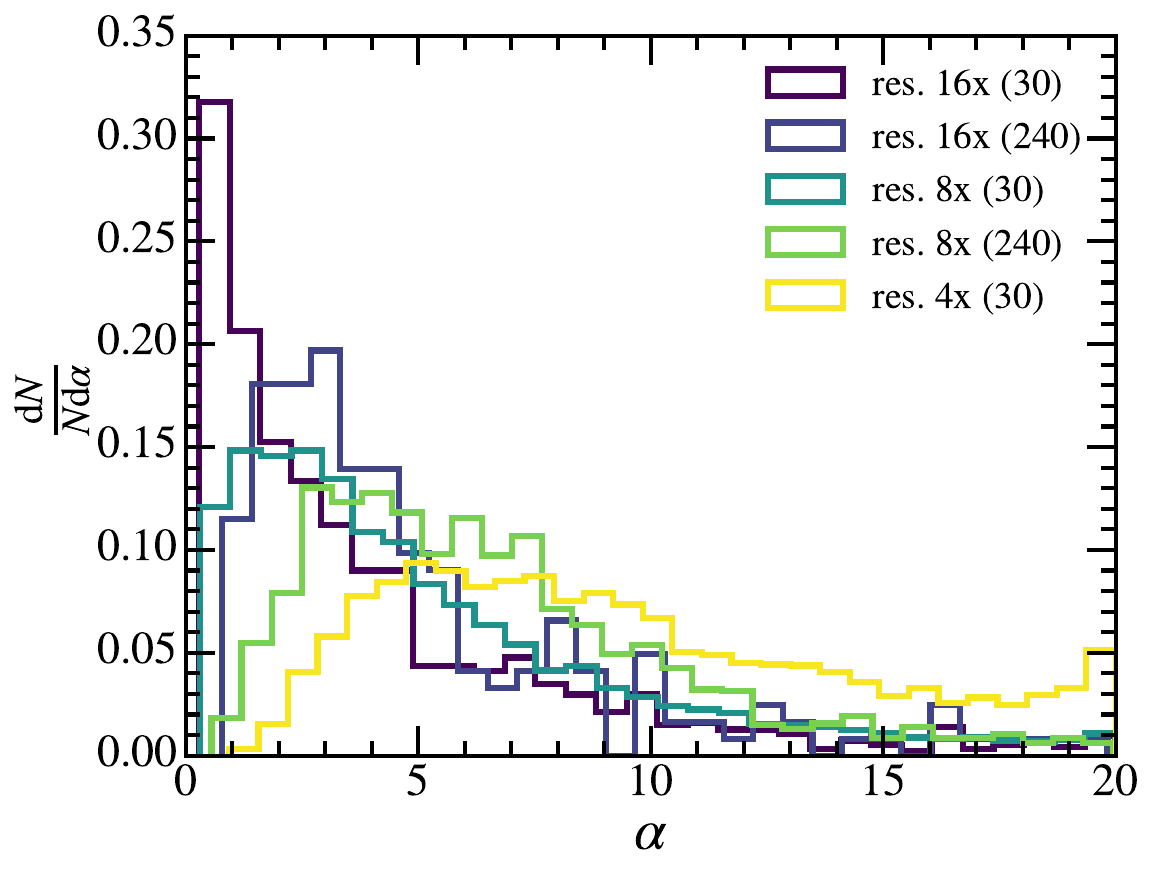}
    \caption{Distribution of GMC virial parameters across different resolution levels, with the numbers in parentheses indicating the minimum number of elements required for a GMC to be considered resolved. Identified GMCs tend to be less virialized at lower resolution. Except for the 4x run, GMC properties are largely insensitive to the maximum $\alpha$ threshold adopted, as the majority of identified GMCs have $\alpha\ll20$.}
    \label{fig:gmc_vp_res}
\end{figure}

In Fig.~\ref{fig:gmc_mf_res}, we present the GMC mass functions across different resolutions at $z \simeq 3-4$. The shape of the mass function is influenced by low-mass GMCs that may not be fully resolved in lower-resolution runs. To ensure a fair comparison, the distributions are normalized by the number of well-resolved GMCs at each resolution level; i.e. $5 \times 10^4\,\msun$ for 8x in the top panel, and $5 \times 10^5\,\msun$ for 4x in the bottom panel. We find that although more low-mass GMCs are resolved at higher resolution, the mass distribution of intermediate- and high-mass GMCs remains well converged across different resolutions, exhibiting a nearly constant slope of $-2.5$.

\begin{figure*}
    \centering
    \includegraphics[width=1\linewidth]{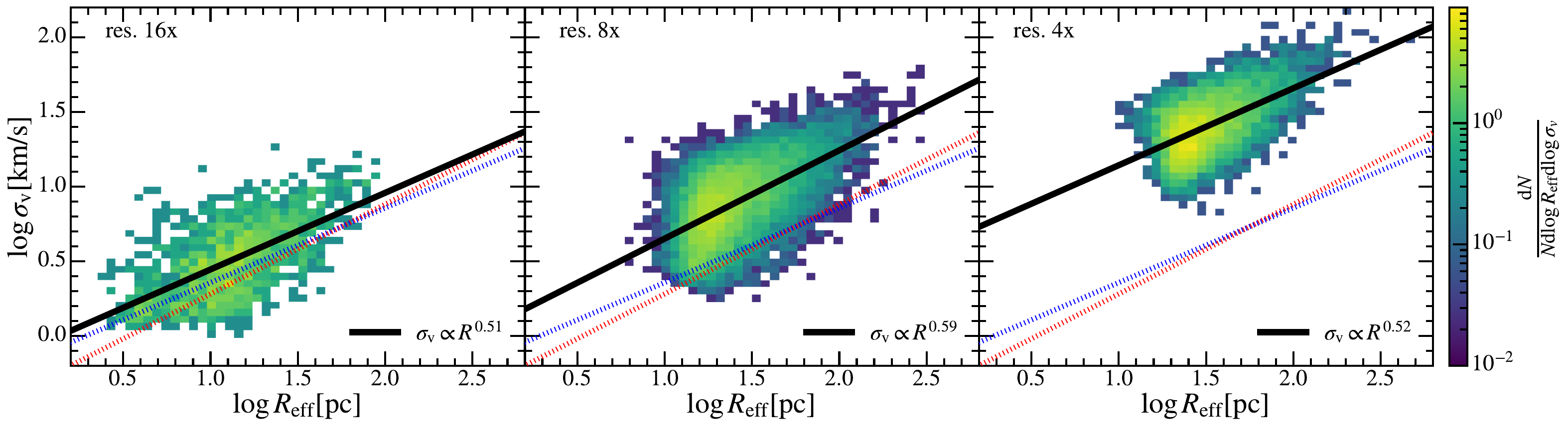}
    \caption{Similar to Fig.~\ref{fig:gmc_larson}, but a comparison between different resolutions. Both the velocity dispersion and effective radius of GMCs decrease with increasing resolution, but the slope of the scaling remains consistent across all resolution levels.}
    \label{fig:gmc_larson_res}
\end{figure*}

In Fig.~\ref{fig:gmc_vp_res}, we show the distribution of the virial parameter for GMCs identified at different resolution levels. As the minimum resolved GMC mass increases with decreasing resolution, the identified clouds tend to appear less virialized. This trend is also evident in Fig.~\ref{fig:gmc_larson_res}, where we present the Larson’s relation across the three resolutions. Both the velocity dispersion and effective radius of GMCs decrease with increasing resolution, but the slope of the scaling remains consistent across all resolutions.

\begin{figure}
    \centering
    \includegraphics[width=1\linewidth]{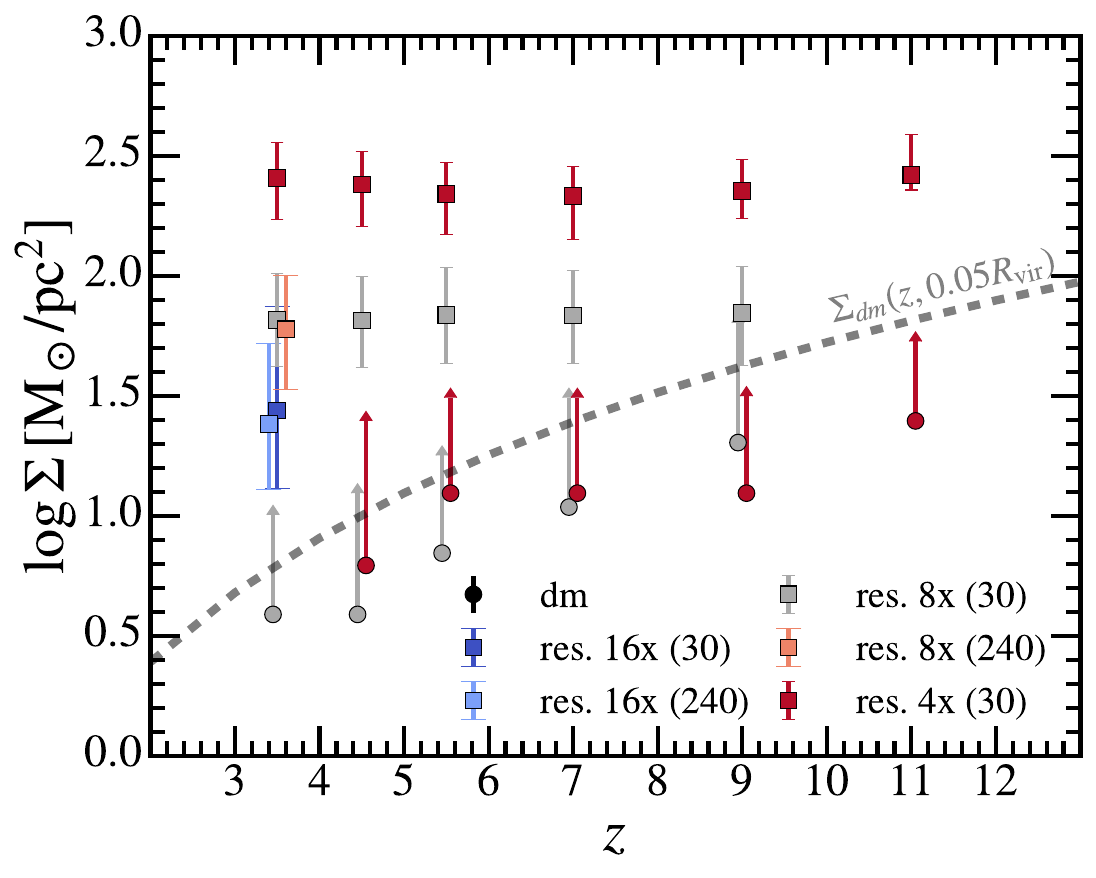}
    \caption{Similar to Fig.~\ref{fig:gmc_sigma}, but a comparison between different resolutions. Squares represent the gas component, circles represent DM, and colors indicate different resolution levels. Neither the gas surface densities of GMCs in the 4x nor the 8x runs show significant redshift evolution. However, both exhibit increasing DM surface density with redshift. The overall surface density decreases with increasing resolution, which likely reflects the non-linear response of GMCs to stellar feedback under different resolution conditions.}
    \label{fig:gmc_sigma_res}
\end{figure}

\begin{figure}
    \centering
    \includegraphics[width=1\linewidth]{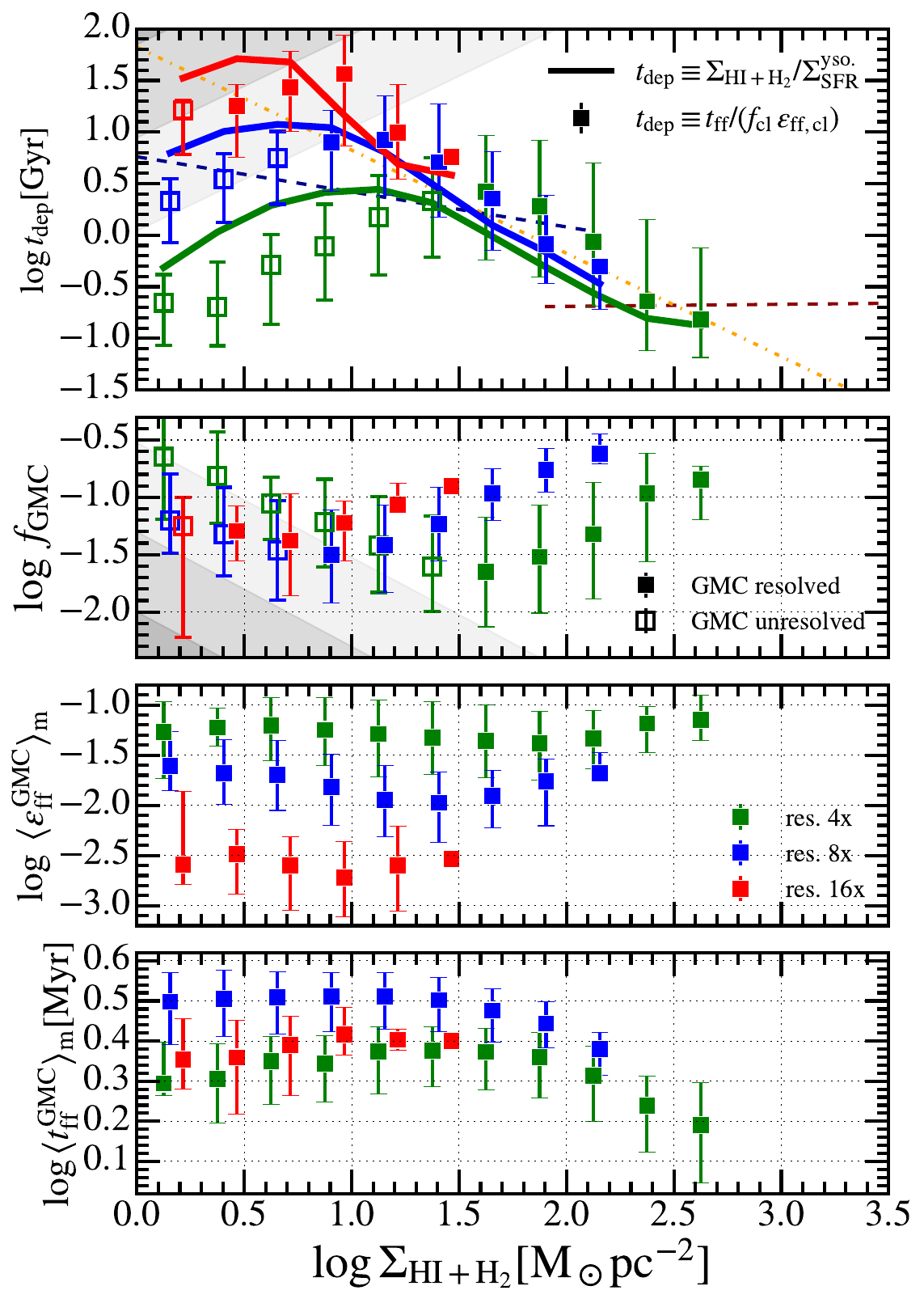}
    \caption{Similar to Fig.~\ref{fig:ks_connection}, but a comparison between different resolutions. Shaded regions indicate resolution limits (darkest for 16x, lightest for 4x). Except for 4x that slightly overestimates the depletion time due to a significant fraction of unresolved GMCs, the global depletion time can be consistently resolved from a microscopic perspective.
    At higher resolution, $\langle\epsilon_{\rm ff}^{\rm GMC}\rangle$ tends to be lower, while $f_{\rm GMC}$ increases. This is likely because gas is converted into stars more rapidly, leading to earlier disruption of GMCs by stellar feedback and consequently reducing the efficiency. Meanwhile, the increased number of short-lived GMCs results in a higher mass fraction. In addition, the non-monotonic variation in $\langle t_{\rm ff}^{\rm GMC}\rangle$ suggests a nonlinear interplay between gravity and stellar feedback.}
    \label{fig:ks_res}
\end{figure}

However, not all GMC properties exhibit monotonic trends with increasing resolution. For example, while higher resolution simulations are expected to resolve GMCs that are more virialized and have higher surface densities, we find that the measured gas surface density of GMCs actually decreases with increasing resolution (see Fig.~\ref{fig:gmc_sigma_res}). This may reflect a nonlinear interplay between stellar feedback and gravity. 

One possible explanation is that in higher-resolution simulations, where hierarchical fragmentation within GMCs can be resolved, star-forming gas collapses more rapidly into stars, leading to earlier and more spatially distributed feedback. Feedback from multiple directions can then suppress further collapse of the host GMC and promote the fragmentation of different segments.
In contrast, in lower-resolution simulations, star-forming gas is typically concentrated near the center of the GMC, and collapses on a longer free-fall timescale. This allows the cloud to contract more globally and reach higher densities before feedback becomes effective.

In Fig.~\ref{fig:ks_res}, we present the connection between global and local SFE as revealed across the three resolution levels. We note that in the 4x run, the depletion time computed from Eq.~(\ref{eq:ks}) exceeds the measured value by a factor of 2, likely due to an underestimation of $f_{\rm GMC}$ caused by a significant fraction of star formation occurring in unresolved GMCs. Overall, our central conclusion that the GMC mass fraction primarily regulates the global SFR–surface density relation holds across all resolutions, although the precise values vary.

Specifically, although it is difficult to fully control for resolution effects because we do not restrict our analysis to only the lowest mass galaxies in the 16x run, we find that the star formation efficiency tends to be lower at higher resolution. This may be because gas is converted into stars more rapidly at higher resolution, leading to earlier disruption of GMCs by stellar feedback and consequently keeping the SFE low. As a result, GMCs at higher resolution are likely more numerous but have shorter lifetimes. Interestingly, the density structure of GMCs does not evolve linearly with resolution. GMCs in the 4x run are denser than those in 8x, consistent with the trend in surface density. In contrast, in 16x, the emergence of a population of very dense gas cells ($n_{\rm H}>10^3\cm^{-3}$) lowers the overall free-fall time, reflecting a shift in the internal structure toward more compact, fragmented components.

In summary, the trends identified in the main text are largely robust against changes in resolution, although some specific numerical values vary nonlinearly. The precise ways in which resolution affects the detailed properties of GMCs remain uncertain, and this aspect has received relatively little attention in previous studies. Further investigation will be necessary to fully understand the resolution dependence of GMC characterization in numerical simulations.


\bsp	
\label{lastpage}
\end{document}